\documentclass[fleqn,usenatbib,useAMS]{mnras}

\usepackage{enumitem}

\usepackage{graphicx}
\usepackage{anysize}
\usepackage{multirow}
\usepackage{makecell} 
\usepackage{epstopdf}
\usepackage{natbib}
\defcitealias{fq12}{FGQ12}
\marginsize{2cm}{2cm}{1cm}{2cm} 

\usepackage{amsmath}
\usepackage{amssymb} 

\usepackage[T1]{fontenc}
\usepackage{ae,aecompl}
\usepackage{courier} 

\newcommand{\pcc}{cm$^{-3}$}

\newcommand{\cmss}{cm\,s$^{-2}$}
\newcommand{\kms}{km\,s$^{-1}$}

\newcommand{\kmsyr}{km\,s$^{-1}$\,yr$^{-1}$}

\newcommand{\caii}{Ca\,{\sc ii}}
\newcommand{\Ncaii}{N_{\rm CaII}}

\newcommand{\ciii}{C\,{\sc iii}}

\newcommand{\civ}{C\,{\sc iv}}

\newcommand{\hei}{He\,{\sc i}}
\newcommand{\nv}{N\,{\sc v}}

\newcommand{\siiv}{Si\,{\sc iv}}

\newcommand{\feii}{Fe\,{\sc ii}}

\newcommand{\mgii}{Mg\,{\sc ii}}

\title[Quasar Outflow Deceleration or Acceleration]{Quasar Outflow Deceleration or Acceleration: Predictions and a Search}

\author[P.\ Hall et\,al.]{P.\ B.\ Hall,$^{1}$\thanks{E-mail: phall@yorku.ca}
E.\ Weiss,$^{1}$
W.\ N.\ Brandt,$^{2,3,4}$
C.\ J.\ Mulholland$^{1}$\\
$^1$Department of Physics and Astronomy,
York University, 4700 Keele St., Toronto, ON M3J 1P3, Canada\\
$^2$Department of Astronomy \& Astrophysics, The Pennsylvania State University, University Park, PA 16802, USA\\
$^3$Institute for Gravitation and the Cosmos, The Pennsylvania State University, University Park, PA 16802, USA\\
$^4$Department of Physics, 104 Davey Lab, The Pennsylvania State University, University Park, PA 16802, USA\\
}

\begin{document}
\label{firstpage}
\maketitle

\begin{abstract}
Quasar winds can shock and sweep up ambient interstellar medium (ISM) gas, contributing to galactic quenching.  
We combine and extend past models of energy-conserving shock bubbles around quasars, investigate model implications from an observational standpoint, and test model predictions using new high-resolution spectroscopic observations of the broad absorption line quasar SDSS J030000.56+004828.0 (J0300). 
Even with constant energy input from the wind, a bubble's expansion decelerates over time as more ISM gas is swept up. 
Our new observations enable a direct search for this deceleration.
We obtain the tightest reported $3\sigma$ limit on the average rest-frame deceleration (or acceleration) of a quasar outflow: $|a|<0.1$ \kmsyr\ ($<3 \times 10^{-4}$ \cmss) in the relatively low-velocity \caii\ outflow of J0300 over 9.65 rest-frame years.
We can satisfy these limits with certain parameter choices in our model, but
the large velocity range of the \caii\ absorption in J0300 rules out the hypothesis that such gas shares the velocity of the swept-up ISM gas in a self-similar shock bubble. 
We investigate the possibility of ram-pressure acceleration of preexisting ISM clouds and conclude that the velocity range seen in \caii\ in J0300 is potentially consistent with such an explanation.
The \caii-absorbing gas clouds in J0300 have been inferred to have high densities by Choi et al., in which case they can only have been accelerated to their current speeds if they were originally at least an order of magnitude less dense than they are today.
\end{abstract}

\begin{keywords}
quasars: absorption lines; shock waves; quasars: general; galaxies: active; galaxies: evolution, quasars: individual: SDSS J030000.56+004828.0
\end{keywords}

\section{Introduction}\label{intro}

Active galactic nuclei (AGN) are actively accreting supermassive black holes at the centres of some galaxies. Quasars are extreme AGN with immensely luminous accretion disks that are visible over very large distances. 
These accretion disks can produce winds that send material flowing outwards \citep[e.g.,][]{mcgv,pk04,dydadavisproga23}.

It is thought that quasar outflows can travel outwards through the host galaxy, shocking and sweeping up ambient interstellar medium (ISM) gas into a large-scale outflow (e.g., \citealt{2018ApJ...857...60A}, 
\citealt{hopkins15},
\citealt{2015ARA&A..53..115K},
\citealt{2012ApJ...745L..34Z}).
These outflows may form a ``shock bubble" consisting of an inner freely flowing wind, a region of hot shocked wind at lower velocities, and an outer shell of shocked, swept-up ISM gas at still lower velocities, as illustrated in Figure \ref{fig:bubble}.  
Such bubbles may contribute to galactic quenching -- in which galaxies are transformed from blue and star-forming to ``red and dead" -- by expelling star-forming ISM gas from the quasar host galaxy \citep[e.g.,][]{2005Natur.433..604D,2011MNRAS.415L...6K,2021ApJ...919..122Vayner}, or by preventing the inflow of external star-forming gas (``strangulation"; e.g., \citealt{1980ApJ...237..692L}; \citealt{Peng+2015}). The collection of such AGN-related quenching processes is known as AGN feedback \citep{Morganti2017}.

\citet[][hereafter FGQ12]{fq12} studied these bubbles from a largely theoretical perspective.
In this work, we connect their model and those of \cite{1977ApJ...218..377W} and \cite{1992ApJ...388..103K} (hereafter W77 and KM92)
and study the implications of the combined model from an observational perspective. 

One observational consequence of these bubbles may be seen in some broad absorption line (BAL) quasars
\citep[e.g.,][]{allenbal,2019MNRAS.483.1808Hamann}.  
In BAL quasars, high-ionization absorbing gas (such as \civ, for example) generally extends to higher velocities than low-ionization gas (such as \mgii, for example), when low-ionization gas is seen. 
In the radio-quiet iron low-ionization `FeLoBAL' quasar SDSS J030000.56+004828.0 \citep[][hereafter J0300]{sb2} the UV continuum source is fully covered by an outflow seen in many ions and extending to at least $v$=10850 \kms.  However, strong absorption in \caii\ --- which must be shielded by a hydrogen ionizing front --- is seen only at the lowest outflow velocities.
Thus, at least in J0300 and a handful of other FeLoBAL quasars with strong, low-velocity \caii\ absorption \citep{felobal1}, the higher-velocity, high-ionization absorbing gas is located closer to the quasar than the lower-velocity, low-ionization gas. 
(Otherwise, higher-velocity gas would also be shielded and show only low-ionization absorption.)
In addition, for BAL quasars with more than one absorption system, the spectral fits of \citet{felobal1} yield a statistical trend for lower-velocity absorbers to be located at larger distances from the quasar (their Figure 12).

The above conclusions raise a question: how can faster-moving outflowing gas close to the quasar co-exist with slower-moving outflowing gas farther from the quasar?
It may be that the lower-velocity absorbing gas is part of a wind launched farther from the quasar than the higher-velocity gas is (e.g., Fig.\ 3 of \citealt{pbhmhd}), with nonzero transverse velocity ensuring that the gas streamlines do not intersect 
\citep{aea99,sb2,2019A&A...630A..94G}.
It is also worth investigating in what circumstances it might be plausible for some or all of the lower-velocity absorbing gas to arise from swept-up ISM gas (or from cooling shocked wind gas) in a shock bubble.
A quasar wind slowing down as it sweeps up gas naturally results in higher-velocity gas being located closer to the quasar than lower-velocity gas. 
Note that we do not discuss herein the detailed origin of the large velocity range seen in high-velocity gas along the line of sight, other than to ensure that our model wind has a terminal velocity exceeding the highest observed outflow velocity in this object.

In the shock-bubble scenario, the absorbing gas may be seen to decelerate over time as more ISM is swept up by the bubble.  
Comparing observed deceleration measurements or limits to model predictions might constrain quasar and ISM parameters as well as the age of the shock bubble. 

In this paper we make such predictions and compare them to observations of SDSS J030000.56+004828.0 taken almost a decade apart in the rest frame. 
We begin in \S~\ref{sec:mod} with an analysis of the model used to describe the shocked gas, and what this model implies about the velocity of the gas and column densities around AGN.
In \S~\ref{sec:j0300} we report deceleration and acceleration limits from comparison of old and new spectroscopy of SDSS J030000.56+004828.
In \S~\ref{sec:discuss} we compare our results to the literature and discuss them in the context of possible explanations for the origin of the \caii-absorbing gas in J0300.
We summarize our conclusions in \S~\ref{sec:conclude}.

\section{Theoretical Model} \label{sec:mod}

\begin{figure}
    \includegraphics[width=8.5
    cm]{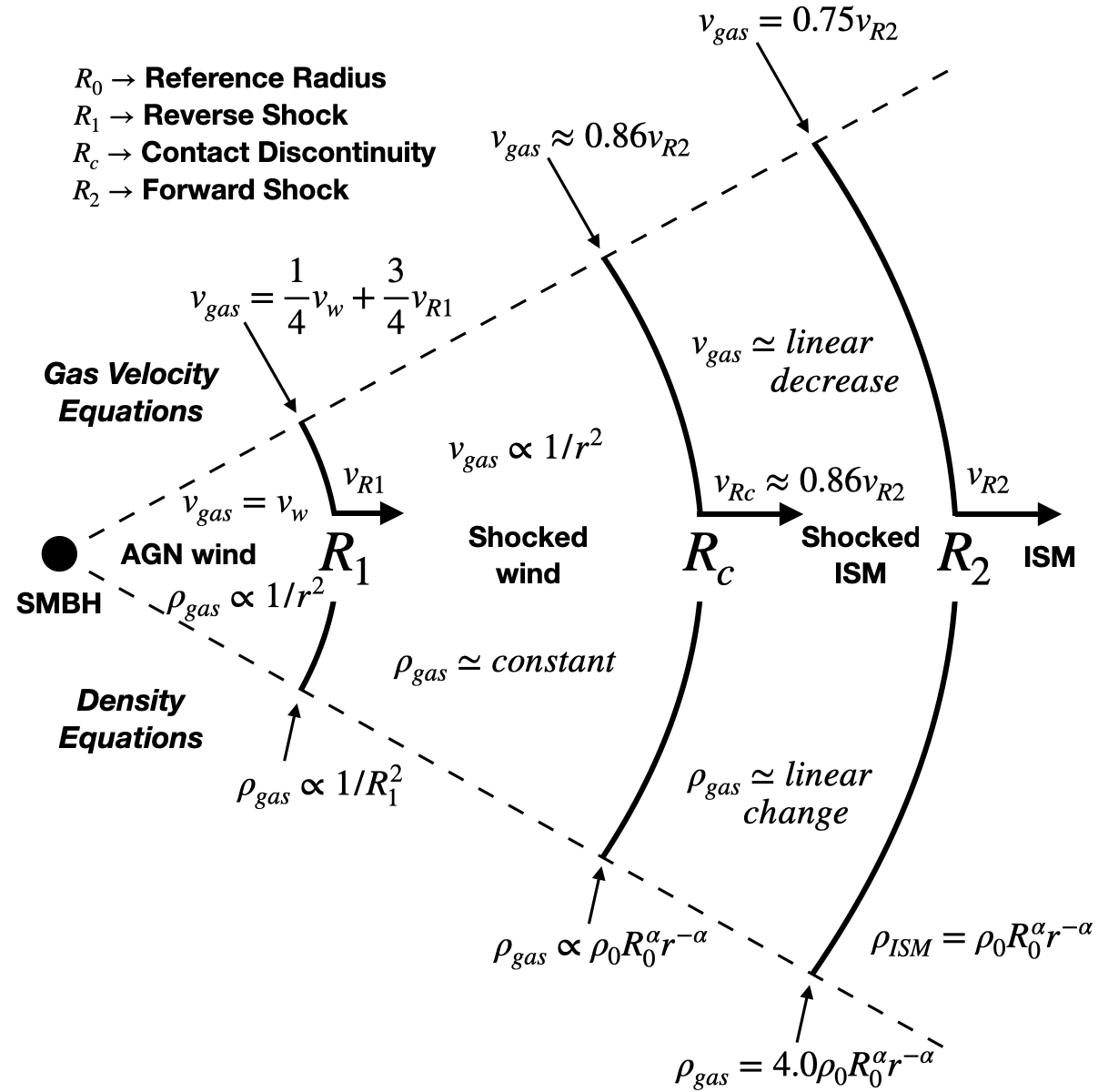}
    \caption{Diagram of different regions of shocked gas within the AGN outflow bubbles considered in this work, including the relevant gas velocity and density relations adopted within each region. The wind from the quasar starts with initial velocity $v_{in}$ and the outer shock front sweeps into the ISM (which has density $\rho_0r_*^{-\alpha}$) at a velocity of $v_{R2}(t)$. 
    The gas velocity drops at $R_1$, is continuous across the contact discontinuity $R_c$, and reaches a minimum of $v_{gas}(t)=0.75v_{R2}(t)$ just inside $R_2$.
    }
    \label{fig:bubble}
\end{figure}

We assume that a high-velocity wind from a quasar accretion disk produces a spherically symmetric shock bubble,
a portion of which is illustrated in Figure \ref{fig:bubble}. 
We assume the wind is accelerated to $v=v_{in}$ at radii much smaller than those at which it might plausibly decelerate, which may be true for gas accelerated off an accretion disk but not for gas accelerated off an obscuring torus \citep{2022SciA....8.3291H}.
Although winds and shock bubbles will deviate from spherical symmetry in reality (see, e.g., \citealt{2018MNRAS.476.2288Hartwig}), in this work we treat the idealized spherically symmetric case for simplicity.
Also for simplicity, we do not consider the possible effects of Rayleigh-Taylor and other instabilities \citep{fqm12,2014MNRAS.439..400Zubovas}.
While we do not directly simulate the sweeping up of preexisting interstellar clouds (\citealt{2020MNRAS.491.4325Z}), we do discuss that possibility in \S~\ref{sec:ismaccel}.
We do not consider the effects of magnetic fields which would be present in the ISM \citep{cb23} or 
if the wind was magnetically driven \citep{ebs92}.
We also do not consider radiative acceleration of gas in the outflow other than its possible role in accelerating the wind to $v=v_{in}$.
We assume that all gas is ideal (adiabatic index $\gamma = 5/3$) and that all shocks are strong shocks (Mach number $\mathcal{M} \gg 1$).

When the wind shocks, the high wind speeds ($v\sim 0.1c$) yield high temperatures. However, protons are heated to much higher temperatures than electrons and under certain conditions this temperature difference may result in inefficient cooling (FGQ12). In this case, the cooling time of the shocked quasar wind (which contains the bulk of the bubble's thermal energy) increases over time; when the cooling time exceeds the age of the outflow, the bubble is described as energy-conserving or adiabatic. 

For many reasonable values of the relevant physical parameters, the shock bubble is initially (or quickly becomes) energy-conserving and expands adiabatically. 
Bubbles that are initially non-adiabatic begin in a radiative (momentum conserving) phase, followed by a partially radiative bubble (PRB) phase. For the parameters explored in this paper, bubbles become energy-conserving after $<25$ yr.\footnote{The exceptions are cases of slow winds (3000 \kms) in an ISM with a steep density gradient ($\rho\propto 1/r$ or steeper); such bubbles become adiabatic after $10^3 - 10^5$ yr. We exclude such parameter combinations from consideration in this paper.}

Our model is intended to provide a first approximation of what signatures of deceleration might be seen from shock bubbles surrounding quasars. 
Our interest is therefore in calculating the radial extents, velocity ranges, and densities of the gas in the shocked wind and the shocked ISM as a function of time.  We will use those quantities and some simple assumptions to calculate upper limits on observable ionic column densities and predicted decelerations as a function of velocity in outflows of a range of ages.

For a given set of parameters including the time $t$ since the start of the outflow, we use the equations in the following subsections to calculate various physical parameters at radii within the different regions of the shock bubble seen in Figure \ref{fig:bubble}.
With regard to velocities, we work in a frame centered on the quasar such that along our line of sight all outflow velocities from the quasar are positive, acceleration of an outflow away from a quasar has a positive sign, and deceleration of an outflow has a negative sign. 

After specifying our initial parameters in \S \ref{sec:mod param}, we review equations describing the time evolution of shock features from several references in \S \ref{sec:sf}.

We then calculate the gas velocity, density, acceleration, and column density in \S \ref{sec:vgas}, \ref{sec:rho}, \ref{sec:accel} and \ref{sec:Nv}, respectively.

We discuss valid combinations of parameters in \S \ref{sec:valid}.

Finally, we present the results of varying our default parameters in \S \ref{sec:vary}.

\subsection{Model Parameters} \label{sec:mod param}

In Table \ref{t_params} we list the parameters of interest in the model, along with brief descriptions and default values.
See \S~\ref{sec:vary} for the values used when we vary those parameters.

The mass outflow rate into the quasar wind is determined by the wind velocity $v_{in}$, the AGN luminosity $L_{AGN}$, and the wind optical depth $\tau_{in}$: 
\begin{equation}\label{eq:dotM}
    \dot{M}_w=\tau_{in} L_{AGN} /cv_{in}. 
\end{equation}
This equation derives from assuming that the wind's momentum flux $\dot{M}_wv_{in}$ is a factor of $\tau_{in}$ times the momentum flux $L_{AGN}/c$ available in the photons radiated by the AGN.
The mass outflow rate
in turn determines the outflow's kinetic luminosity, which appears explicitly in the derivations of the W77 and KM92 models:
\begin{equation}\label{eq:Lin}
    L_{in} = \frac{1}{2}\dot{M}_wv_{in}^2.
\end{equation}

Another useful way of writing the wind optical depth $\tau_{in}$ can be found as follows. We define the AGN's dimensionless radiative efficiency $\eta_r$ as the ratio between its luminosity and its mass-energy accretion rate:
\begin{equation}\label{eq:eta}
    \eta_r\equiv L_{\rm AGN}/\dot{M}_{acc}c^2,
\end{equation}
where $\dot{M}_{acc}$ is the mass accretion rate into the black hole. We can write the AGN luminosity in several ways:
\begin{equation}
    L_{\rm AGN} \equiv f_{Edd}L_{Edd} \equiv f_{Edd}\eta_r \dot{M}_{Edd}c^2 = \eta_r \dot{M}_{acc}c^2
\end{equation}
where $\dot{M}_{Edd}$ is the Eddington mass accretion rate, $L_{Edd}=1.47\times 10^{38}(M_{BH}/M_\odot)$ erg~s$^{-1}$ is the Eddington luminosity for the AGN's black hole mass $M_{BH}$, and $f_{Edd}\equiv \dot{M}_{acc}/\dot{M}_{Edd}$ is the Eddington fraction.
We can also write $L_{AGN}$ in terms of the mass outflow rate $\dot{M}_w$ by rearranging Equation \ref{eq:dotM} to yield $L_{AGN}=\dot{M}_wv_{in}c/\tau_{in}$.  Equating the  expressions for $L_{AGN}$ in terms of $\dot{M}_{acc}$ and $\dot{M}_w$ and solving for $\tau_{in}$, we find:
\begin{equation}\label{eq:tau}
    \tau_{in}=\eta_r^{-1}(v_{in}/c)(\dot{M}_w/\dot{M}_{acc}).
\end{equation}
We adopt $\eta_r=0.175$, approximately the maximum value of $\eta_r$ predicted for disks dominated by magnetic turbulence \citep{Shapiro2005}.
For our default parameter values of $v_{in}$, $L_{AGN}$ and $\tau_{in}$ given in Table \ref{t_params}, equations \ref{eq:dotM} and \ref{eq:tau} yield values $\dot{M}_w=2.7$ $M_\odot$ yr$^{-1}$ and $\dot{M}_{acc}=1$ $M_\odot$ yr$^{-1}$, meaning that we assume 2.7 times more mass is ejected in a wind from the disk and/or torus than accretes onto the black hole \citep{2012MNRAS.426..656S}.

We consider times from $10^3 - 10^6$ yr. However, 
we make note of parameter combinations where the shocked ISM 
cools faster than the flow timescales and therefore 
consists of a thin shell.
The shocked ISM cooling timescale decreases/increases with time below/above a threshold $\alpha_{\rm shell}$ (= 1.4 for free-free cooling in a one-temperature plasma).
The time for which the cooling and flow timescales are equal is denoted by $t_{\rm shell}$. 
If $\alpha < \alpha_{\rm shell}$ 
(which is the case for almost all values of $\alpha$ we consider, including our default $\alpha=0$),
at $t = t_{\rm shell}$ the shocked ISM region begins to collapse (increasing $R_c$ and decreasing $R_2$; e.g., \citealt{1975A&A....43..323F}).
If $\alpha > \alpha_{\rm shell}$, at $t = t_{\rm shell}$ the newly shocked ISM gas begins to form an adiabatic region outside a thin shell of cooled gas at $r>R_c$.
Our default parameters from Table \ref{t_params} yield
$t_{\rm shell} \simeq 10^{4.0}$ yr.\footnote{We calculate $t_{\rm shell}$ following KM92 \S~5.2. We use a cooling time $t_{\rm cool} = x_tkT/n_H\Lambda$ for the shocked ambient region (KM92 \S~2.2), but with
a free-free cooling function $\Lambda$ more appropriate for our higher shock velocities $v_{R2} \gtrsim 1000$ \kms. We adopt $\Lambda = \Lambda_1 T^{1/2}$, with $\Lambda_1 = 1.4 \times 10^{-26}$ erg s$^{-1}$ cm$^{3}$ K$^{-1/2}$, as used in \cite{2020MNRAS.491.4325Z}.  
We use $x_t = 1.1$ particles per hydrogen nucleus for a two-temperature plasma (i.e. neglecting electron energy due to differential shock heating), but note that our value of $\Lambda_1$ reflects the assumption of a one-temperature plasma. This serves to lower the shell formation time, giving us a conservative estimate of $t_{\rm shell}$. Two-temperature effects (e.g., FGQ12, \citealt{Mayer2007}) may significantly increase $t_{\rm shell}$ and/or decrease $\alpha_{\rm shell}$.
}

\begin{table}
\centering 
\begin{tabular}{c c c} 
\hline\hline  
\multicolumn{3}{c}{Quasar Wind Bubble Model Parameters}\\
\hline\hline  
{} & \textbf{Time Epochs} & {}\\
\hline\hline
\textit{Parameter} & \textit{Description} & \textit{Default Values}\\  
\hline 
$\log t$ & $\log$(time in yr) & $3,3.5,4,5,6$\\ 
\hline\hline  
{} & \makecell[c]{\textbf{ISM Density} \\ $\rho_{ISM}(r)=\rho_0(r/R_0)^{-\alpha}$} & {}\\
\hline\hline
\textit{Parameter} & \textit{Description} & \textit{Default Value}\\  
\hline
$\alpha$ & power law exponent & 0\\
$R_0$ & reference radius & 100 pc\\
$\rho_0 = \mu m_p n_0$ & mass density at $R_0$ & See below\\
$n_0$ & $\text{H}$ nucleus \# density at $R_0$ & 100 cm$^{-3}$\\
$\mu$ & mean atomic mass / particle & 1.4\\
$m_p$ & proton mass & constant\\
\hline\hline  
{} & \makecell[c]{\textbf{Momentum Flux} \\ $\dot{M}_wv_{in} = \tau_{in}L_{AGN}/c$} & {}\\
\hline\hline
\textit{Parameter} & \textit{Description} & \textit{Default Value}\\  
\hline
$v_{in}$ & quasar wind outflow velocity & 20,000 \kms\\
$L_{AGN}$ & AGN bolometric luminosity & $10^{46}$ erg s$^{-1}$\\
$\tau_{in}$ & quasar wind optical depth & 1\\
\hline 
\end{tabular}
\caption{Model parameters for the quasar wind bubble model.}
    \label{t_params}
\end{table}

\subsection{Shock Features} \label{sec:sf}

Our model builds on the work of W77, KM92, and FGQ12. 
KM92 studied the general case of a wind with a power-law energy injection rate $L_{in}\propto t^{\eta_{in}-1}$ expanding into an ISM with density profile $\rho=\rho_0 (R/R_0)^{-\alpha}$ (note that KM92 use $k_\rho$ instead of $\alpha$).
We consider only the case of constant energy injection $(\eta_{in}=1)$.

The bubble consists of a freely expanding wind region which abruptly slows at an inner shock at $r=R_1$, a very hot shocked wind region between $R_1$ and a contact discontinuity at $r=R_c$, and a hot shocked ISM region between $R_c$ and an outer shock at $r=R_2$.
These three radii are all increasing functions of time whose derivatives (i.e., the velocities of these boundaries within the bubble) decrease over time. That is, the shock bubble expands outward into the ISM at a rate which continually decelerates. 

Because these radii and their corresponding velocities appear explicitly in the equations derived from the model, we generally refer to them explicitly rather than as functions of time; i.e., we write $R_2$ and $v_{R2}$ instead of $R_2(t)$ and $v_{R2}(t)$.

Regarding our notation: we follow W77's convention using $R_1$, $R_c$ and $R_2$ for the boundary radii and FGQ12's convention in our use of the terms $\alpha$, $A_E$ and $\beta_E$ (see below), where the subscript $E$ represents the energy-conserving aspect of the bubble. In particular, $\alpha$ and $\beta_E$ replace KM92's $k_\rho$ and $\eta$.

\subsection{Outer shock} \label{sec:R2}

Under the above assumptions, the radius of the outer shock is (KM92 Eq.\ 3.1):
        \begin{equation} \label{eq:R2}
            R_2=A_Et^{\beta_E}
        \end{equation}
with
\begin{equation} \label{eq:AE1}
            A_E=\left[ \frac{(3-\alpha)\Gamma_r\xi L_{AGN}\tau_{in}v_{in}}{6c\rho_0 R_0^{\alpha}} \right]^{\frac{1}{5-\alpha}}
\end{equation}
and
\begin{equation} \label{eq:BE1}
            \beta_E=\frac{3}{5-\alpha}
\end{equation}
and in which $\Gamma_r$ is defined as the fraction of the injected energy still in the bubble and $\xi$ as a dimensionless numerical factor $\leq 0.55$, with their product given by KM92 Eq.\ 3.10: 
    \begin{equation} \label{eq:Grxi}
\Gamma_r\xi \equiv \frac{2(5-\alpha)^3}{3\pi\lambda_c^3(11-\alpha)(7-2\alpha)}
    \end{equation}
in which $\lambda_c$ is the ratio between the radius of the contact discontinuity and the radius of the outer shock (KM92 Eq.\ 3.7):
    \begin{equation} \label{eq:lamc}
\lambda_c \equiv \frac{(123-8\alpha)}{(143-8\alpha)}.
    \end{equation}
Dimensional analysis can be used to show that Eq.\ 1 has the correct units for all $\alpha<5$.
In the range $0\leq \alpha \leq 1.5$, we have $0.860>\lambda_{c}>0.847$ (see KM92 Table 3) and $0.541>\Gamma_{rad}\xi>0.394$.\footnote{Equation \ref{eq:Grxi} for $\Gamma_r\xi$ is KM92 Eq.\ 3.10 with $\eta_{in}=1$, $\eta=\beta_E$, and ratios of specific heats in the shocked wind and in the shocked ambient medium $\gamma_{sw}=\gamma_{sa}=5/3$.
Equation \ref{eq:lamc} for $\lambda_c$ comes from KM92 Eq.\ B8b using our assumed $\eta_{in}=1$.
Note that FGQ12 approximated $\Gamma_r\xi\simeq(5-\alpha)^2/12\pi$, an overestimate of a factor of $\sim$1.22, which led FGQ12 to overestimate $A_E$ by a factor of $\sim$1.04.}
Note that we consider only adiabatic shock bubbles with $\Gamma_r\equiv 1$.

The velocity of the outer shock (called $v_s$ by FGQ12) is $v_{R2}\equiv\dot{R}_2=\beta_E R_2/t$:
\begin{equation} \label{eq:vR2}
    v_{R2} = \frac{3}{5-\alpha} \left( \frac{(3-\alpha)\Gamma_r\xi L_{\rm AGN}\tau_{in}v_{in}}{6c\rho_0 R_0^\alpha} \right) ^\frac{1}{5-\alpha} t^\frac{\alpha-2}{5-\alpha}
\end{equation}
which can be written explicitly in terms of $R_2$ as:
\begin{equation} \label{eq:KMvR2}
    v_{R2} = \left[ \frac{9(3-\alpha)\Gamma_r\xi L_{\rm AGN}\tau_{in}v_{in}}{2(5-\alpha)^3c \rho_0 R_0^{\alpha}} \right]^{\frac{1}{3}} R_2^{\frac{\alpha-2}{3}}.
\end{equation}

\subsection{Contact discontinuity} \label{sec:Rc}

The contact discontinuity is located at radius
\begin{equation}
    R_c(t)=\lambda_c R_2(t)
\end{equation}
and the velocity of the contact discontinuity is simply
\begin{equation}
    v_{Rc}(t) = \lambda_c v_{R2}(t)
\end{equation}
where $\lambda_c$ is defined in Eq.\ \ref{eq:lamc}.

At very early times, before the mass in the wind equals the swept-up ambient medium mass, $R_c$ expands more quickly than given above (KM92 Eq.\ 5.1a and 5.1b). We account for this simply by setting $R_c=R_1$ and $v_{Rc}=v_{R1}$ if $R_c<R_1$ at any early time.

\subsection{Inner shock} \label{sec:R1}

The radius of the inner shock ($R_1$) at all stages of an adiabatic shock bubble's expansion can be found by setting the ram pressure of the unshocked wind equal to the thermal pressure of the shocked wind (e.g., W77 Eq.\ 55), which results in a cubic equation for $R_1$ (KM92 Eq.\ 5.4).

We follow KM92 Eq.\ 2.4 and define the fiducial radius $R_f$ as the radius at which the wind density $\dot{M}_w/4\pi R_f^2 v_{in}$ equals the average ambient medium density within $R_f$, $3\rho_0R_0^\alpha/(3-\alpha)R_f^\alpha$.  Substituting $\dot{M}_w=2L_{in}/v_{in}^2$ and solving for $R_f$:
\begin{equation} \label{KM:Rf}
R_f = \left[ \frac{(3-\alpha) L_{in}}{6\pi\rho_0R_0^{\alpha}v_{in}^3} \right]^\frac{1}{2 - \alpha}
= \left[ \frac{(3-\alpha) L_{\rm AGN} \tau_{in}}{12c\pi\rho_0R_0^{\alpha}v_{in}^2} \right]^\frac{1}{2 - \alpha}.
\end{equation}
Note that this equation is not valid for $\alpha=2$. The case of $\alpha=2$ is discussed in KM92 Appendix A, but we limit our range of study to $0\leq \alpha \leq 1.5$ (see \S~\ref{sec:valid}).
 
\begin{figure}
        \includegraphics[width=8
        cm]{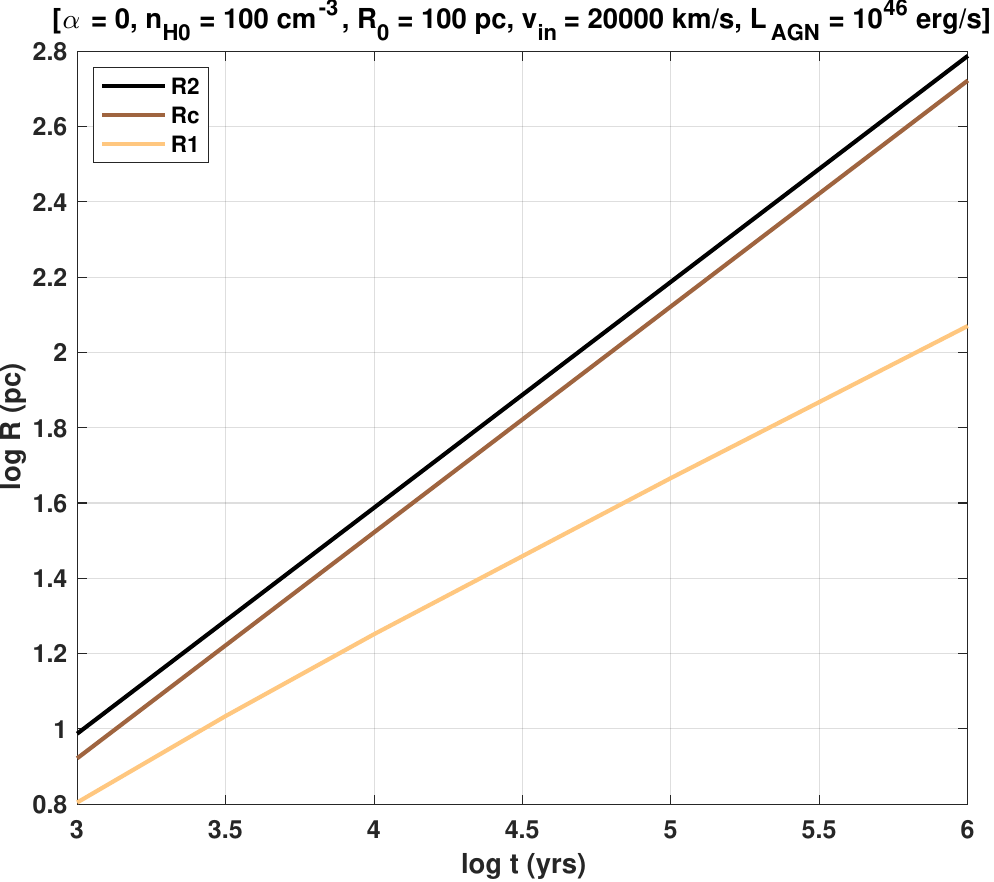}
            \caption{Radii of the different shock features relative to the quasar vs.\ time. 
            Note that our adiabatic assumption will break down for swept-up ISM gas at $t \gtrsim t_{\rm shell}$, when that gas will collapse and $R_c$ and $R_2$ will converge to an intermediate value. For our parameters, we have a conservative lower limit of $t_{\rm shell} \geq 10^{4.0}$ yr (\S\,\ref{sec:mod param}).
            }
        \label{fig:rvst}
    \end{figure}

The solution for $R_1$ can be written in terms of $R_2$, $R_f$, and $\alpha$ (KM92 Eq.\ 5.4):
\begin{equation} \label{KM:R1}
    R_1=\frac{\lambda_{c0} R_2}{[1+\Upsilon (R_2/R_f)^{(1-\alpha/2)}]^{1/3}} 
\end{equation}
where 
\begin{equation} \label{upsilon}
    \Upsilon \equiv \frac{4}{3} \left(\frac{15}{16}\right)^{15/4} \frac{(7-2\alpha)^{1/2}}{11-\alpha} \left(\frac{\lambda_{c0}}{\lambda_c}\right)^3
\end{equation}
and $\lambda_{c0}$ from KM92 Eq.\ 5.1a simplifies in our case to
\begin{equation} \label{eq:lamc0}
    \lambda_{c0} = \frac{215-54\alpha}{235-54\alpha}.
\end{equation}
Note that $R_1$ increases with time less quickly than either $R_c$ or $R_2$ (Figure \ref{fig:rvst}).

In the range $0\leq \alpha \leq 1.5$, we have $0.303>\Upsilon>0.239$ and $0.915>\lambda_{c0}>0.870$.

The inner shock velocity $v_{R1}$ is the time derivative of $R_1$:
\begin{equation} \label{eq:vR1}
    v_{R1} = \lambda_{c0} \left(\frac{1+[(\alpha+4)/6]\Upsilon(R_2/R_f)^{(1-\alpha/2)}}{[1+\Upsilon(R_2/R_f)^{(1-\alpha/2)}]^{4/3}} \right) v_{R2}. 
\end{equation}

\subsection{Gas velocity equations} \label{sec:vgas}

Our equations for the velocity of the gas within the different regions are presented below and then explained.
        \begin{equation} \label{eq:vgas}
        \begin{aligned}
            v_{gas}(r<R_1)&=v_{in}\\
            v_{gas}(r=R_1)&=0.25v_{in} + 0.75v_{R1}\\
            v_{gas}(R_1<r<R_c)&=\left[ \frac{11-\alpha}{5(5-\alpha)} \right] \frac{R_c^3}{r^2t} + \left( \frac{\alpha +4}{5(5-\alpha)} \right) \frac{r}{t}\\
            v_{gas}(r=R_c)&=\lambda_c v_{R2}=v_{Rc}\\
            v_{gas}(R_c<r<R_2)&=
            \left[C_{v0}+2C_{v1}\frac{r}{R_2}
            +C_{v2}\frac{r^2}{R_2^2}\right]
            \frac{3}{4}v_{R2}\\
            v_{gas}(r=R_2)&=0.75v_{R2} \\
            v_{gas}(r>R_2)&=0
        \end{aligned}
        \end{equation}
with
        \begin{equation} \label{eq:vgasC_vx}
        \begin{aligned}
            C_{v0}&=\dfrac{90(1-\lambda_c) + (33-8\alpha)}{60(1-\lambda_c)}\\
            C_{v1}&=-\left[\dfrac{15(1-\lambda_c) + (33-8\alpha)}{60(1-\lambda_c)}\right]\\
            C_{v2}&=\dfrac{33-8\alpha}{60(1-\lambda_c)}          
        \end{aligned}
        \end{equation}
    
\begin{figure}
        \begin{center}
        \includegraphics[width=8 cm]{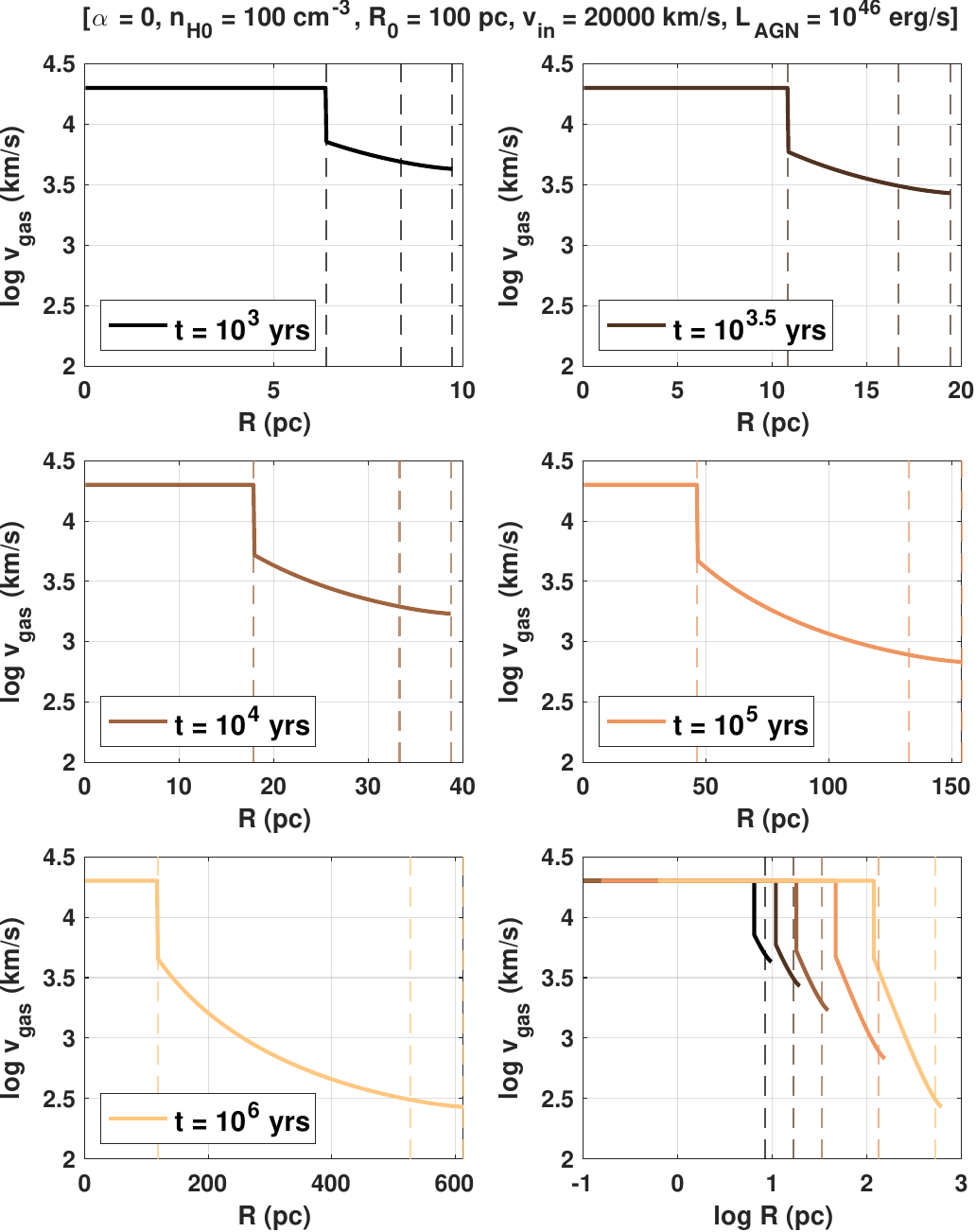}
            \caption{Velocity of the outflowing gas vs.\ distance from the quasar, with the different regions of the shock bubble separated by dashed lines corresponding to $R_1$, $R_c$, and $R_2$ seen in Figure \ref{fig:bubble}. Later epochs in time are shown in each subplot with all epochs shown in the last subplot. All subplots have the same log(velocity) scale, but different linear distance scales (or log(distance) in the last subplot). An initial quasar wind velocity of 20,000 \kms\ and an ISM density of 100 cm$^{-3}$ is used for all epochs. As the age of the outflow increases, so do the radii of the expanding shock bubble, and the velocity of the gas within decreases.
            Note that our adiabatic assumption will break down for swept-up ISM gas at $t \gtrsim t_{\rm shell}$, when $R_c$ and $R_2$ will converge to an intermediate value and the gas velocity will increase at $R_c$ and decrease near $R_2$ (\S\ \ref{sec:vrange}).
            For our parameters, we have a conservative lower limit of $t_{\rm shell} \geq 10^{4.0}$ yr (\S\,\ref{sec:mod param}).
            }
        \label{fig:vVSr}
        \end{center}
    \end{figure}

The gas velocity is plotted in Figure \ref{fig:vVSr} as a function of radius for our five time epochs.

The gas is initially outflowing as the quasar wind with constant velocity $v_{in}$. 
The inner shock moves outward at velocity $v_{R1}$ and in that frame, the gas slows from velocity $v_{in}-v_{R1}$ to $(v_{in}-v_{R1})/4$ as it crosses the shock.
In the frame of the quasar the shocked wind gas has $v_{gas}=v_{in}/4 + 3v_{R1}/4$, which approaches $v_{in}/4$ at late times. 
    
At $r>R_1$ the gas velocity decreases approximately as $1/r^2$ until the discontinuity at $R_c$ between the shocked wind and the shocked ISM, at which it has a value of $\lambda_cv_{R2}$.
The equation above for the region $R_1<r<R_c$ (equivalent to KM92 Eq.\ B10) reproduces the W77 case of constant ambient ISM density ($\alpha$=0).

Outside $R_c$, we model $v_{gas}$ as a second-order function of $(r/R_2)$ using the constraints 
$v_{gas}(R_c)=\lambda_cv_{R2}$, 
$v_{gas}(R_2)=0.75v_{R2}$, and the velocity gradients at $R_c$ and $R_2$ given by KM92 Eq.\ B7a and B7b.
Comparison to the exact self-similar solution from Equation B5 of KM92, to their Figure 4b, and to the discussion in their Appendix B shows this approximation to be an excellent one for our assumed constant $L_{in}$. 

\subsection{Gas density equations} \label{sec:rho}

Our equations for the density of the gas within the different regions are presented below and then described.
        \begin{equation} \label{eq:rhogas}
        \begin{aligned}
            \rho(r<R_1) &= \frac{\dot{M}_w}{4\pi r^2v_{in}} \\
            \rho(r=R_1) &= \frac{\dot{M}_w}{\pi r^2 v_{in}} \\
            \rho(R_1<r<R_c) &= 
        \frac{5(5-\alpha)}{4\pi(11-\alpha)\lambda_c^3}
        \left[1-\frac{r^3}{R_c^3(t)}\right]^{-\frac{2(4+\alpha)}{3(11-\alpha)}}\\
        &\times \left[\frac{6\rho_0R_0^\alpha\dot{M}_w^{\frac{2-\alpha}{3}}}{(3-\alpha)v_{in}^2\Gamma_r\xi}\right]^{\frac{3}{5-\alpha}}
        t^{\frac{-4-\alpha}{5-\alpha}} \\
            \rho(r=R_c) &= (1.60+2.04\alpha^2)\rho_0[R_2/R_0]^{-\alpha} \\
            \rho(R_c<r<R_2) &= \rho(R_c) +
            \left[\frac{\rho(R_2) - \rho(R_c)}{R_2 - R_c}\right](r - R_c) \\
            \rho(r=R_2) &= 4\rho_0[R_2/R_0]^{-\alpha} \\
            \rho(r>R_2) &= \rho_0[r/R_0]^{-\alpha}
        \end{aligned}
        \end{equation}

We assume the mass outflow rate and the velocity in the quasar wind are constant in time, which sets the density of the gas in the unshocked wind at $r<R_1$. 

We assume the inner and outer shocks are strong shocks, so that the density jumps by a factor of four at $r=R_1$ and $r=R_2$.

We have extended the work of KM92 section 3 and appendix B to find the above expression for the shocked wind gas density at $R_1 < r < R_c$ for the case of $\alpha > 0$.
The density formally becomes infinite at $R_c$ (unless $\alpha>11$, which is implausible), but the integrated mass is finite for all 
$\alpha<5$.
For $\alpha=0$, our expression yields the same exponents as in W77 Eq.\ 16 and a scaling factor only 0.8\% lower (0.623 vs.\ 0.628).

For the density of the shocked ISM at most radii $R_c<r<R_2$, a density profile that changes linearly with radius is a reasonable approximation (see KM92 Figure 4b).
However, the slope of that profile varies with $\alpha$ because the density at the contact discontinuity increases with increasing $\alpha$.
Thus, the density profile of the shocked ISM can be increasing or decreasing with radius, depending on the value of $\alpha$.
Using the results in KM92 Table 3 and Figure 4b, we approximate the density at the contact discontinuity as
    \begin{equation}
\rho(R_c)=(1.60+2.04\alpha^2)\rho_0(R_2/R_0)^{-\alpha}.
    \end{equation}
The ratio of the shocked wind densities at $R_c$ and $R_2$ is found by dividing the above equation by $4\rho_0(R_2/R_0)^{-\alpha}$. That ratio ranges from 0.4 for $\alpha=0$ (meaning, the shocked ISM density increases with radius), to 1.0 for $\alpha=\sqrt{20/17}\simeq 1.085$ (constant shocked ISM density), to 2.44 for $\alpha=2$ (shocked ISM density decreases with radius). Linear density profiles constructed using those ratios at $R_c$ and a ratio of unity at $R_2$ are acceptable approximations to the numerical density profiles shown in KM92 Figure 4b, although they do underestimate the density by up to a factor of $\sim$1.5 in $\sim$5\% of the shocked ISM region just outside $R_c$ as $\alpha\rightarrow 2$.

\begin{figure}
        \begin{center}
        \includegraphics[width=8
    cm]{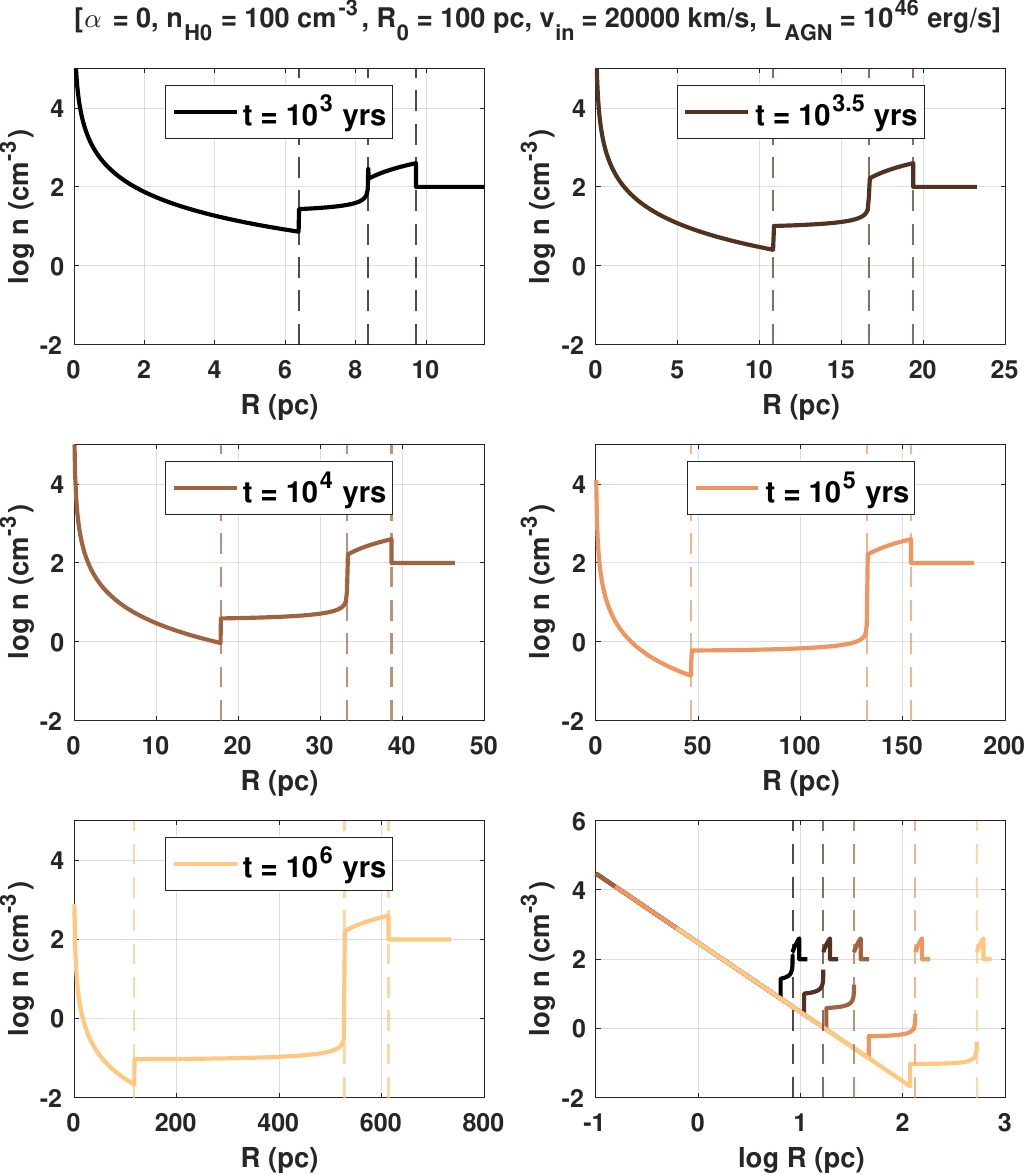}
            \caption{Density of the outflowing gas vs.\ distance from the quasar. Format and initial conditions are the same as in Figure \ref{fig:vVSr} (constant-density ISM). As the age of the outflow increases, so does the radius of the expanding shock bubble. The density of the shocked wind decreases with time. The density profile of the swept-up ISM is self-similar. 
            Note that our adiabatic assumption will break down for swept-up ISM gas at $t \gtrsim t_{\rm shell}$, at which point $R_c$ and $R_2$ will converge to an intermediate value and the density between them will increase substantially (\S~\ref{sec:cool}).
            For our parameters, we have a conservative lower limit of $t_{\rm shell} \geq 10^{4.0}$ yr (\S\,\ref{sec:mod param}).
            }
        \label{fig:nVSr}
        \end{center}
    \end{figure}

\begin{figure}
        \begin{center}
        \includegraphics[width=8
    cm]{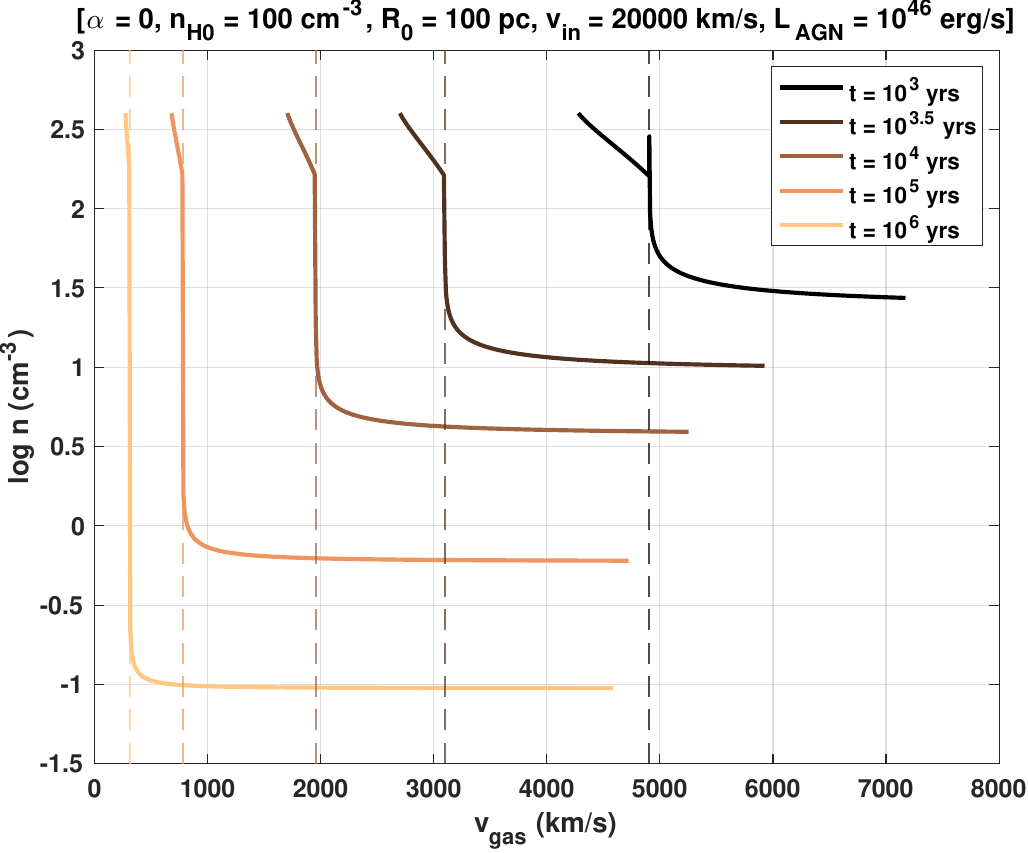}
            \caption{Density of the shocked gas vs.\ its velocity. Format, initial conditions, and caveats are the same as in Figure \ref{fig:vVSr}.}
        \label{fig:nVSv}
        \end{center}
    \end{figure}

The density is plotted for our five time epochs in Figure \ref{fig:nVSr} as a function of radius and in Figure \ref{fig:nVSv} as a function of gas velocity.

\subsection{Gas acceleration equations} \label{sec:accel}

    \begin{figure}
        \begin{center}
        \includegraphics[width=8
    cm]{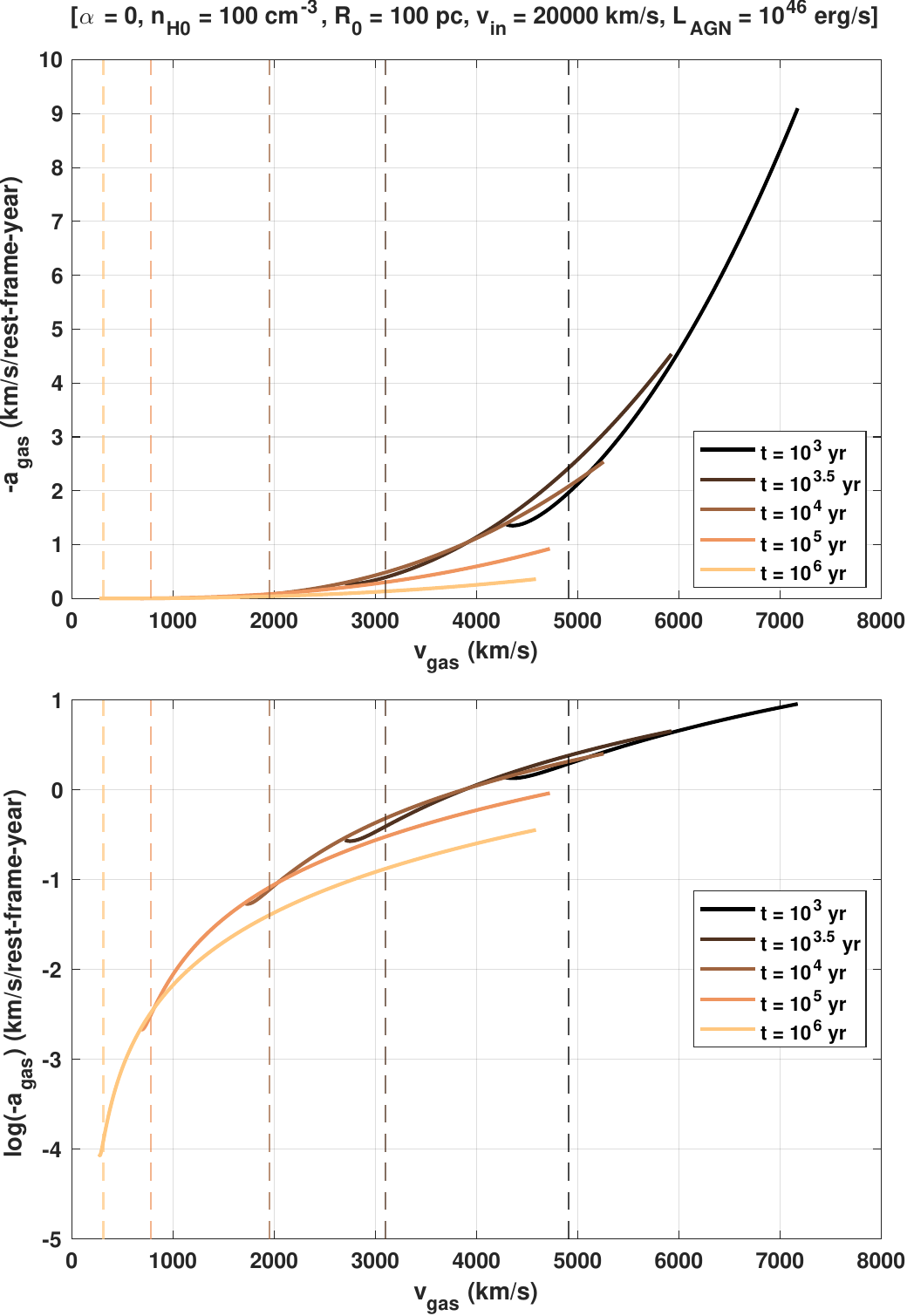}
            \caption{Deceleration (decrease in gas outflow speed per rest-frame year) vs.\ outflow speed of the shocked gas 
            at different epochs 
            with the same initial conditions and caveats as in Figure \ref{fig:vVSr}. Shown in linear scale (top panel) and log scale (bottom panel).
            The largest deceleration occurs in the youngest outflow.  
            }
        \label{fig:dvdtVSv}
        \end{center}
    \end{figure}
    
For the acceleration of fluid parcels in the gas, we must consider the Langrangian derivative of the velocity:
\begin{equation} \label{eq:agas}
a_{gas}=v_{gas}\frac{\partial v_{gas}}{\partial r} + \frac{\partial v_{gas}}{\partial t}.
\end{equation}

For the shocked wind, we recall that $\beta_E=3/(5-\alpha$) so that (writing $v_{sw} = v_{gas} = v$ and $a_{sw} = a_{gas}$):
\begin{eqnarray} \label{eq:a_shwind}
v_{sw} = k_1 R_c^3 r^{-2} t^{-1} + k_2 r t^{-1} ~~~~~~~~\nonumber\\
~ = k_1 \lambda_c^3 A_E^3 r^{-2} t^{3\beta_E-1} + k_2 r t^{-1}\\
{\rm where} \quad k_1 = \beta_E(11-\alpha)/15; \quad k_2 = \beta_E(4 + \alpha)/15\\
\frac{\partial v_{sw}}{\partial r} = -2k_1\lambda_c^3A_E^3r^{-3}t^{3\beta_E-1} + k_2 t^{-1}\\
\frac{\partial v_{sw}}{\partial t} = (3\beta_E-1)k_1\lambda_c^3A_E^3r^{-2}t^{3\beta_E-2} - k_2 r t^{-2}\\
a_{sw}=v_{sw}\frac{\partial v_{sw}}{\partial r}+\frac{\partial v_{sw}}{\partial t}=(k_1 \lambda_c^3 A_E^3 r^{-2} t^{3\beta_E-1} + k_2 r t^{-1})\nonumber\\
\times~(-2k_1\lambda_c^3A_E^3r^{-3}t^{3\beta_E-1} + k_2 t^{-1})\nonumber\\
+~(3\beta_E-1)k_1\lambda_c^3A_E^3r^{-2}t^{3\beta_E-2} - k_2 r t^{-2}
\end{eqnarray}
Note that $A_E$ has units of distance/time$^{\beta_E}$, which ensures that the quantities above have the correct units.

For the shocked ambient ISM, we use 
$R_2=A_Et^{\beta_E}$ and $v_{R2}/R_2=\beta_E/t$ to write (using $v_{sa} = v_{gas}$ and $a_{sa} = a_{gas}$):
\begin{eqnarray} \label{eq:a_shISM} 
v_{sa}&=&\frac{3\beta_E}{4}\big[
C_{v0}A_Et^{\beta_E-1}
+2C_{v1}rt^{-1}
+\frac{C_{v2}}{A_E}r^2t^{-1-\beta_E}
\big]\nonumber\\
~\\
\frac{\partial v_{sa}}{\partial r} &=& 
\frac{3\beta_E}{4}\big[
2C_{v1}t^{-1}
+\frac{2C_{v2}}{A_E}rt^{-1-\beta_E}
\big]\\
\frac{\partial v_{sa}}{\partial t} &=& 
\frac{3\beta_E}{4}\big[
(\beta_E-1)C_{v0}A_Et^{\beta_E-2}
-2C_{v1}rt^{-2}\nonumber\\
&~& ~~~~~~-(\beta_E+1)\frac{C_{v2}}{A_E}r^2t^{-2-\beta_E}
\big]\\
a_{sa} &=& v_{sa}\frac{\partial v_{sa}}{\partial r}+\frac{\partial v_{sa}}{\partial t}
\end{eqnarray}
with $C_{v0}$, $C_{v1}$, and $C_{v2}$ defined in Eq.\ \ref{eq:vgasC_vx}.

Negative values of the acceleration $a$ correspond to deceleration of the outflow.
Figure \ref{fig:dvdtVSv} shows the deceleration $-a_{gas}$ vs.\ outflow speed of the shocked gas at our five time epochs.

\subsection{Column density per unit velocity} \label{sec:Nv}

The column density $N$ (units of cm$^{-2}$) is the particle density $n$ integrated along the line of sight.
Our column density differential is $dN(r) = -n(r)~dr$ (the negative sign appears because we integrate along the line of sight towards the centre of the bubble, from large $r$ to $r=0$ in steps of $-dr$).
What is actually observed in a spectrum is the column density per unit velocity $N_v$, with units of (cm$^{-2}$)/(\kms).
Because in our model each position has a unique velocity, we can write $N_v$ as the velocity derivative of $dN$: $N_v(r) = -n(r)~dr/dv$. 
Using the velocity differential $dv = dr(\partial v/\partial r) + dt(\partial v/\partial t)$
with $dt=0$ for a single moment in time, we have
\begin{equation}\label{eq:Nv}
N_v(r) = -n(r)\frac{dr}{dv} = -\frac{n(r)}{\partial v/\partial r}.
\end{equation}
At each time epoch and radius $R_1\leq r\leq R_2$ within the shock bubble, we calculate $N_v(r)$ using 
the above expression.

\begin{figure}
        \begin{center}
        \includegraphics[width=8
    cm]{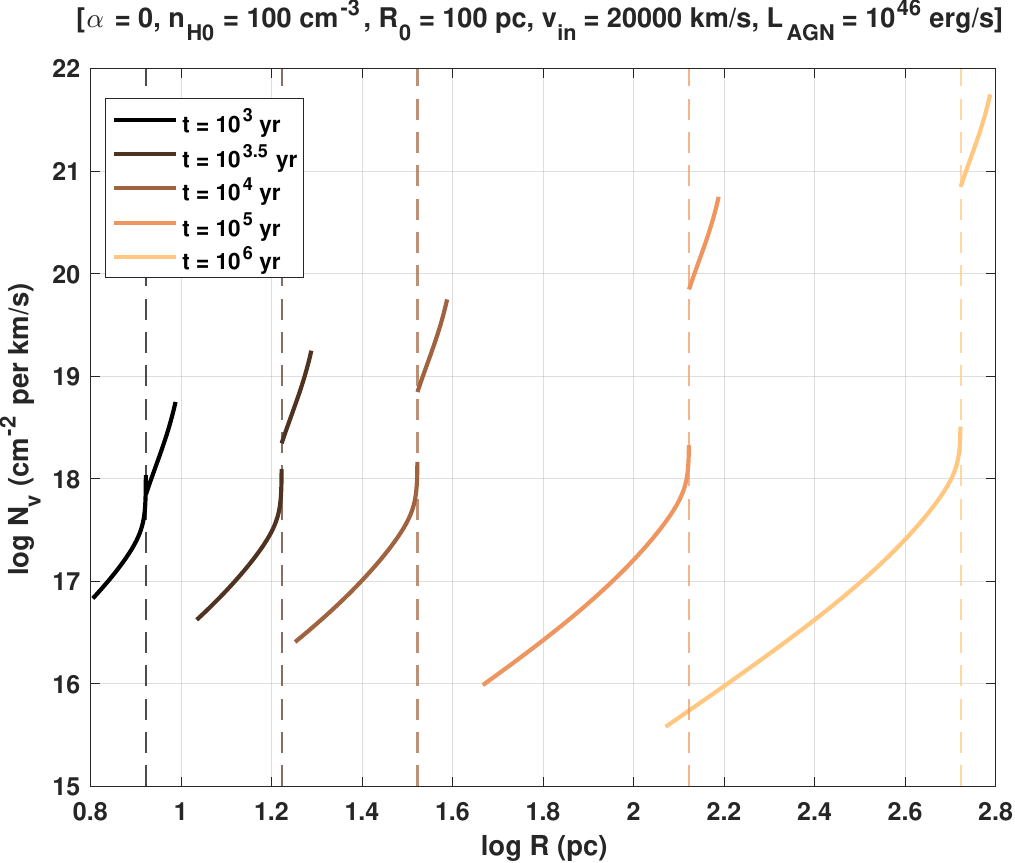}
            \caption{Column density per unit velocity of the shocked gas vs.\ distance from the quasar. The initial conditions and caveats are the same as in Figure \ref{fig:vVSr}. As the age of the outflow increases, the average value of $N_v$ increases near and outside $R_c$ (shown by the dotted line), and decreases inside it.}
        \label{fig:NvVSr}
        \end{center}
    \end{figure}

\begin{figure}
        \begin{center}
        \includegraphics[width=8
    cm]{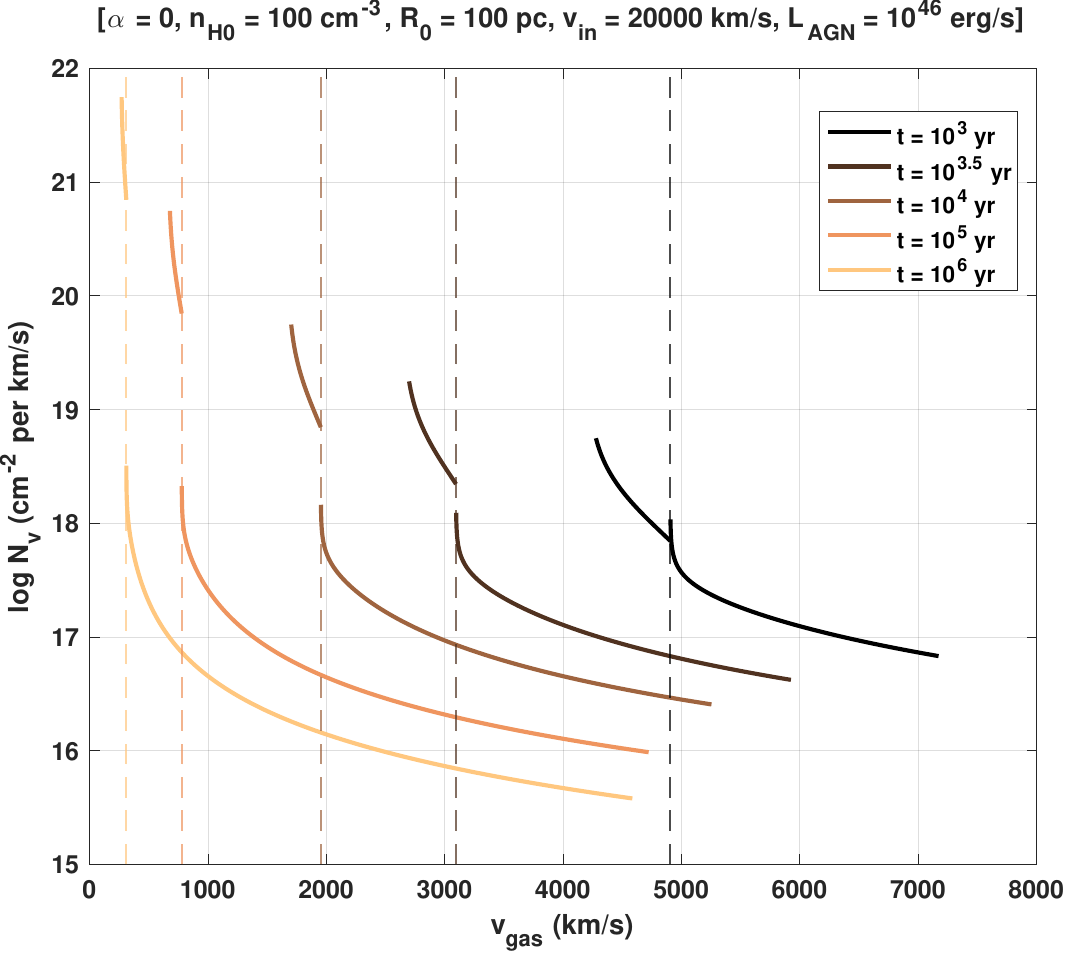}
            \caption{Column density per unit velocity of the shocked gas as a function of gas velocity. Format, initial conditions, and caveats are the same as in Figure \ref{fig:vVSr}.}
        \label{fig:NvVSv}
        \end{center}
    \end{figure}

The column density per unit velocity is shown in Figures \ref{fig:NvVSr} and \ref{fig:NvVSv} as a function of radius and gas velocity, respectively, for each of our five time epochs.
The values of $N_v$ near and outside $R_c$ increase with time, and the values are largest in the shocked ISM region at $R>R_c$.
The total column density of shocked gas can be calculated by integrating $N_v$ over velocity.

\subsubsection{Unshocked Quasar wind total column density} \label{sec:unshockedcolden}

For completeness, we present an equation for $N_w$, the column density through the unshocked quasar wind. 
$N_w$ is the number density in the wind $n_w(r)$ integrated from $R_1$ inward to the radius at which it enters our line of sight.
We assume that the wind enters our line of sight 
at $r_{LOS}=10r_L$, where $r_L$ is the launch radius. 
At that point, it would have reached 89\% of its terminal velocity according to the model of \cite{mc97}.
We adopt $r_L=0.02$~pc, the radius of the 3900\,\AA\ continuum emission region estimated in \S \ref{sec:ismaccel}, so that our assumed $r_{LOS} = 0.2$~pc.

We write the number density in the wind as
\begin{equation}
n_w(r) = \dfrac{\rho_w(r)}{\mu m_p} = \dfrac{\dot{M}_w}{4\pi r^2 v_{in}\mu m_p} = \dfrac{\tau_{in} L_{AGN}}{4\pi\mu m_p r^2 v_{in}^2c}.
\end{equation}

Integrating inward along the line of sight to get the wind column density, we obtain:
\begin{equation}
N_w = \int_{R_1}^{r_{LOS}} n_w(r) (-dr) = \dfrac{\tau_{in} L_{AGN}}{4\pi\mu m_p v_{in}^2c}\left(\dfrac{1}{r_{LOS}} - \dfrac{1}{R_1}\right).
\end{equation}
With $r_{LOS}=10r_L$, we have $r_{LOS} \ll R_1$, and so
\begin{equation}
N_w \approx 4.6 \cdot 10^{21} \text{ cm}^{-2} 
\left[ \dfrac{\tau_{in} L_{AGN}/10^{46} \text{ erg s}^{-1}}{(v_{in}/20000 \text{ \kms})^2}\right] \left(\dfrac{r_{LOS}}{0.2 \text{ pc}}\right)^{-1}
\end{equation}
which overestimates $N_w$ by only 2\% for $R_1=10$~pc.
All else being equal, the value of $N_w$ will increase by a factor of 10 if the wind enters our line of sight at $r_{LOS}=r_L$ instead of 10$r_L$, and similarly will decrease if $r_{LOS}$ is larger than we have assumed.

\subsection{Valid parameter combinations} \label{sec:valid}

For valid predictions to result from a particular combination of model parameters (including the age $t$ of the outflow), our model assumptions must remain valid.

Our assumption of velocity varying parabolically with depth in the shocked ISM leads to a mixture of deceleration and acceleration in that region for 
$\alpha>1.625$. 
At $\alpha>1.625$, the absolute value of $a_{gas}$ is small throughout the region and approaches zero everywhere as $\alpha \rightarrow 2$, at which $v_{R2}$ is constant; see KM92 Appendix A.
Nonetheless, to simplify our analysis we only consider a range of exponents for the ISM density of $0\leq \alpha \leq 1.5$.
This is consistent with the range of values $0.5<\alpha<1.5$ for parsec-scale gas profiles around AGN found in the study of \cite{2020MNRAS.492..444Y}, who note that values of $\alpha\simeq 1$ have been inferred from observations of several nearby AGN. 

In our model it must also be plausible that the shock bubble is adiabatic for a given parameter combination.
The adiabatic phase is defined by the condition $t_{cool} > t_{flow} = R_2/v_{R2}$, where $t_{cool}$ is the proton cooling time of the shocked wind (FGQ12 Eq. 17). This condition can be expressed in the form $t > t_{PRB}^E$, where $t_{PRB}^E$ is the partially-radiative-to-adiabatic transition time. 
Furthermore, the bubble structure only forms after a period of free expansion lasting until the bubble reaches size $R_f$ (Eq.\ \ref{KM:Rf}) at time $t_f= R_f/v_{in}$. Thus, for our purposes we must ensure that both $t_{PRB}^E$ and $t_f$ are $\ll 10^3$ yr. 
As mentioned in \S\,\ref{sec:mod}, $t_{PRB}^E < 25$ yr for all parameter combinations explored in this paper.
The only case we consider in which $t_f>10^3$ yr is in the top panel of Figure \ref{fig:varyparam} for $v_{in} = 3000$ \kms, which has $t_f \sim 10^{3.57}$ yr.

\subsection{Varying Default Parameters} \label{sec:vary}

    \begin{figure*}
        \begin{center}
        \includegraphics[width=10
    cm]{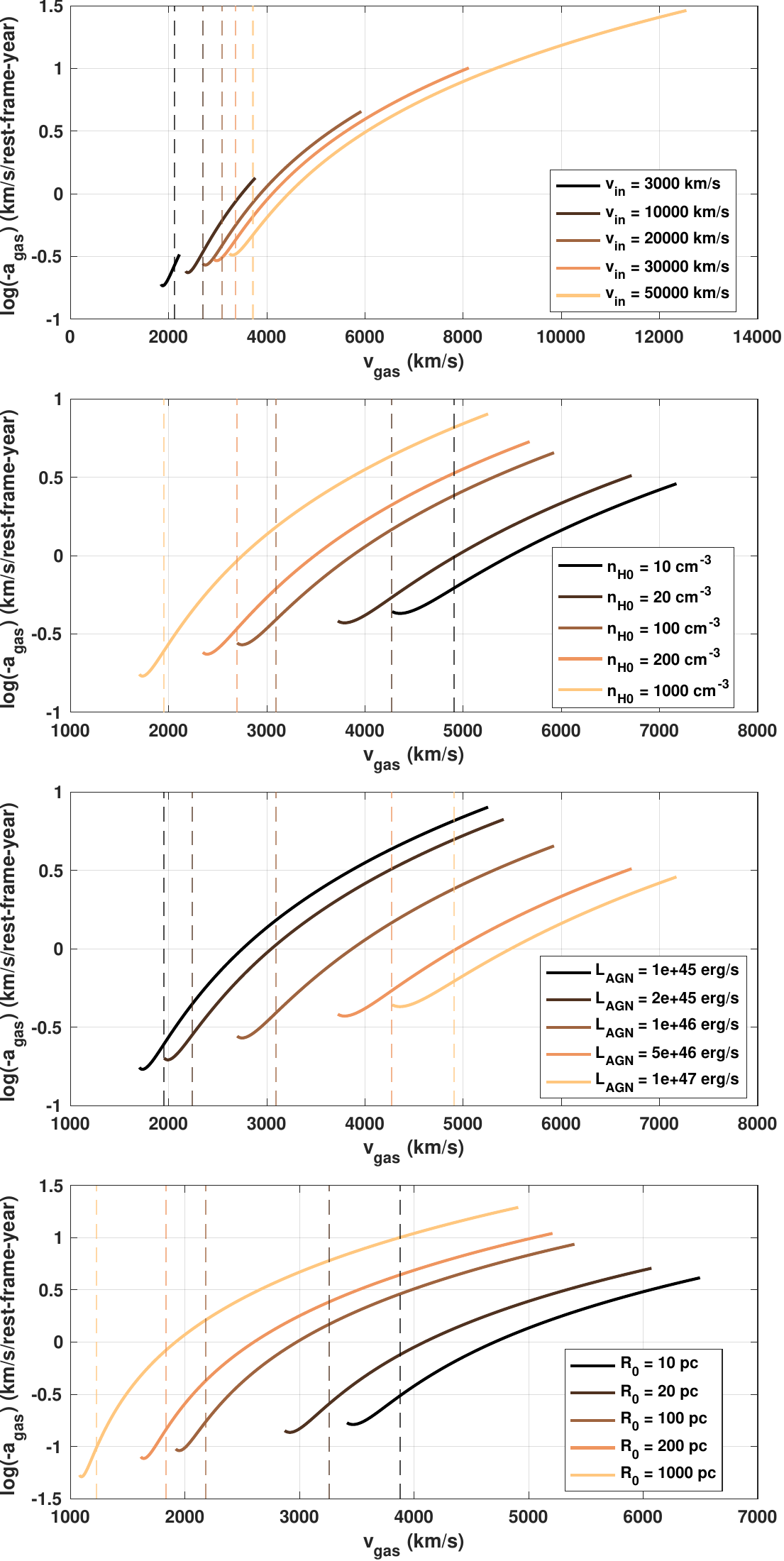}
            \caption{Variation of gas deceleration profiles ($-a_{gas}$ vs.\ $v_{gas}$) with individual model parameters at $t = 10^{3.5}$ yr. The dashed vertical lines show the location of $R_c$ for each profile. Model parameter varied, from top to bottom panel: outflow velocity, reference density, AGN luminosity, reference radius. The other fixed parameters take on their default values, with the exception of the bottom panel, which uses $\alpha = 1$ since only $R_0^\alpha$ terms appear in the gas equations. We note the following cases with $t_{\rm shell} < 10^{3.5}$~yr: (second panel) $n_{0} = 1000$ cm$^{-3}$, $t_{\rm shell} = 10^{3.2}$ yr; (bottom panel) $R_0 = 100$ pc and $R_0 = 1000$ pc, $t_{\rm shell} = 10^{2.7}$ yr and $t_{\rm shell} = 10^{0.3}$ yr respectively. In the top panel, the 3000 \kms\ case may have underestimated gas velocities and overestimated gas deceleration, as the free expansion time in this case is $t_{\rm f} = 10^{3.57}$ yr.
            }
        \label{fig:varyparam}
        \end{center}
    \end{figure*}
    
    \begin{figure}
        \begin{center}
        \includegraphics[width=8
    cm]{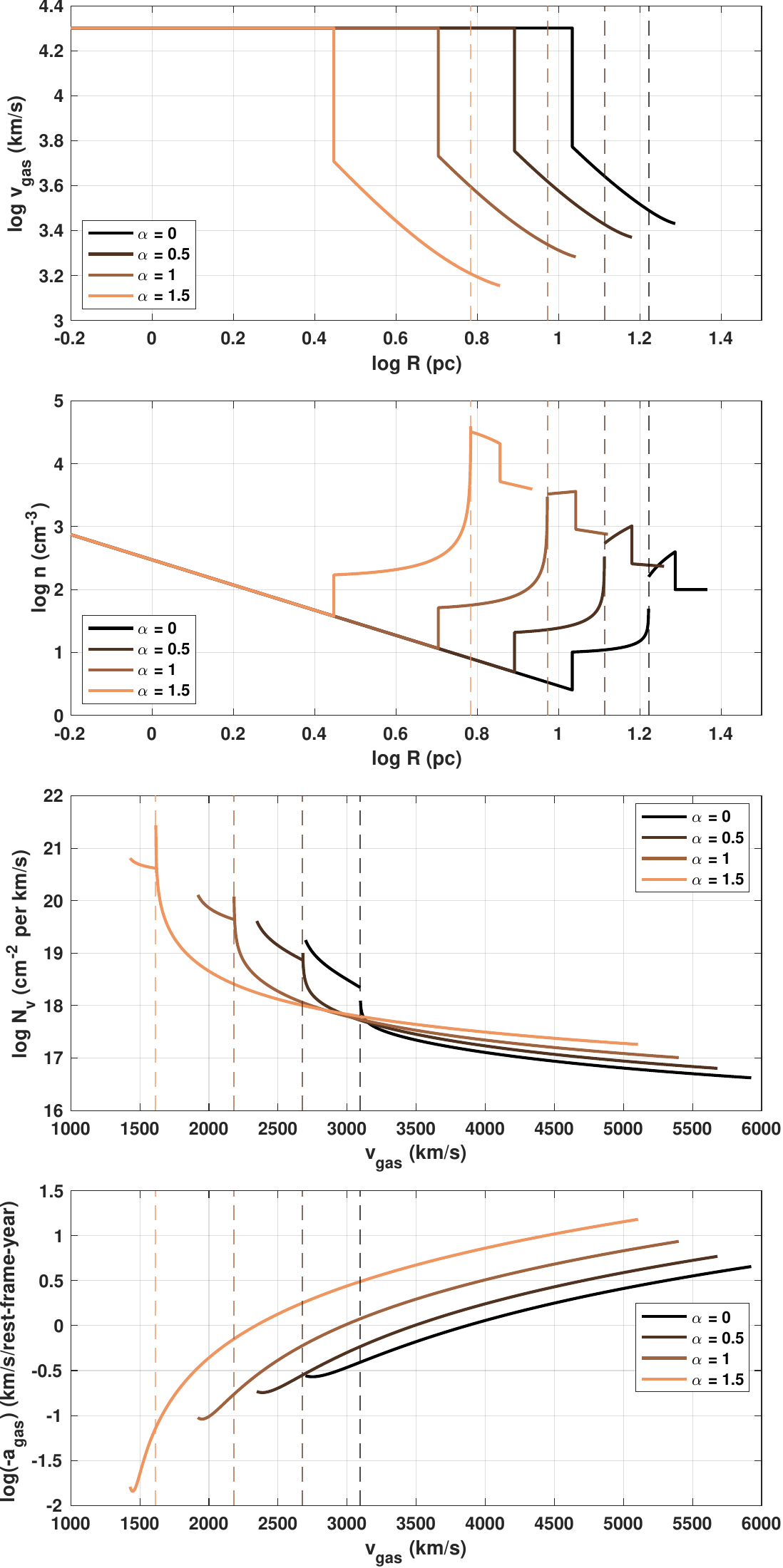}
            \caption{Profiles of various gas properties at $t = 10^{3.5}$ yr vs.\ radius or $v_{gas}$  for different values of $\alpha$ (all other model parameters are set to their default parameters). From top to bottom: velocity, density, column density per unit velocity, and deceleration. The dashed vertical lines show the location of $R_c$ for each profile. In the second subplot, the density of a small region of undisturbed ISM is shown outside $R_2$ for each profile. 
            See \S~\ref{sec:vary} for a discussion of shell formation timescales.
            }
        \label{fig:varyalpha}
        \end{center}
    \end{figure}

The trends of the observable quantities within the model were examined by varying: 
\begin{itemize}
        \item the initial quasar wind velocity ($v_{in}$), ranging from 3000 \kms\ to 50,000 \kms\ (default 20,000 \kms)
        \item the density of the ambient ISM ($n_{0}$) ranging from 10 cm$^{-3}$ to 1000 cm$^{-3}$ (default 100 cm$^{-3}$)
        \item the bolometric luminosity of the AGN ($L_{\rm AGN}$), ranging from $10^{45}$ erg/s to $10^{47}$ erg/s (default $10^{46}$ erg/s)
        \item the normalizing radius for the gas density profile ($R_0$), ranging from 10 pc to 1000 pc (default 100 pc)
        \item the exponent of the density profile variation with radius ($\alpha$), ranging from 0 to 1.5 (default 0)
\end{itemize}
    As a cautionary note, while there may be physical intuition behind the trends presented below, the trends are not necessarily simple from a mathematical perspective, and as such trends in other regions of parameter space may be non-intuitive.
    
    How the deceleration changes at $t=10^{3.5}$ yr when the above parameters are varied individually can be seen in Figure \ref{fig:varyparam}, which shows the deceleration $dv/dt$ as a function of the velocity of the gas $v_{gas}$, similar to Figure \ref{fig:dvdtVSv}. Note that the gas deceleration does not depend on $R_0$ for $\alpha = 0$, and thus we use $\alpha = 1$ when varying $R_0$ (last subplot in Fig.\ \ref{fig:varyparam}). All other fixed parameters are set to their default values.
    
    The subplots show that deceleration in the shocked ISM region increases with the quasar wind velocity $v_{in}$ and luminosity $L_{AGN}$, and decreases with the reference density $n_0$ and reference radius $R_0$. The trends for the shocked wind region in these cases are non-trivial, except perhaps in the case of the gas at the inner shock $R_1$, due to the large variation of the gas velocity range with the varying parameters. In the case of the gas at $R_1$, we see that the deceleration increases with $v_{in}$, $n_0$ and $R_0$ and decreases with $L_{AGN}$.
    
    More predictably, the curves shift toward higher velocities with increasing $v_{in}$ and $L_{AGN}$, and to lower velocities with increasing $n_0$ and $R_0$ (since increasing $R_0$ while keeping $n_0$ fixed increases the density at $r < R_0$). Finally, we note that the range of deceleration values increases with the range of velocities, which in turn increases with $v_{in}$, $n_0$ and $R_0$ and decreases with $L_{AGN}$.

Fig.\ \ref{fig:varyalpha} shows how the gas velocity, density, column density and acceleration vary at $t = 10^{3.5}$~yr when $\alpha$ is varied between 0 and 1.5, with all other model parameters fixed at their default values. 

Note that the cases of $\alpha = 0$, $\alpha = 0.5$, $\alpha = 1$ and $\alpha = 1.5$ respectively have shell formation times $t_{\rm shell} = 10^{4.0}$ yr, $t_{\rm shell} = 10^{3.7}$ yr, $t_{\rm shell} = 10^{2.7}$ yr and $t_{\rm shell} = 10^{2.4}$ yr. 

For $\alpha < 1.4$, the shocked ambient cooling time decreases over time and the swept-up ISM will collapse by $t \simeq 1.2 t_{\rm shell}$ \citep{2018MNRAS.478.3100Richings}.
Thus, the true $\alpha=1$ case at $10^{3.5}$~yr will have a swept-up ISM density significantly larger than shown in Fig.\ \ref{fig:varyalpha}, and a swept-up ISM $N_v$ larger than shown over a narrower velocity range near $v_{Rc}$.

For $\alpha = 1.5$, the swept-up ISM region is not adiabatic until $t\gtrsim t_{\rm shell}$ and so the true $\alpha=1.5$ case at $10^{3.5}$~yr will have a narrower swept-up ISM region and a higher average swept-up ISM density than are shown in Fig.\ \ref{fig:varyalpha}.

With other model parameters held fixed, at $t=10^{3.5}$~yr as $\alpha$ is increased from 0 to 1.5 the entire shock bubble becomes smaller, with $R_1$, $R_c$, and $R_2$ all decreasing: the scale of the bubble is less than $R_0$, and so increasing $\alpha$ increases the mass of ISM into which the bubble expands.
At this time at which the bubble is smaller than $R_0$, we see the shocked gas velocity drop with increasing $\alpha$, though the relative velocity range increases in the shocked wind region. The shocked gas density and column density both increase with $\alpha$, while deceleration only increases in the inner regions of the shocked wind. Deceleration decreases with increasing $\alpha$ for shocked ISM and shocked wind near the contact discontinuity, likely due to an increase with $\alpha$ in the mass of swept-up ISM causing significant deceleration to have occurred before $t = 10^{3.5}$ yr.

Note that the behaviour of the density profile in the shocked ambient gas region changes from increasing with $r$ for $\alpha=0$ to decreasing with $r$ for $\alpha \gtrsim 1.1$. This behaviour arises from the solutions to the hydrodynamic equations assuming self-similarity (see Fig.\ 4b of KM92 for the exact behaviour). Qualitatively, this is a consequence of higher-density gas being swept up earlier and piling up near the contact discontinuity, leading to the swept-up gas density decreasing with radius for sufficiently steep density profiles.

\section{Comparison to Observations of J0300} \label{sec:j0300}

Our theoretical model predicts the magnitude of the deceleration expected to be experienced by the shocked gas in quasar outflow bubbles.
Even with constant energy input, for all plausible density profiles the expansion of the bubble will slow with time as the host galaxy ISM is swept up (FGQ12).
Yet deceleration of quasar outflows has rarely been observed (see \S \ref{sec:complit}).

Here we search for deceleration directly at relatively low velocities in low-ionization gas in a quasar outflow.
 
\subsection{Choice of Target} \label{sec:target}

Our target, SDSS J030000.56+004828.0 (\citealt{sdss123}), was observed at high spectral resolution in 2001 \citep{sb2}.
We adopt the same systemic redshift of $z=0.89185$ used in that paper.
%
J0300 
has the highest equivalent width \caii\ BAL trough known. 
Strong \caii\ $\lambda\lambda$3934,3969 absorption is seen outflowing at $v\simeq 2000-4000$~\kms\ with weaker \caii\ absorption at $v\simeq 1700-2000$~\kms\ and $v\simeq 4000-5660$~\kms\ (Fig.\ \ref{fig:2spec}, and Figs.\ 3-6 of \citealt{sb2}). 

We chose this target for its outflow in which lower velocity gas is inferred to be at greater distances (\S\,\ref{intro}) and for its existing high-resolution spectrum.
Also, unlike low-velocity Mg\,{\sc ii} or C\,{\sc iv} in many BAL quasars, the unsaturated Ca\,{\sc ii} absorption in J0300 has considerable structure in velocity space which should aid in detecting velocity shifts.
Lastly, the relatively high velocities of  Ca\,{\sc ii} in J0300 suggest that if the outflow's current phase does arise from a shock bubble, the bubble is relatively young, which is when any deceleration is expected to be largest and most easily detectable.

\subsection{Previous High-Resolution Spectrum (UVES)} \label{oldJ0300}
    
\begin{figure}
        \begin{center}
        \includegraphics[width=8cm]{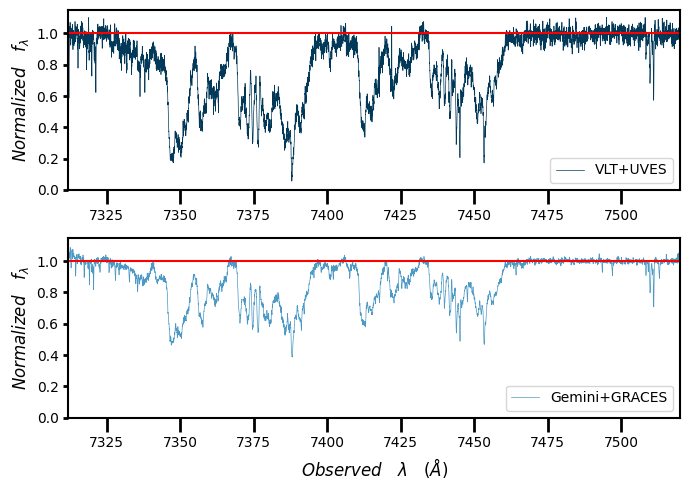}
            \caption{Top: Ca\,{\sc ii} $\lambda\lambda$3934,3969 absorption in the continuum-normalized 2001 UVES spectrum of J0300. At far left, below 7340\,\AA\ there is weak high-velocity $\lambda$3934 absorption. At far right, the narrow absorption at 7510\,\AA\ is $\lambda$3969 absorption at the quasar systemic redshift. In between is  absorption in both transitions at wavelengths corresponding to outflow velocities $v$=1700-4000 \kms. Bottom: Ca\,{\sc ii} $\lambda\lambda$3934,3969 absorption in the continuum-normalized 2019 GRACES spectrum of J0300. The absorption is shallower in the GRACES spectrum because that spectrum was obtained through an optical fiber which includes light from the sky.}
        \label{fig:2spec}
        \end{center}
    \end{figure}

Observations of J0300 were obtained on UT 10-12 Aug 2001 (MJD 52131-52133) using the ESO Very Large Telescope (VLT) Unit 2 (Kueyen) and Ultra-Violet Echelle Spectrograph (UVES).
A 1\arcsec\ slit was used, which in combination with subarcsecond seeing yielded a resolution $R \simeq 52,000$ at our wavelengths of interest, with 1.75 \kms\ pixel$^{-1}$.
A depolarizer was also used for all observations.
We used the SQUAD \citep{SQUAD1} reductions of these observations, downloaded from the ESO Science Archive.
The SQUAD reductions produce continuum-normalized  spectra at vacuum heliocentric wavelengths.
We found that the wavelength solution of the 
UVES spectrum used in \cite{sb2} was such that features in it appear at wavelengths $2.1\pm 0.1$ \kms\ to the red as compared to the same features in the SQUAD spectrum.
The wavelength solution of the SQUAD reduction is consistent with the wavelength solution of the new observations discussed in the next section.
The spectrum has an average signal-to-noise ratio per \kms\ in the wavelength region of interest of SNR/(\kms) = 25.5 and is shown in the top panel of Figure \ref{fig:2spec}.

\subsection{New High-resolution Spectrum (GRACES)} \label{newJ0300}
New observations of J0300 were obtained on the night of November 13th 2019 (MJD 58800) using the GRACES instrument \citep{GRACES} on the Gemini-North telescope. GRACES consists of a fiber optic cable which takes the light collected with Gemini-North through a 1.2 arcsecond diameter aperture on the sky and feeds it to the ESPaDOnS spectrograph \citep{2006ASPC..358..362D} at the Canada-France-Hawaii Telescope (CFHT).
 
We used the one-fiber observing mode to obtain a spectrum of resolution $R\simeq 60,000-65,000$ with 2.6 \kms\ pixel$^{-1}$. 
We obtained 4 exposures of 2400s each.

Each exposure was reduced individually using the OPERA pipeline \citep{OPERA}. 
OPERA is based on the Libre-ESpRIT package \citep{1997MNRAS.291..658D}. 
OPERA places the individual GRACES exposures on an heliocentric atmospheric wavelength scale, corrected using telluric lines.
(The telluric corrections range from $-0.079$\,\kms\ to $+0.045$\,\kms, and were applied despite having quoted uncertainties of $\pm 0.33$\,\kms.)
We converted these wavelengths to vacuum heliocentric wavelengths using the formula of \cite{mor91}. 
We describe how we combined these exposures in the next subsection.
    
\subsection{Processing and Cross-Correlation} \label{xcor}

To measure the velocity shift over the time between the UVES and GRACES observations, we cross-correlated the spectra to find the shift that gives the maximum correlation. The processing prior to this is described below.

First, we visually compared the \caii\ broad absorption troughs in the two spectra. No dramatic changes in the broad absorption profile shapes occurred between the two epochs.

The wavelength region of interest was isolated as 
7312.000\,\AA\ to 7520.026\,\AA\ (vacuum heliocentric).
We have four separate GRACES exposures, and in each exposure there are two orders which overlap in wavelength near the center of the above wavelength range.
Because the central wavelengths for each pixel in each exposure were slightly different, we treated each order from each exposure as a separate spectrum and combined all spectra at a given wavelength together as discussed below.

In order to increase the velocity resolution of the cross-correlation, we upsampled the UVES and four GRACES spectra to a logarithmic wavelength scale of 0.1 \kms\ pixel$^{-1}$ using fifth-order polynomial interpolation (for 84100 pixels total over our wavelength range).
To combine the GRACES spectra, we used the scombine task in IRAF.\footnote{The Image Reduction and Analysis Facility 
was distributed by the National Optical Astronomy Observatories.} 
We created a weighted average spectrum using the uncertainty values at each pixel from each exposure.
The weighted average GRACES spectrum has a SNR/(\kms) ranging from 22 to 47 in the region of interest.

We confirmed that pixels with spurious values (from cosmic rays, e.g.) were not affecting the weighted average spectrum by comparing to a median spectrum (with the median of an even number of pixels computed as the average of the two central values); no significant differences were found.
The GRACES spectrum does, however, contain a number of sky emission-line and absorption-line features, particularly below 7350~\AA.
We manually created pixel masks used to interpolate over emission-line features, but did not do the same for absorption-line features.


Finally, we normalized the GRACES spectrum by an estimate of the underlying quasar continuum consisting of a fourth-order polynomial fit to five wavelength regions in which the UVES spectrum had an average normalized flux of 0.98 or higher. 
The normalized GRACES spectrum is shown in the bottom panel of Figure \ref{fig:2spec}.

For purposes of estimating cross-correlation uncertainties, we created 100 simulated UVES spectra and 100 simulated GRACES spectra, each with added noise.
In both cases the simulated flux in a given pixel, $f_{\lambda\rm sim}$, was randomly drawn from the normal distribution $\mathcal{N}(f_\lambda,\sigma_\lambda^2)$, where $f_\lambda$ and $\sigma_\lambda$ are the measured flux and corresponding uncertainty at that pixel.
For UVES, the original spectrum was used as $f_\lambda$, and the simulated spectrum was then resampled in the same fashion as the actual spectrum.
For GRACES, each order from each exposure was used as $f_\lambda$, and the resulting simulated input spectra were processed in the same fashion as the actual spectra to produce a simulated output spectrum.

\subsubsection{Wavelength calibration check} \label{wavecal}

To check for any shift in the wavelength calibration between the UVES and GRACES spectra, we compared two regions of strong sky absorption lines using vacuum geocentric wavelengths.
For the UVES spectrum, this meant removing the heliocentric correction that had been applied. 
For each individual GRACES exposure, we removed its heliocentric correction before converting to vacuum wavelengths and combining all exposures in the same manner as the science exposures. 
The two regions cover vacuum geocentric wavelengths of 7590\,\AA\ to 7640.004\,\AA\ (in the atmospheric A band) and 6868\,\AA\ to 6923.002\,\AA\ (in the B band), both using 0.1\,\kms\ pixels.

We cross-correlated the spectra for each sky line region separately. Before doing so, following the cross-correlation scheme of \cite{1979AJ.....84.1511T}, 
we zeroed their continua by subtracting 1 and then tapered their edges (5\% of the total pixels at each edge of the spectra) by multiplying them with a half-cosine bell function to de-weight any abrupt cutoff.

We estimated uncertainties using 100 simulated spectra of each region for each instrument, created as described at the end of \S \ref{xcor}. We cross-correlated all 10,000 possible pairs of such spectra and adopted the standard deviations of the resulting shift distributions as our uncertainties.

This procedure yields shifts of 8 $\pm$ 0.5 and 6 $\pm$ 0.3 pixels for the A and B bands respectively (features in the GRACES spectrum appear at longer wavelengths than in the UVES spectrum). The A band result has a larger uncertainty due to broader sky lines in that spectral region at our resolution. 

The temperatures during the nights of our observations were not unusually high or low, so we rule out temperature effects as a significant source of systematic wavelength calibration error. 
Regardless of the origin of the wavelength calibration offset, we must account for it to determine any velocity shift in the J0300 outflow between our two epochs of observation.

A weighted average of the A and B band results yields a shift of 6.5 $\pm$ 0.5 pixels, but for simplicity we adopt the B band integer-pixel shift and acknowledge the different values measured in the two bands by adopting a systematic uncertainty of 1 pixel.
We remove the offset of 0.6 $\pm$ 0.1 \kms\ from the GRACES spectrum and place both spectra on the UVES wavelength scale.
This shift has the advantage of placing the associated \caii\ H absorption at the same velocity in each spectrum (see \S~\ref{pythoncc} and Table \ref{xcortable}).
It is also within the expected accuracy of the GRACES wavelength solution of $1-2$~\kms\ (K.\ Chiboucas, personal communication).

\subsubsection{Cross-Correlation} \label{pythoncc}

\begin{figure}
        \begin{center}
        \includegraphics[width=8
        cm]{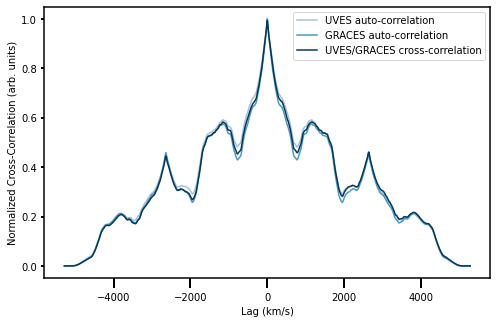}
            \caption{Normalized UVES/GRACES cross-correlation function for the full \caii\ absorption profile (see Table \ref{xcortable}), using the weighted masked average GRACES spectrum. The autocorrelations of both spectra are overlaid for comparison.}
        \label{fig:crosscorr}
        \end{center}
    \end{figure}

The velocity shift between the UVES and GRACES \caii\ spectra were measured and uncertainties estimated in the exact same manner as for the wavelength calibration check. The resulting cross-correlation function can be seen in Fig.\ \ref{fig:crosscorr}, with autocorrelations of both spectra for comparison.
Taking into account the wavelength calibration, we find the cross-correlation maximized at a shift of $0.3 \pm 0.3$ \kms\ (statistical) $\pm 0.1$ \kms\ (systematic) in the quasar reference frame, corresponding to a statistically insignificant bulk acceleration of the outflow.
The maximum normalized cross-correlation coefficient is 0.988.

Using the same method, we performed separate cross-correlations on eight select subsections of the spectra. A description of these subsections, along with their wavelength ranges and measured shifts, can be found in Table \ref{xcortable}. 

We cross-correlated the \caii\ H and K absorption regions separately, and also cross-correlated low-velocity (1899-3084 \kms) and high-velocity (3084-4540 \kms) partitions of these regions in order to check for differential shifts as a function of velocity. We find that in each case the shifts of the H and K cuts in a given velocity range agree to within the uncertainties, and that the full H and K profile shifts agree with the full spectrum shift. We also note a substantial but statistically insignificant ($<3\sigma$) difference between the shifts of the low- and high-velocity cuts, with the low-velocity cuts decelerating and the high-velocity cuts accelerating. As expected, the bulk shift lies between the low- and high-velocity shifts, but closer to the low-velocity shifts due to their sharper features (which yield lower uncertainties). 

The associated \caii\ H absorption has zero shift by expectation and by design (\S\,\ref{wavecal}). 
The low-velocity \caii\ H wavelength region is contaminated by associated absorption from \caii\ K.  We tested whether this affected the cross-correlation shift in the region by interpolating over the contaminating absorption and recalculating the cross-correlation. We found that the shift was smaller by 1 pixel, well within the uncertainty of $\pm$2.6 pixels, so we conclude that the contamination does not significantly affect the cross-correlation signal in this region.

\begin{table}
\centering 
\begin{tabular}{c c c} 
\hline\hline  
{} & Cross-Correlations & {}\\
\hline\hline  
{} & \textbf{Sky Lines} & {}\\
\hline\hline
\makecell[c]{\textit{Description} \\ \textit{of Spectrum}} & \makecell[c]{\textit{Wavelength Range} \\ (\AA)} & \makecell[c]{\textit{Velocity Shift} \\ (\kms)}\\  
\hline 
A Band & 7590.000$-$7640.004 & 0.8$\pm$0.05\\
B Band & 6868.000$-$6923.002 & 0.6$\pm$0.03\\
\hline\hline  
{} & \textbf{\caii\ Absorption} & {}\\
\hline\hline
\makecell[c]{\textit{Description} \\ \textit{of Spectrum}} & \makecell[c]{\textit{Wavelength Range} \\ (\AA)} & \makecell[c]{\textit{Velocity Shift} \\ (\kms)}\\  
\hline
Full Profile & 7312.000$-$7520.026 & +0.3$\pm$0.30$\pm$0.1\\
Associated \caii\ H & 7500.000$-$7520.026 & +0.0$\pm$0.45$\pm$0.1\\
\caii\ H \& K & 7332.127$-$7462.447 & +0.3$\pm$0.28$\pm$0.1\\
\caii\ H & 7397.000$-$7462.447 & +0.2$\pm$0.38$\pm$0.1\\
\caii\ K & 7332.127$-$7397.000 & +0.5$\pm$0.46$\pm$0.1\\
\caii\ H (low $v$) & 7433.000$-$7462.447 & $-$0.1$\pm$0.26$\pm$0.1\\
\caii\ K (low $v$) & 7367.811$-$7397.000 & $-$0.4$\pm$0.44$\pm$0.1\\
\caii\ H (high $v$) & 7397.000$-$7433.000 & +1.2$\pm$1.27$\pm$0.1\\
\caii\ K (high $v$) & 7332.127$-$7367.811 & +1.9$\pm$0.72$\pm$0.1\\%
\hline 
\end{tabular}
\caption{Velocity shifts of various spectra (GRACES with respect to UVES) as measured by cross-correlation maxima. The skyline velocity shifts are given in the geocentric frame as $v_{\rm shift}\pm\sigma_{\rm statistical}$, while the \caii\ absorption shifts are adjusted for the calibration offset of \S \ref{wavecal} and given in the quasar rest frame as $v_{\rm shift}\pm\sigma_{\rm statistical}\pm\sigma_{\rm systematic}$. The low and high velocity ranges are respectively 1899-3084 \kms\ and 3084-4540 \kms\ for both the H and K lines.}
\label{xcortable}
\end{table}

\subsection{Summary of Observational Results} \label{sec:sumobs}

As a reminder, we work in a frame centered on the quasar such that along the line of sight from the quasar to us all outflow velocities from the quasar are positive, acceleration of an outflow away from a quasar has a positive sign, and deceleration of an outflow has a negative sign (as in Table \ref{xcortable}). 

    Including our systematic uncertainty, we place a $3\sigma$ limit of $|\Delta v|<1$ \kms\ on any bulk change of the velocity of the $v=1700-4000$~\kms\ \caii\ outflow in J0300 between our two observations.
    At a redshift of $z=0.89185$, our observations are separated by 9.65 rest-frame years. Given that, our observations place a $3\sigma$ limit of $|a|<0.1$ \kmsyr\ or $<3 \times 10^{-4}$ \cmss\ on the average magnitude of any bulk deceleration (or acceleration) of the \caii\ outflow between our two observations.

\section{Discussion} \label{sec:discuss}

\subsection{Comparison with the Literature} \label{sec:complit}

Our deceleration limit is tighter than the reported $-0.2\pm 4$ \kmsyr\ 
found by \cite{XuQ0059} for the 1450 \kms\ \feii\ outflow in the FeLoBAL quasar Q0059$-$2735 over 4.6 years in its rest frame.

It is also stronger than the $3\sigma$ limits on deceleration of 0.7 km s$^{-1}$ yr$^{-1}$ found by \cite{2019ApJ...870...68M} in {\sl narrow} intrinsic absorption lines of \civ, \siiv, and \nv\ in six bright quasars. 
    
However, our limit is more than three orders of magnitude smaller than the possible detection of $-$190 \kmsyr\ deceleration over one rest-frame year in the 11,400 \kms\ component of the He\,I* outflow in the FeLoBAL quasar Mrk 231 \citep{2014ApJ...788..123L}.
Such large deceleration in Mrk 231, if confirmed, would require some combination of a very fast wind, a dense surrounding ISM, and a very young outflow. 

\cite{gea03} reported apparent variable deceleration of $a=-0.08\pm 0.02$ to $-0.18\pm 0.03$ cm s$^{-2}$ [$-24\pm 6$ to $-57\pm 9$ \kmsyr] over spans of 13 and 9 months in a narrow-line outflow at $v=3000$ \kms\ in NGC 3783 between 2000 and 2002.
\cite{2019A&A...621A..12Kriss} report that this `Component 1a' absorption has continued to show variable apparent deceleration in 2011, 2013 and 2016 (most obviously in \siiv\ and \ciii*), but also that absorption near the original velocity of this component reappeared in 2016.  Given the velocity stability of other features, they raise the possibility that the apparent velocity shifts might instead be due to absorbing structures crossing our line of sight. (For example, gas in a rotating filament with a radial velocity gradient might be observed as apparently accelerating gas.) 

\citet{2014MNRAS.442..862J} found cases of apparent $a=-220$ to $-630$ \kmsyr\ [$-0.7\pm 0.01$ to $-2.0\pm 0.1$ \cmss] at $v$=20,000 \kms\ over rest-frame time-spans of 3.11 and 2.34 yr in two \civ\ BAL quasars.  However, both of those systems also show significant wavelength-dependent BAL trough profile depth variability.  
The observed absorption variability could either be due to velocity-dependent variations in the absorbing column (whether due to ionization variability or transverse motion of absorbing gas), or to such variations plus deceleration of the gas.  This is a generic limitation of searching for deceleration or acceleration in BAL troughs in the presence of velocity-dependent depth variability (including in NGC 3783 above).  
It may be significant that acceleration at similar values of $|\Delta v/\Delta t|$ is sometimes seen (e.g., \citealt{q0242}),
as velocity-dependent depth variability might be expected to mimic acceleration as often as deceleration.

\citet{2016ApJ...824..130G} found a case of apparent deceleration 
$a=-260$ \kmsyr\ [$-$0.83 \cmss] at $v$=23,000 \kms\ over 3.891 rest-frame years, with minimal profile variability,
and two cases of apparent variable acceleration at similar $|a|$ in a sample of 151 C\,{\sc iv} BAL quasars. 
The deceleration is stronger between the first pair of three observations than between the second pair, but the inferred deceleration values are consistent at $<2\sigma$.
\citet{2016ApJ...824..130G} also found the vast majority of BAL troughs studied to be stable to within 3\% of their mean velocities on rest-frame timescales of $2.7-5.5$ years.

\citet{2019ApJ...871...43J} found a case of apparent $a=-500$ \kmsyr\ [$-$1.6 \cmss] over 4.15 rest-frame years at $v$=14,000 \kms, in one object in a sample of 10 X-ray bright C\,{\sc iv} BAL quasars. Only mild profile variability is seen in this case, and the inferred deceleration is consistent within $2-3\sigma$ between most observations. However, a lack of deceleration significant at approximately 5$\sigma$ is seen between the latest two spectra of the object presented in that paper, arguing against the presence of constant long-term deceleration in this object.

\subsection{Potentially Relevant Properties of J0300 and other Ca II BAL quasars}\label{sec:properties}

J0300 is part of the sample of \citet{felobal1}, who summarize past observations of this quasar in their Appendix C. 
We highlight a few points regarding J0300 here and compare to their results further in \S~\ref{sec:compare-Choi}.

As reported in \citet{sdss123}, J0300 forms a wide binary with SDSS J025959.69+004813.5, a non-BAL quasar located 19.5$''$ (projected 153~kpc) away at $z = 0.894 \pm 0.001$ ($\Delta v = 340 \pm 160$~\kms).  
In {\em Hubble Space Telescope} imaging obtained by \citet{2019MNRAS.483.2441Villforth}, SDSS J025959.69+004813.5 is seen to reside in a disk galaxy.

\citet{2019MNRAS.483.2441Villforth} found J0300 to have a luminous host galaxy ($M_{160W}=-25.13$) with best-fit effective radius of 0.93$\pm$0.01 arcsec (7.4 kpc at $z=0.89185$) for an exponential disk fit, but with no morphological analysis possible due to contamination from the quasar PSF.
J0300 is also seen to have a galaxy at a separation of 1.5 arcsec (projected 12 kpc at $z=0.89185$) in that imaging and in the {\em HST} imaging of \citet{2018MNRAS.475.3213Lawther}, who estimate a 98\% chance of physical association between the galaxy and the quasar.

\citet{sb2} note that the lack of absorption from low-energy excited states of \feii\ accompanying the strong \caii\ absorption at $2000<v<4000$~\kms\ in J0300 means that the gas at those velocities must have either a density or a temperature too low to significantly populate such states. In practice, this means $n_e<10^3$ cm$^{-3}$ or $T\lesssim 1100$~K or both.

\citet{2011NewA...16..128R} compared X-ray and UV absorption properties of J0300 and found the absorbing column toward the X-ray emitting region along our line of sight is $N_H\geq 1.8\times 10^{24}$ cm$^{-2}$.
Although the X-ray and UV absorption in J0300 might not arise in the same gas, they found that many properties of the UV absorption in \mgii\ and \feii* at $4000<v<10850$~\kms\ can be matched with a slab of gas of constant density $n_e=10^6$ cm$^{-3}$ with ionization parameter $\log U = -0.5$ at its ionized face (implying a distance of $\sim$60 pc from the black hole) and a thickness of $\sim 3\times 10^{18}$ cm ($\sim$1 pc).

The only other quasar with outflowing \caii\ absorption that has been studied in detail is Mrk 231, which has \caii\ outflowing at $v\simeq 4000-5000$~\kms\ \citep{2014ApJ...788..123L}. 
\citet{2014ApJ...788..123L} find that a constant-pressure slab can explain the observed \caii\ in Mrk 231 as arising in a region with density $n\gtrsim 10^6$~cm$^{-3}$ and temperature $T\lesssim 10^3$~K.
They find that a jump in density at the hydrogen ionization front is needed to simultaneously explain the observed columns of higher-ionization gas at higher velocities (traced by \hei*) and lower-ionization gas such as \caii\ at lower velocities. 
They interpret this finding as evidence for a high-velocity wind impacting, compressing, and accelerating preexisting dusty gas in the host galaxy ISM.
In their models, the absorption is produced in a region of thickness $\simeq$0.027 pc at a distance of 13-230 pc from the quasar.

\subsection{Summary of Theoretical Considerations} \label{sec:sumth}

\subsubsection{Cooling of shocked ISM gas} \label{sec:cool}

Our model does not treat the cooling of the shocked ISM or wind gas. However, low-ionization gas such as \caii\ will be seen in absorption only if the outflowing gas has cooled sufficiently for such gas to exist in it or if clouds swept up by the outflow have survived and ended up with physical conditions allowing such gas to exist in them.
We discuss the first possibility in this section.

Low-ionization gas formed {\em in situ} will form first in high density regions that cool off and self-shield from ionization. 
As seen in Figures \ref{fig:nVSr} and \ref{fig:varyalpha}, for our model's spherically symmetric density distribution, these regions lie within the swept-up ISM.  
(Note that a very thin shell of high density gas will also be present just inside the contact discontinuity at $R_c$; see Equation \ref{eq:rhogas}.) 
In this scenario, when low-ionization gas forms it will form from just inside $R_c$ out to $R_2$ and will serve as a tracer of the velocity of the earliest shocked wind gas and the swept-up ISM gas, at the lowest velocities in the outflow.
(Note, however, that \citealt{Nguyen+23} find that the velocities of clouds which radiatively cool out of a hot wind decelerating through a sonic point due to mass-loading do not exactly match those of the wind (their Fig.\ 10); see also \citealt{TanOhGronke}.) 

\citet{2012MNRAS.420.3174Sharma} show that local thermal instability (TI) can cause cold filaments to condense out of a hot medium when the ratio of its TI timescale to the free-fall (dynamical) timescale is 
$t_{\rm TI}/t_{\rm dyn}\lesssim 10$; see also \citet{2012ApJ...745..148Joung}. The TI timescale can range from slightly to a few times longer than the cooling time of the hot gas.
\citet{2014MNRAS.439..400Zubovas} argue that cold gas clouds form naturally in AGN outflows through mixing of gas across the Rayleigh-Taylor unstable contact discontinuity (see also \citealt{2018MNRAS.480L.111Gronke}). 
\cite{2015ApJ...804..137P} show that clouds can form and evaporate naturally from thermal instabilities in a radiatively accelerated flow, but
\citet{2016ApJ...833...46Ferrara} have a more skeptical take on the formation and survival of cold clumps in AGN outflows. 

\citet{2022ApJ...931..134Waters} discuss TI in the context of ionized AGN outflows. They show that a time-variable radiation field can place gas in a TI zone, that clump formation occurs after the wind is accelerated, and that the usefulness of $t_{\rm TI}/t_{\rm dyn}$ as a diagnostic is limited. Namely, clump formation can occur even when $t_{\rm TI}\gg t_{\rm dyn}$ because the TI growth rate depends on more than just that ratio.

\citet{2018MNRAS.474.3673Richings,2018MNRAS.478.3100Richings} find that the shocked ISM in outflows in a similar range of parameter space to ours mostly cool within $10^6$ yr, and show that such gas will host molecules, not just low-ionization gas.
Once the swept-up ISM has cooled and formed a shell, the shell has a radial extent $\sim$1\% of the shock bubble radius (their Figure 1).
Prior to shell formation, $R_c\simeq 0.86R_2$ (\S~\ref{sec:Rc}), so this corresponds to a bulk decrease in volume and an increase in density of a factor $\sim 8-9$ if the shell forms over 20\% of the outflow age (their Figure A2).
However, the cooling and density increase is inhomogeneous. A maximum density increase of a factor of $10^5$ (to $n=10^6$~cm$^{-3}$) is seen in \citet{2018MNRAS.474.3673Richings}, yielding gas with $T=400$~K. \caii\ absorption could be produced in such conditions.

\subsubsection{Velocity ranges of shocked ISM gas} \label{sec:vrange}

There is a fundamental limitation in how well absorption from uniform shocked ISM gas in self-similar wind bubble models can match the observations of J0300.
The velocity range of the strong Ca\,II-absorbing gas in J0300 is about a factor of 2, from 2000 \kms\ to 4000 \kms\ \citep{sb2}, and the velocity range of detectable Ca\,II is about a factor of 3.3, extending from 1700 \kms\ to 5660 \kms.
Neither range can be matched in self-similar wind bubble models in the shocked ISM region alone (from $R_c$ to $R_2$). 
The relationship cited after \cite{1992ApJ...388..103K} Eq.\ (B6) states, in our notation, that $v_{gas}(R_c)/v_{gas}(R_2) \leq 4/3$ for $\gamma=5/3$, because $\lambda_c=R_c/R_2 \leq 1$.
This might match numerous low-velocity quasar outflows, but $v_{gas}(R_c)/v_{gas}(R_2) = 2$ is required for 
the velocity range of shocked ISM in our model to match the observations of J0300. (Note that $v_{gas}(Rc)=v_{Rc}$, but that $v_{gas}(R_2)=0.75v_{R2}$.)

We now discuss possible ways around this limitation. All but the last two appear to be ruled out in J0300.
    
    \citet{clm02} 
    state that a weak shock can produce $R_c/R_2 \simeq 0.45$, which might lead to $v_{Rc}/v_{gas}(R_2) \simeq 1/0.45 = 2.2$.  A weak shock in this case is defined as $\Delta E/E \simeq 0.3$, where $\Delta E$ is the injected energy and $E$ is the preexisting thermal energy of the ISM.  It seems unlikely that a weak shock could result from a quasar outflow with an initial velocity of 10,000 \kms\ or higher unless a new outflow was injected into an old shock bubble that had not yet cooled substantially.

Might magnetic fields alter the expected velocity range?
\citet{1975MNRAS.172...55F} finds that for weak fields ($<20~\mu$G), ``the main effect of the field is to increase the effective ratio of specific heats [$\lambda_{sa}$]. This leads to a faster expansion and, in this case at least, is more important than the tension along the field lines which ought to retard the expansion." Even for strong fields, a realistic maximum value of $\lambda_{sa}=2$ (Figure 2 of \citealt{ChaoWiskerchen1974}) as compared to $\gamma_{sa}=5/3$ results in a value of $\lambda_c$ only 4.2\% smaller and a value of $A_E$ only 2.60\% larger, yielding only a $\sim$2\% larger range of velocities in the shocked ISM.  

However, \citet{2010ApJ...717L..98Bautista} argue that turbulence from supersonic ionization fronts caused by variations in a quasar's luminosity can produce velocity structure in BAL gas that scales roughly with the Alfven velocity in the gas. The observed \caii\ velocity span of 2000~\kms\ in J0300 would require magnetic field strengths of up to order 40 mG.  Strengths of up to 10 mG have been postulated in models of BAL outflows; see  \citet{2010ApJ...717L..98Bautista} \S~4.

We investigated the possibility is that J0300 is observed in the brief phase around the shell formation time $t_{\rm shell}$ when the shocked ISM bubble is losing pressure support and collapsing.  According to section 5 of
\citet{1975A&A....43..323F}, during this phase ``the [outer] shock decelerates and the contact discontinuity accelerates'' (as seen between times 2 and 4 in their Figure 3d). 
Because the gas just behind the outer shock is initially moving more slowly than the gas near the contact discontinuity, the above deceleration and acceleration combine to yield a larger velocity spread in the shocked ISM than either before or after the collapse.
Figure 3d of \citet{1975A&A....43..323F} shows a maximum velocity range of $\sim 2.4$ during this collapse phase, sufficient to reproduce that of strong \caii\ in J0300.
If this explanation for the velocity range of the \caii\ in J0300 is correct, then differential changes in velocity across the outflow are expected: first, a larger deceleration at the lowest velocities and a smaller deceleration or even acceleration at the highest velocities, followed by an acceleration at the lowest velocities and a deceleration at the highest velocities as the shocked ISM velocities converge to the range $0.75v_{R2} \leq v \leq v_{R2}$ (ignoring any effects on $v$ or ionization from shocks in the collapsing shell). Most of the gas will have cooled and accelerated to $v_{R2}$; only recently swept-up gas will have $0.75v_{R2} \leq v < v_{R2}$.

Our data do offer an intriguing, but not statistically significant, suggestion of deceleration at low velocities and acceleration at high velocities in the outflow.
However, the values found are not consistent with the above hypothesis, as we now show.

The timescale of shell formation is found to be $t_{\rm collapse}\simeq 0.2 t_{\rm shell}$ in the simulations of \citet[][their Figure A2]{2018MNRAS.478.3100Richings}, consistent with Case A of Falle 1975 (their Figure 3).  
In this scenario gas should shift from a velocity range $v_{Rc}/v_{gas}(R_2)=4\lambda_c/3$ to $v_{Rc}/v_{gas}(R_2)=2.4$, and then to $v_{Rc}/v_{gas}(R_2)=4/3$, each in less than 
$t_{\rm collapse}/2\simeq 0.1 t_{\rm shell}$
(and possibly much less).  
We consider an example in which the gas shifts from a range of $v_{gas}(R_2)=2600$~\kms\ and $v_{Rc}=3000$~\kms\ prior to shell formation, to $v_{gas}(R_2)=1900$~\kms\ and $v_{Rc}=4540$~\kms\ during shell formation, to $v_{gas}(R_2)=2200$~\kms\ and $v_{Rc}=2900$~\kms\ after shell formation.
For $t_{\rm shell}=10^{4}$~yr, the velocity adjustments should occur over $<10^{3}$~yr. Initially, this means decelerations of $<-0.7$~\kmsyr\ at low velocities and accelerations of $>+1.5$~\kmsyr\ at high velocities. 
At the end of the shell formation phase, this means accelerations of $>+0.3$~\kmsyr\ at low velocities and decelerations of $<-1.6$~\kmsyr\ at high velocities.
Those values are all above our $3\sigma$ observational upper limit of $|a|<0.1$~\kmsyr\ and can be ruled out.

Finally, we can think of two ways 
to explain the velocity range of the \caii\ outflow in J0300 in the context of our model.

First, the \caii-absorbing gas could be found in the shocked wind region as well as the shocked ISM region. The gas can thus span a larger velocity range, albeit at lower column densities (e.g., Fig.\ \ref{fig:nVSv}).

Second, preexisting relatively dense clouds of gas in the ISM may been accelerated from rest to their observed velocities in J0300 by the ram pressure of the surrounding shocked low-density ISM gas. We consider this further in \S \ref{sec:ismaccel}.

\subsection{Comparison to Choi et al.} \label{sec:compare-Choi}

J0300 has also been studied in \citet{felobal1} as one of 50 FeLoBAL quasars modeled using the spectral-synthesis code {\sl SimBAL} \citep{simbal} to solve for the physical conditions in the gas producing the observed absorbing columns in those quasars.
\citet{felobal1} find that the best fit for J0300 has the absorbing gas at $1800<v<10600$~\kms\ located at $12.0\pm 0.3$ pc from J0300 with an ionization parameter $\log U=-1.85$, a density of $\log (n/{\rm cm}^{-3}) =7.96^{+0.01}_{-0.02}$, and a column density $\log (N_H/{\rm cm}^{-2}) =22.31$ (implying a thickness $dr=7.3\times 10^{-5}$ pc).
Notably, the UV-absorbing outflow in J0300 has the highest density, lowest ionization parameter, and largest distance inferred among the 8 overlapping-trough FeLoBALs studied by \citet{felobal1} and in their full sample of 50 objects it has the largest velocity width and ties for the highest density.
J0300 is also an exception to the finding of \citet{felobal1} that \caii\ absorption is generally found in BAL outflows located at kpc scales.
However, while the \citet{felobal1} {\sl SimBAL} model for J0300 does an excellent job of fitting \mgii\ and \feii\ absorption, it overpredicts the level of \caii\ absorption at $v>4000$~\kms\ and does not predict the strong \caii\ at $2000<v<4000$~\kms.
[Note added in proof: We are grateful to Dr.\ Choi for pointing out that the model absorption we presumed to be \caii\ at $v>4000$~\kms\ is in fact \ion{He}{1}* absorption at lower velocity. Such absorption is absent or at best very weak in the observed spectrum.]
The latter is likely due to the use of a {\sl SimBAL} model with a single ionization parameter and a single density. 

As a fiducial bolometric luminosity for J0300, we adopt $L_{AGN}=1.32\times 10^{47}$ erg s$^{-1}$ as estimated by \citet{felobal1} 
using a bolometric correction from a rest-frame 3~$\mu$m flux value found by interpolating WISE data.
We assume a black hole mass of $3\times 10^9 ~M_\odot$, appropriate for luminous quasars at this redshift \citep{2011ApJS..194...42R}, which yields an Eddington ratio of $f_{Edd}=0.3$, $\dot{M}_{Edd}=45$ $M_\odot$ yr$^{-1}$, and $\dot{M}_{acc}=13$ $M_\odot$ yr$^{-1}$. 
For an initial wind velocity $v_{in}$=20,000~km\,s$^{-1}$, J0300 would have $\dot{M}_w=35\tau_{in}$ $M_\odot$ yr$^{-1}$.
This yields a larger mass outflow rate from the disk/torus than the accretion rate through them for $\tau_{in}> 0.37$. 

The outflow in our model includes both mass in the wind and in the swept-up ISM. 
The total mass of swept-up ISM gas is given approximately by FGQ12 Eq.\ (A2):
\begin{equation}
M_{sa}(t)=\frac{4\pi\mu m_p n_0 R_0^\alpha}{3-\alpha}R_2^{3-\alpha}.
\end{equation}
Anticipating results later in this section, for comparison to \cite{felobal1} we further adopt $R_2=12$~pc, $R_0=10$~pc, $n_0=10^3$~cm$^{-3}$ and $\alpha=0.8$, for which $M_{sa}=2.6\times 10^5$~$M_\odot$ as compared to $M_w=5.1\times 10^4$~$M_\odot$ in the 1450~yr needed for such an outflow to reach $R_2=12$~pc.
The rate at which shocked ambient medium gas is swept up by the outflow is
\begin{equation}
\dot{M}_{sa}(t) = 4\pi \mu m_p n_0 R_0^\alpha R_2^{2-\alpha} v_{R2}
\end{equation}
and the instantaneous kinetic luminosity of newly swept-up shocked ambient medium gas is
\begin{equation}
L_{k,sa}(t) = \frac{1}{2}\dot{M}_{sa} v^2_{gas}(R_2)  = \frac{9}{8}\pi \mu m_p n_0 R_0^\alpha R_2^{2-\alpha} v_{R2}^3.
\end{equation}
For the above outflow, 
$v_{R2}=5800$~\kms, $v_{gas}(R_2)=4400$~\kms,
$\dot{M}_{sa}=320\tau_{in}^{(3-\alpha)/(5-\alpha)}$ $M_\odot$~yr$^{-1}$ 
and 
$L_{k,sa} = (1.9\times 10^{45})\tau_{in}$ erg~s$^{-1}$,
which for $\tau_{in}=1$ is 1.4\% of the adopted bolometric luminosity of J0300.
These quantities depend on more than just $\tau_{in}$, of course, but we quote the $\tau_{in}$ dependencies to compare them with $\dot{M}_{w}\propto\tau_{in}$ and $L_{in}\propto\tau_{in}$.
The kinetic luminosity of the quasar wind in the above outflow is $L_{in}=4.4\times 10^{45}$ erg s$^{-1}$ for $\tau_{in}=1$, which is expected since approximately half the wind kinetic energy goes into the kinetic motion of the swept-up gas (W77, KM92, FGQ12).

\cite{felobal1} use their fitted $N_H$, $R$ and $v$ to infer an mass outflow rate in J0300 of 95($\Omega$/0.2) $M_\odot$ yr$^{-1}$ and a kinetic luminosity of $L_{KE}=(1.7\times 10^{45})(\Omega/0.2)$ erg s$^{-1}$, and a total outflowing mass of $M=8.3\times 10^4$ ($\Omega$/0.2) $M_\odot$, where $\Omega$ is the fraction of $4\pi$ steradians covered by the outflow. 
Scaling to $\Omega=1$ to match our comparison outflow, the \cite{felobal1} parameters yield a 1.4$\times$ larger mass outflow rate, a 1.6$\times$ larger total mass, and a 3.6$\times$ larger kinetic luminosity than our values for $\tau_{in}=1$, consistent with the larger value of $v$ they use.

\subsubsection{Searching for matching parameter choices} \label{sec:search}

We investigated whether our model could match the outflow properties inferred by \citet{felobal1} for J0300.
We ran $10^7$ calculations of the predictions of our model for randomly sampled values of $L_{AGN}$, $t$, $v_{in}$, $\tau_{in}$, $n_0$, $R_0$, and $\alpha$ over the ranges considered in this work (\S~\ref{sec:vary}).
We then extracted for comparison all calculations with $R_c$, $v_{Rc}$, and $L_{AGN}$ matching the values estimated by \citet{felobal1} with a maximum total deviation of a factor of two:
$[\log (R_c/12 \text{ pc})]^2 + [\log (v_{Rc}/3000 \text{ \kms})]^2 + [\log (L_{AGN}/10^{47.12} \text{ erg s}^{-1})]^2 \leq [\log 2]^2.$
Recall that in our model for J0300, the gas absorbing in \caii\ (primarily at $2000 < v < 4000$~\kms) is assumed to share the velocity and deceleration of the shocked ISM gas at $R_c<r<R_2$ due to having been condensed out of that gas (or, alternatively, having been accelerated to match its velocities).

We then further filtered our results by including only those parameter combinations in which the immediate post-shock acceleration $a_{post}$ lies below our observed upper limit, which excluded all but a small fraction of parameter combinations. Here, $a_{post}$ serves to analytically approximate the minimum deceleration (the deceleration closest to zero) found in the shocked ambient region.
For our assumptions, we find the maximum deviation to be only $1.7$\% at $\alpha = 0$, down to $0.03$\% at $\alpha = 1.5$. Among the parameter combinations passing this filter of $a_{post}>-0.1$~\kmsyr, we find a median $a_{post}=-0.081$~\kmsyr\ and a minimum deceleration of $a_{post}=-0.013$~\kmsyr.

The filtered comparison cases which most closely match
$R_c = 12$~pc and $v_{Rc}=3000$~\kms\ have $\log (t/{\rm yr}) =3.55 \pm 0.05$,
$\tau_{in} < 0.1$,
$n(r)\simeq 4\times 10^3$~cm$^{-3}$ and
$\alpha \simeq 0.8 \pm 0.1$,
implying (e.g.) $n_0 \simeq 10^{3}$~cm$^{-3}$ at $R_0 = 10$~pc.
We do find some filtered comparison cases with $t_{\rm shell}\simeq 10^{3.6}$~yr; such cases have higher $\tau_{in}$ and larger $R_2$ than the closest-match cases discussed above.

Gas cooling is required to explain the fitted value of $n=10^8$~\pcc\ found by \citet{felobal1} for the J0300 outflow. 
That density cannot be reached only through shock compression of preexisting ISM clouds, which have $n\sim 10^{2-3}$~\pcc\ with rare cores reaching $n\sim 10^{4-6}$~\pcc\ (\S~III.B.\ of \citealt{f01}).
Shock compression by itself will only cause at most a factor of four increase in gas density. 
Any further significant density increase must come as a result of cooling and cooling-induced compression (see, e.g., \citealt{FIREclumps23}).
Recall from \S~\ref{sec:cool} that density increases after shell collapse of up to a factor of $10^5$ were found by \citet{2018MNRAS.474.3673Richings}.

Overall, for our model in which the gas absorbing in \caii\ is assumed to share the velocity and deceleration of the shocked ISM gas, we can find parameter choices that simultaneously match most of the observed and inferred properties of the J0300 outflow, with the exception of the density inferred by \citet{felobal1}.
However, only a small subset of parameter choices matching some constraints match the rest. 
Given that and the fact discussed in \S\,\ref{sec:vrange} that the shocked ISM in our model lacks a sufficiently large velocity range to match the velocity range spanned by \caii\ in this object, we conclude that our initial model of \caii\ absorption arising in the shocked ISM region of a self-similar outflow does not accurately represent the J0300 outflow.

\subsubsection{Observed vs.\ Model Column Densities} \label{sec:obsmodelcolden}

The parameters in the previous section can be used to calculate the swept-up ISM column density for this outflow at age 1450~yr by summing up $N(v)$ values calculated following \S~\ref{sec:Nv}.
We find a total $\log (N_H/{\rm cm}^{-2})=22.21$, in good agreement with the \citet{felobal1} fit value of $\log (N_H/{\rm cm}^{-2}) =22.31$. 
For the column density through the shocked quasar wind, we find $\log (N_{sw}/{\rm cm}^{-2})=21.47$.

We can also check that our model predicts a total Ca\,{\sc ii} column density consistent with the values measured for this target in \citet{sb2}:
$\Ncaii=(7.1\pm 1.1)\times 10^{14}$ cm$^{-2}$ ($\log N=14.85$) over $1697<v<5657$ \kms.
The relative solar abundance of Ca is $\rm\log (Ca/H)=-5.66$
\citep{2009ARA&A..47..481A}.
We subtract that value from $\Ncaii$ to find the absolute lower limit of $\log N_H$ in the shocked ISM gas: $\log N_H=20.51$. 
This is a lower limit mainly because only a small fraction of calcium will be in the \caii\ ionization stage even at high densities and column densities. E.g.,  \citet{2011NewA...16..128R} find a maximum \caii\ ionization fraction of $\sim 1$\%, suggesting $\log N_H\simeq 22.51$ as a more likely value -- again in reasonable agreement with the estimate of \citet{felobal1}.

For the column density through the unshocked quasar wind $N_w$ (\S~\ref{sec:unshockedcolden}), we find $\log (N_w/{\rm cm}^{-2})=22.78$ for $\tau_{in}=1$.
Note that the unshocked wind must be sufficiently ionized such that no absorption from it is seen at line-of-sight velocities it holds along our line of sight.
The exact level of ionization in the unshocked wind is not well constrained for J0300 because of the overlapping troughs shortward of rest-frame 2800\,\AA. The only transitions in which we might see absorption at high velocity are \caii\ and H$\beta$. \caii\ does not show evidence for any absorption at $v>5700$~\kms, and the rest-frame optical spectrum shows strong \feii\ emission but no sign of H$\beta$ absorption (Figure 2 of \citealt{sb2}).

\subsection{Acceleration of preexisting ISM clouds} \label{sec:ismaccel}

It is possible that the velocity range of the low-ionization BAL gas in J0300 is larger than our model predicts because our model does not account for the survival of sufficiently dense preexisting clouds of gas in the ISM. Such clouds will undergo a combination of acceleration and disruption
\citep[e.g.,][]{fqm12,2020MNRAS.491.4325Z,cb23,ChenOh23}.
The gas in such clouds will require time to accelerate to the velocity of the surrounding shocked lower-density ISM, potentially resulting in some absorbing gas having lower velocities for a given outflow age than are predicted by our model.

\citet{fqm12} show that FeLoBAL absorber properties can be explained by preexisting ISM clouds with the right physical properties shocked by a quasar outflow.
We define $n_{cl,i}$ and $n_0(r)$ as the initial cloud and surrounding ISM densities and we define their ratio as $\chi=n_{cl,i}/n_0(r)$.
When a shock wave with velocity $v_{R2}$ in the surrounding ISM overtakes a preexisting cloud at radius $r$, the shock speed inside the cloud, $v_{sh,cl}$, is lower by a factor $\chi^{1/2}$ and the post-shock temperature inside the cloud is lower by a factor $\chi$: $T_{cl}=T_{gas}/\chi$, with $T_{gas}=6 \times 10^8 {\rm ~K~} (v_{gas}/5000 {\rm ~km~s^{-1}})^2$ (\citealt{fqm12} Eq.\ 2).

The evolution of such clouds is governed by several timescales.
Our discussion of these timescales below is based on that of \citet{fqm12}.

The cloud-crushing timescale $t_{cc}$ for a cloud of radius $r_{cl}$ is the time for the shock to travel $r_{cl}$:
\begin{eqnarray} \label{eq_tcc}
    t_{cc}&=&\frac{r_{cl}}{v_{sh,cl}} = \frac{r_{cl}}{v_{R2}}\sqrt{\chi}
    = \frac{3r_{cl}}{4v_{gas}(r)}\sqrt{\chi}\\
    &=& 44~{\rm yr} \left[\frac{r_{cl}}{\rm 0.01~pc}\right] \left[\frac{v_{gas}(r)}{\rm 5000~km~s^{-1}}\right]^{-1} \left[\frac{\chi}{900}\right]^{1/2}\nonumber
\end{eqnarray}
using $v_{gas}=3v_{R2}/4$.  
Even neglecting ram pressure acceleration, after one cloud-crushing timescale we expect that the entire cloud has $n_{cl}=4n_{cl,i}$ and is moving at least at velocity $v_{cl}(t_{cc})=0.75v_{R2}\sqrt{\chi}$.

The cloud will be accelerated by ram pressure to roughly the speed of the hot gas on the timescale over which the 
(assumed spherical)\footnote{An oblate cloud will have a shorter $t_{cc}$ and $t_{drag}$ than a spherical cloud of equal mass and density. E.g., they are 3.2 and 3.5 times shorter, respectively, in the limit of a face-on disk-shaped cloud of diameter ten times its thickness.} 
cloud intercepts a mass of hot gas equal to its initial mass:
\begin{eqnarray}
    t_{drag}&=&\frac{\frac{4}{3}\pi r_{cl}^3 n_{cl,i}}{4n_0\pi r^2_{cl} v_{gas}(r)}
    =\frac{r_{cl}\chi}{3v_{gas}(r)}
    =\frac{4\sqrt{\chi}}{9}t_{cc}\\
    &=&{\rm 600~yr}~\left[\frac{r_{cl}}{\rm 0.01~pc}\right] \left[\frac{v_{gas}(r)}{\rm 5000~km~s^{-1}}\right]^{-1} \left[\frac{\chi}{900}\right].\nonumber
\end{eqnarray}

A cloud which has not matched speeds with the hot gas is expected to be destroyed by velocity shear via the Kelvin-Helmholtz instability on a timescale 
\begin{equation} \label{eq_tKH}
    t_{\rm KH} = \frac{8}{3}\kappa_{\rm KH} t_{cc}
\end{equation}
where $1 \lesssim \kappa_{\rm KH} \lesssim 10$. Greater post-shock cooling increases $\kappa_{\rm KH}$.

Finally, the post-shock cloud will cool on a timescale 
\begin{equation}
    t_{cool,cl}={\rm 1.5\times 10^4~yr~} \left(\frac{T_{gas}/\chi}{\rm 10^6~K}\right)^{1.6} \left(\frac{4n_{cl,i}}{10~{\rm cm^{-3}}}\right)^{-1}
\end{equation}
(Eq.\ 14 of \citealt{fqm12}) where $T_{gas}$ is the temperature of the surrounding shocked gas.

For preexisting clouds to end up visible as FeLoBAL absorbers containing relatively cool and low-ionization gas, 
\citet{fqm12} posit that they must match speeds with the hot gas before being disrupted ($t_{drag}<t_{\rm KH}$)
and must cool by the time the shock passes through the cloud ($t_{cool,cl}<t_{cc}$), so that they leave cool material behind after their disruption.\footnote{\cite{2018MNRAS.480L.111Gronke,2020MNRAS.492.1970G} find that sufficiently large clouds ($r_{cl}\gtrsim 2$~pc) can mix their cool gas with entrained hot gas which cools and enables the clouds to survive, accelerate, and grow as elongated structures with a range of densities (their Fig.\ 2). 
\cite{Xu+23} confirmed this prediction for a sample of starburst galaxies.
} 

The requirement $t_{cool,cl}<t_{cc}$ yields a lower limit on the cloud hydrogen column density $N_{cl}=2r_{cl}n_{cl,i}$ (\citealt{fqm12} Eq.\ 15) for clouds to survive:
\begin{equation}\label{Ncl-vgas-chi}
   N_{cl}\gtrsim 1.4\times 10^{20}~{\rm cm}^{-2} \left(\frac{v_{gas}(r)}{\rm 5000~km~s^{-1}}\right)^{4.2}
   \left(\frac{\chi}{900}\right)^{-2.1}.
\end{equation}
That is, a factor of two increase in wind velocity will result in clouds a factor of sixteen higher in column density not cooling before disruption, unless the clouds also have at least a factor of four higher relative density.
The \citet{felobal1} fit value of $\log N_H = 22.31$ and the observed values of $v_{gas}=3000\pm 1000$~\kms\ imply a range of $\chi>30^{+24}_{-16}$ to satisfy the predicted column density lower limit for J0300.  

\citet{fqm12} assume that the original clouds will eventually fragment into cloudlets comoving and in pressure equilibrium with the hot gas at temperatures $T_{cl}\simeq 10^4$~K and densities
\begin{equation}
   n \simeq 3000~{\rm cm}^{-3} \left(\frac{n_0(r)}{0.02~{\rm cm}^{-3}}\right) \left(\frac{v_{gas}}{\rm 5000~km~s^{-1}}\right)^2.
\end{equation}
Note that our comparison model of J0300 satisfies that prediction for $T_{cl}\simeq 10^4$~K with $n=10^8$~\pcc, $n_0=10^3$~\pcc, and $v_{gas}=4400$~\kms. 
Our comparison model has $n_0=10^3$~\pcc\ and $T_{gas}=4.6\times 10^8$~K, so pressure equilibrium with a cloud of density $n=10^8$~\pcc\ would be achieved at $T_{cl}=4600$~K.

The requirement $t_{drag}<t_{\rm KH}$ yields an upper limit on $\chi = \frac{n_{cl,i}}{n_0(r)}$, because clouds that are sufficiently denser than their surroundings will be disrupted before they reach the hot gas velocity. 
However, as pointed out by \cite{he10}, shocked clouds expand laterally and increase their effective area for ram pressure acceleration (their Fig.\ 1).
In the limit of a cloud that deforms its shape at constant density and increases its lateral radius by a factor of $f_r$ in time $t_{\rm KH}$ and therefore decreases its $t_{drag}$ by a factor of $f_r^2$, the values of $\chi$ that produce $t_{drag}<t_{\rm KH}$ are
\begin{equation}\label{chi-kappa}
    \chi 
    \lesssim 36\kappa_{\rm KH}^2f_r^2 
    ~~{\rm or}~~ 
    \chi \lesssim (36-3600)f_r^2.
\end{equation}
A cloud with any value of $\chi$ that maintains that $\chi$ while increasing its lateral radius by a factor of $f_r>\sqrt{\chi^{0.5}/6\kappa_{\rm KH}}$ in time $t_{\rm KH}$ will satisfy $t_{drag}<t_{\rm KH}$. 
A cloud expanding laterally at a fraction $f_v$ of the shock speed inside the cloud will reach that value of $f_r$ for $\chi<1800f_v\kappa_{\rm KH}^6$; i.e., $\chi<1.8f_v\times (10^3 - 10^9)$.  
That weak constraint suggests it is possible that clouds will expand and match speeds with the hot gas before being disrupted.

More recently, \cite{2020MNRAS.491.4325Z} have simulated in 2-D the interaction between 
a preexisting cloud of size $\sim 1$~pc and $n_{cl,i}=10^6$~cm$^{-3}$ embedded in gas of density $n_0=2.5$~cm$^{-3}$ (yielding $\chi=4\times 10^5$)
with a quasar wind of kinetic luminosity $10^{47}$ erg~s$^{-1}$, density $\simeq 10$~\pcc\ (their Fig.\ 2), and velocity 30,000~\kms. 
Although this is a larger velocity contrast between the cloud and the surrounding flow than assumed in our model, in their simulation a bow shock forms between the cloud from the wind and decelerates gas in the wind to $3000 \lesssim v \lesssim 9000$~\kms\ in the immediate environs of the cloud (M.\ Zeilig-Hess, personal communication).

The cloud studied by \cite{2020MNRAS.491.4325Z} fragments into cloudlets of size $\sim 0.01-0.10$~pc (0.03~pc on average, with a resolution limit of 0.01 pc; their Fig.\ 6) by time $\sim 4t_{cc}$ (65,000 yr; their Fig.\ 2).
By that time the distribution of cloudlets has expanded laterally by a factor $\simeq 5-10$ beyond the initial size of the cloud, although this expansion has occurred simultaneous with the cloud's disruption instead of preventing it. 
\cite{2020MNRAS.491.4325Z} identify these cloudlets as producing BAL troughs with significant velocity structure in their Fig.\ 8. 

The cloudlets have lower densities than the original cloud, though this is partly due to an artificial limit on achieving high densities through cooling (the simulation has a floor temperature of $3\times 10^5$~K).
With smaller sizes and lower densities, $t_{drag}$ for the cloudlets is less than or equal to $2t_{cc}$ for the cloud as given by Eq.\ \ref{eq_tcc},\footnote{\cite{2020MNRAS.491.4325Z}
define $t_{cc}$ to be twice our value.} so the cloudlets can accelerate on the timescale on which they are produced.  By a time $t=4t_{cc}$ for $r_{cl}=1~{\rm pc}$ in the simulation ($t=65,000$~yr),
most cloudlets have accelerated to velocities 1000~\kms\ to 3000~\kms, with ``much slower'' acceleration at later times (their \S\,3.2 and Fig.\ 4) and ``somewhat smaller'' velocities in a simulation with wind velocity 9,000~\kms\ instead of 30,000~\kms. 
Note that the cloudlets in the simulations of \cite{2020MNRAS.491.4325Z} achieve essentially terminal velocities which are smaller than the flow in which they are embedded by a factor of $\sim$3.

In summary, it is reasonable to expect that a combination of lateral expansion and acceleration of cloud fragments (cloudlets) can result in cool gas surviving acceleration to be seen at velocities of thousands of ${\rm km~s}^{-1}$.
This mechanism has been proposed by \cite{Guillard+2009} to explain observations of Stephan's Quintet (see their \S 3 and Fig.\ 1), wherein small-scale cold $H_2$ structures are found embedded in warm gas behind shocks of $v<1000$~\kms\ thought to originate from galaxy-galaxy interactions; see also \cite[\S 5.1 and Fig.\ 13 of][]{Appleton+23}.
Fragmentation of clouds into cool cloudlets with a range of bulk velocities can explain the velocity structure of multiple narrow absorption features over a wide range of velocities seen in some BAL outflows such as the \caii\ outflow of J0300 (Figure \ref{fig:2spec}).

\subsubsection{Comparison to the J0300 acceleration upper limit}\label{sec:compare-accel}

\begin{table*}
\centering 
\begin{tabular}{c c c} 
\hline\hline  
\multicolumn{3}{c}{J0300 Comparison Model Outflow Parameters}\\
\hline\hline  
\multicolumn{3}{c}{\textbf{Observationally Inferred Parameters}}\\
\hline 
$L_{AGN}$ & bolometric luminosity & $10^{47.12}$ erg s$^{-1}$\\
$R_2$ & shock bubble outer radius & $12.0$~pc\\ 
$n_{cl}$ & final cloud \# density & 10$^8$ cm$^{-3}$\\
\hline\hline  
\multicolumn{3}{c}{\textbf{Observational Comparison Parameters}}\\
\hline 
$\alpha$ & ISM density power law exponent & 0.8\\
$R_0$ & ISM density reference radius & 10 pc\\
$n_0$ & ISM $\text{H}$ nucleus \# density at $R_0$ & 10$^3$ cm$^{-3}$\\
$v_{in}$ & wind launch velocity & 20,000 \kms\\
$\tau_{in}$ & wind optical depth & 1\\
$\dot{M}_w$ & wind mass loss rate & 35 $M_\odot$ yr$^{-1}$\\
$L_{in}$ & wind kinetic luminosity & $10^{45.64}$ erg s$^{-1}$\\
\hline\hline  
{} & \textbf{Assumed Parameters} & {}\\
\hline 
$M_{BH}$ & black hole mass & $3\times 10^9$~$M_\odot$\\ 
$\eta_r$ & accretion disk radiative efficiency & 0.175\\
$r_{cl}$ & cloud radius & 0.005 pc\\
$\chi$ & initial cloud overdensity & $10^4-10^5$\\
\hline\hline  
\multicolumn{3}{c}{\textbf{Resultant Parameters}}\\
\hline 
$\lambda_c$ & ratio $R_c/R_2$ & 0.854\\ 
$R_f$ & outflow free expansion radius & $0.52$~pc\\ 
$t_f$ & outflow free expansion time & $25$~yr\\ 
$t$ & outflow age & $1450$~yr\\ 
$R_1$ & wind shock radius & $7.62$~pc\\ 
$R_c$ & contact discontinuity radius & $10.3$~pc\\ 
$t_{csa}$ & shocked ambient medium crossing time & $360$~yr\\ 
$v_{gas}(R_2)$ & gas velocity at $R_2$ & $4400$~\kms\\ 
$T_{gas}$ & shocked ISM temperature at $R_2$ & $4.6\times 10^8$~K\\ 
$t_{cc}$ & cloud crushing time & $80-270$~yr\\ 
$t_{\rm KH}$ & cloud disruption time ($\tau_{\rm KH}=10)$ & $2200-7100$~yr\\ 
$t_{drag}$ & cloud drag time & $3800 - 38,000$~yr\\ 
%
\hline 
\end{tabular}
\caption{Parameters for our model of the outflow in J0300. Note that in the text we also explore results for larger values of $r_{cl}$ and smaller values of $\chi$.}
    \label{t_j0300params}
\end{table*}

We now explore further whether the acceleration of preexisting ISM clouds can explain our observations of J0300.
We discuss parameter value combinations that would produce ram pressure acceleration below our observed acceleration upper limit.
We estimate various timescales for the gas clouds and compare them to the age of the outflow inferred to exist in J0300.
Finally, we estimate the velocities that swept-up gas could have when it exits the shocked-ambient-medium region of the shock bubble.

The parameters of our comparison model of J0300 are summarized in Table \ref{t_j0300params}.
Observationally inferred parameters are taken from \citet{felobal1}.
Observational comparison parameters, assumed parameters, and resultant parameters are discussed in \S \ref{sec:compare-Choi}, \S \ref{sec:search}, and in this section.
Note in particular that the age at which our comparison model of J0300 has reached $R_2=12$~pc is $t=1450$~yr. 
Also note that we assume $\tau_{in}=1$ despite most comparison cases having $\tau_{in}<0.1$.

First, we estimate a value of $r_{cl}$ for cloudlets observed in J0300 (which may or may not be the same size as the original clouds). 
This value must be comparable to or smaller than the size of the continuum-emitting region at 3900\,\AA\ rest frame, since the \caii\ absorption covers $\sim30$\%$-90$\% of that region (Fig.\ 6 of \citealt{sb2}).
For a radiative efficiency $\eta_r=0.175$, the predicted half-light radius of the emission region at 3900\,\AA\ in J0300 is 0.005 pc \citep{2015ApJ...798...95B}.
We therefore adopt a cloudlet radius of 0.005~pc.

Microlensing studies indicate that quasars may have half-light radii a factor of $\simeq$4 larger than predicted by thin disk theory \citep{ad}. 
That sets either an upper limit on the individual cloud size or a requirement that more than one cloudlet is present at most velocities.
On the other hand, the cold outflowing gas could reside in a `mist' of even smaller cloudlets: \cite{McCourt+17} find a minimum cold gas length scale $\ell_{\rm shatter}=c_{s,cl}t_{cool,cl}$; cloudlets of this length scale or smaller do not shatter further.
This length scale is only 500 km for gas with $n=10^8$~cm$^{-3}$ and $T=10^4$~K.

For comparison with our results, the ram pressure acceleration expected for a stationary cloud (in the rest of this section we do not need to distinguish between clouds and cloudlets) of radius $r_{cl}$ surrounded by shocked ambient gas moving at velocity $v_{gas}(r)$ is given by
\begin{eqnarray}\label{eq_aram}
    a_{ram} &=& \frac{v_{gas}(r)}{t_{drag}}\\
    &=& 8.3 {\rm~km~s^{-1}~yr^{-1}}\nonumber\\
    &\times&
    \left[\frac{v_{gas}(r)}{\rm 5000~km~s^{-1}}\right]^2
    \left[\frac{\rm 0.01~pc}{r_{cl}}\right] 
    \left[\frac{900}{\chi}\right].\nonumber
\end{eqnarray}
For a cloud moving with velocity $v_{cl}$, the above becomes
$a_{ram}\propto (\Delta v)^2$, where $\Delta v = v_{gas}(r)-v_{cl}$.

For ram pressure acceleration to lie below a given acceleration upper limit requires 
\begin{equation}\label{eq_Deltav}
   \Delta v <  1740 \left[
   \frac{a_{ram}}{\rm 1~km~s^{-1}~yr^{-1}}
   \frac{r_{cl}}{\rm 0.01~pc}
   \frac{\chi}{900}\right]^{1/2} {\rm ~km~s^{-1}}.
\end{equation}
In our case, assuming gas clouds that originated with $n=10^8$~cm$^{-3}$, we have a most likely value in the range $10^4\lesssim\chi\lesssim 10^5$ (\S~\ref{sec:search}). To match our observed limit of $|a|<0.1~{\rm~km~s^{-1}~yr^{-1}}$ requires $\Delta v<4100$~\kms\ for $r_{cl}=0.005$~pc and $\chi=10^5$. This is not a particularly useful constraint, but it does confirm that sufficiently dense gas clouds can be effectively coasting even when surrounded by fast-moving gas.
The disruption timescale for such a cloud would be $t_{\rm KH}>(750-7500)$~yr. Such clouds are unlikely to have been fully disrupted yet in this young outflow, though that will likely happen eventually and partial disruption can occur in the meantime (e.g., \cite{2020MNRAS.491.4325Z} Fig.\ 2, top right and bottom left panels). 

In our comparison model of J0300, assuming $r_{cl}=0.005$\,pc and $\chi=10^5$ yields $t_{cc}=270$~yr, $t_{\rm KH}=710-7100$~yr, and $t_{drag}=(3.8\times 10^4)f_r^{-2}$~yr.
If such a cloud undergoes lateral expansion by a factor of at least $f_r=2-7$ in radius, then it will have $t_{drag}\leq t_{\rm KH}$.
However, the drag time for $10^4\lesssim\chi\lesssim 10^5$ is much greater than the age of the outflow, meaning that the clouds in J0300 could not have been accelerated to their observed speeds at their inferred densities and distance.
Increasing $v_{in}$ cannot alleviate this problem because in such cases the time needed for the outflow to reach 12~pc decreases by the same factor by which $v_{gas}(R_2)$ increases and the above timescales decrease, meaning that the drag time remains long relative to the age of the outflow.

For swept-up ISM clouds to have survived to cause \caii\ absorption in J0300 at a distance of 12 pc from the central black hole, they must have begun accelerating as clouds with low $\chi$ before reaching their observed velocities, their low observed accelerations, and their inferred large $\chi_{cl}$, where $\chi_{cl}$ is the instantaneous overdensity of the cloud.
For the clouds considered above, $t_{cool,cl}<1$~yr, so cooling and contraction can reasonably increase the clouds' overdensities relative to their surroundings.
Note that gas clouds of constant mass which contract by a factor of $g_r$ in radius will have larger $\chi_{cl}\propto g_r^{-3}$, longer $t_{\rm KH}\propto g_r^{-1/2}$, and longer $t_{drag}\propto g_r^{-2}$.

Note that it would be possible for preexisting clouds of density $n\sim 10^8$~\pcc\ to reach the velocities observed in J0300 if the outflow is older than we infer, with age $t \simeq t_{drag} \simeq 3800 - 38,000$~yr. In that scenario the clouds are located at 12~pc but the outer edge of the shock bubble is located much farther away. The only potential origin for such dense clouds would be the quasar accretion disk because the highest ISM density observed outside of AGN accretion disks is only $\simeq 10^5$\,cm$^{-3}$ \citep{hq10} and even proto-stellar cores in molecular clouds only reach gas-phase densities of $\simeq 10^7$\,cm$^{-3}$ \citep{2007prpl.conf...17D}.

\subsubsection{Acceleration in the shocked ambient medium region}\label{sec:accel-csa}

Following on the results above, we explore a few parameter combinations including smaller $\chi$ (by at least an order of magnitude) that might make it possible for the \caii-absorbing gas to have been accelerated to its observed velocities during its passage through the shocked ambient medium region of our model outflow.
After that time, the ram pressure acceleration will drop due to the lower density in the shocked-wind region.  

We define the shocked-ambient-medium crossing time as the time required for the shocked-ambient-medium part of the outflow to cross a given fixed radius in space:
$t_{csa}=t_2-t_1$, where we find $t_2$ and $t_1$ by solving
\begin{equation}
    R_2(t_1) = A_E t_1^{\beta_E} = \lambda_c A_E t_2^{\beta_E} =R_c(t_2)
\end{equation}
which yields $ t_2 = t_1\lambda_c^{-1/\beta_E} $.
For $\alpha=0.8$, $\beta=5/7$ and $\lambda_c=0.854$ so $t_2=1.25t_1$ and $t_{csa}=0.25t_1$.
This is the length of time that preexisting gas which does not experience significant ram pressure acceleration will spend in the shocked ambient medium region of the outflow after entering the outflow at time $t_1$.
For our model outflow at age $t_2=1450$~yr, $t_{csa}=360$~yr.  
 
We have estimated the terminal velocities of gas clouds after their passage through the shocked ambient medium region of our model for the J0300 outflow.
We launch clouds at $r=R_2(t)$ and $v_{cl}=0$ at several different times $t$ and in each case recalculate $a_{ram}$, $v_{cl}$, and $r$ in timesteps of one year until we reach a time $t'$ when $r(t')=R_c(t')$.
We thus account for the motion of the clouds in radius due to their acceleration in that region but we neglect acceleration in the shocked wind region.\footnote{Acceleration in the shocked wind region may be significant for clouds swept up at early times, when the density in that region is highest. The timescale for evaporation in that region is $t>3000$~yr for $r_{cl}>0.0015$ pc (Eq.\ 6 of \citealt{fqm12}).}
We find that gas clouds of fixed size and overdensity swept up at the smallest radii generally end up at smaller velocities.
At earlier times, the ram pressure acceleration is larger but the time spent undergoing acceleration is much less, so the final cloud velocity is lower. 

For $r_{cl}=0.005$~pc and $\chi=900$, we obtain a velocity range from 700~\kms\ for gas swept up at $t=t_f=25$ yr to $2000\pm 400$~\kms\ for gas swept up at $t\geq 200$~yr; note that $t_{\rm KH}=670$~yr for such clouds.
For even smaller clouds with $r_{cl}=0.0015$~pc but still $\chi=900$, we obtain a velocity range from 1700~\kms\ for gas swept up at $t=25$ yr to $4000\pm 200$~\kms\ for gas swept up at $t\geq 200$~yr; note that $t_{\rm KH}=200$~yr for such clouds.  
For larger and less overdense clouds with $r_{cl}=0.01$~pc and $\chi=90$, we obtain a velocity range from 3000~\kms\ for gas swept up at $t=25$ yr to $4900-4200$~\kms\ for gas swept up at $t\geq 200$~yr; note that $t_{\rm KH}=320$~yr for such clouds.
For those parameters, clouds swept up earlier have higher terminal velocities due to being accelerated by higher-velocity shocked ISM when the outflow was younger.
Such clouds swept up at $t=800$~yr only pass through the contact discontinuity at $t\simeq 1430$~yr, meaning that such gas clouds from $\sim$70\% of the volume swept up by the outflow would be located between $R_c=0.854R_2$ and $R_2$ when the outflow is 1450~yr old.
Such clouds will survive for a time $t_{\rm KH}$ before being disrupted, but cloudlets produced by the cloud's disruption can survive long after that if they are in pressure equilibrium with the surrounding gas.

In summary, the velocity range seen in \caii\ in J0300 is potentially consistent with the velocities of cloudlets formed out of some swept-up clouds from the ISM.  
The terminal velocity, current velocity, and current radius of a cloudlet will depend on its initial size, initial overdensity, and the time when its parent cloud was swept up.  Thus, a one-to-one relation between observed velocity and current radius is not expected.
It is also worth emphasizing that not all clouds swept up by a quasar outflow will produce cool cloudlets moving at high velocity, nor will all the gas in clouds that do produce cloudlets end up in them.

\section{Conclusions} \label{sec:conclude}

In this paper we have studied models of energy-conserving outflows around quasars and their observational implications for understanding certain low-ionization absorption troughs seen in BAL quasars.
The main points of this paper are the following:

\begin{enumerate}[label=(\roman*)]

\item We presented equations for shock bubble radii, velocities, densities, and other potentially observable quantities for self-similar energy-conserving outflows into environments with ISM density profile slopes $\alpha \neq 0$ (\S~\ref{sec:mod}). For the first time in the literature to our knowledge, we include an expression for the shocked wind gas density at $R_1 < r < R_c$ (\S~\ref{sec:rho}), expressions for the deceleration of the outflowing gas (\S~\ref{sec:accel}), and expressions for the gas column density along the line of sight (\S~\ref{sec:Nv}).

\item We presented plots of the variations of these shock bubble quantities in space and time for our default parameter values and for a range of input velocities, ambient ISM densities and $\alpha$ values, and AGN luminosities (\S~\ref{sec:vary}).
For a model in which low-ionization absorbing gas is assumed to share the velocity and deceleration of the shocked ISM gas, the expected decelerations have magnitudes as large as $|a|=2$~\kmsyr.

\item We compared new and previous high-resolution spectroscopic observations of the quasar SDSS J030000.56+004828.0 (J0300) and placed a $3\sigma$ rest-frame limit of $|a|<0.1$ \kmsyr\ or $<3 \times 10^{-4}$ \cmss\ on the average bulk deceleration or acceleration of its \caii\ outflow over 9.65 rest-frame years. This is the tightest limit on velocity changes in a quasar outflow reported to date.

\item We discussed a key limitation of our model in matching the \caii\ outflow in J0300: the prediction of too small a range of outflow velocities in the shocked ISM (\S~\ref{sec:vrange}).
Our model predicts $v_{max}\leq\frac{4}{3}v_{min}$, but we observe $v_{max}\gtrsim 2v_{min}$.
A larger velocity range could arise if at least some absorption occurs at velocities matching those of the shocked wind gas, or arises in swept-up preexisting clouds that are being (or have been) accelerated from rest.
It is also true that 3-D hydrodynamic effects (e.g., \citealt{cb23}) may yield a velocity outflow range distinct from that predicted by our simple 1-D analysis.

\item We searched for parameter combinations for our model that could match the physical conditions for the J0300 outflow inferred by \citet{felobal1}. 
We found parameter choices for a comparison model that simultaneously match most of the observed and inferred properties of the J0300 outflow (Table \ref{t_j0300params}), but only a small subset of parameter choices matching some constraints matched the rest. 
Although this reinforced our conclusion that our initial model of \caii\ absorption sharing the velocity and deceleration of the shocked ISM cannot explain the J0300 outflow, our comparison model remains useful for exploring possible origins for the \caii\ absorption (\S~\ref{sec:compare-Choi}).

\item We investigated the possibility that ram-pressure acceleration of preexisting ISM clouds could explain the observations of J0300 (\S~\ref{sec:ismaccel}).
We argued that a combination of lateral expansion and acceleration of cloud fragments (cloudlets) could result in cool gas reaching speeds of thousands of ${\rm km~s}^{-1}$.
However, we found that the clouds in J0300 could not have been accelerated to their observed speeds (as opposed to forming out of gas already moving at those speeds) at their inferred densities and distance. Any such acceleration must have occurred at lower densities followed by cooling and compression. 
We noted that sufficiently dense gas clouds can be effectively coasting even when surrounded by gas moving at larger relative velocities (\S~\ref{sec:compare-accel}).

\item We estimated the acceleration of clouds in the shocked ambient medium region of a self-similar shock bubble (\S~\ref{sec:accel-csa}).
We concluded that the velocity range seen in \caii\ in J0300 is potentially consistent with the velocities of cloudlets formed out of some swept-up clouds from the ISM.  
We found that the terminal and current velocity and current radius of a cloudlet will depend on its initial size and overdensity and the time when its parent cloud was swept up; thus, swept-up clouds may not have monotonically declining velocities with distance from the ionizing source. 

\end{enumerate}

\subsection{Future directions}\label{sec:future}

Our model of the low-ionization absorbing gas in BAL outflows sharing the velocity and deceleration of shocked ISM gas, with higher-velocity gas located closer to the quasar, cannot match all parameters of the J0300 outflow but it may still match other BAL quasar outflows. 
Long-term high-resolution spectroscopy of FeLoBAL quasars with narrow absorption features can test this and other models for absorbing gas acceleration/deceleration, but to date such data have been obtained only for J0300 and Q0059$-$2735 \citep{XuQ0059}.

In addition to considering clouds forming at the velocities of swept-up gas, explaining the range of velocities of low-ionization absorption in BAL quasars requires considering the possible acceleration of preexisting gas clouds.
More detailed simulations similar to those of \citet{2020MNRAS.491.4325Z} would be helpful in this regard.

For J0300 specifically, a more refined picture of the outflow may result from refining the {\sl SimBAL} modeling of \citet{felobal1} to better match the distinct inferred physical conditions and observed velocities for \feii* and \caii\ and other ions in its high-resolution spectrum, including allowing a density jump at the hydrogen ionization front \citep{2014ApJ...788..123L} 
which was suggested in that reference to arise in swept-up ISM clouds.
The \feii* absorption seen over the wide velocity range of $4000<v<10600$~\kms\ in this object may arise in low-density gas from disrupted ISM clouds still being accelerated to match the velocity of the surrounding shocked or unshocked quasar wind. 
If such an explanation proves untenable, a model in which the quasar wind initially consists of low-ionization gas accelerated due to radiation pressure on dust may be preferred (e.g., \citealt{Naddaf23}).  

Finally, note that if acceleration of preexisting clouds is significant in low-ionization quasar outflows, then we expect observable accelerations only at times $t \lesssim t_{drag}$ after the clouds enter the shock bubble, after which the clouds will be coasting.
Thus, if clouds survive for a long time in their coasting phase, then the fraction of low-ionization outflows with observable acceleration might be small even if such acceleration happens in all of them.

\section{Acknowledgments}

We thank J.\ Chu, J.\ Roediger, and K.\ Chiboucas at Gemini, C.\ Kielty and K.\ Venn for GRACES advice, and M.\ Singha and H.\ Choi for discussion.
PH, EW and CM acknowledge support from the Natural Sciences and Engineering Research Council of Canada (NSERC), funding reference numbers 2017-05983 and 2023-05068.
WNB acknowledges support from NSF grant AST-2106990.

Based on observations obtained through the Gemini Remote Access to CFHT ESPaDOnS Spectrograph (GRACES). ESPaDOnS is located at the Canada-France-Hawaii Telescope (CFHT), which is operated by the National Research Council of Canada, the Institut National des Sciences de l’Univers of the Centre National de la Recherche Scientifique of France, and the University of Hawai’i. ESPaDOnS is a collaborative project funded by France (CNRS, MENESR, OMP, LATT), Canada (NSERC), CFHT and ESA. ESPaDOnS was remotely controlled from the international Gemini Observatory, a program of NSF’s NOIRLab, which is managed by the Association of Universities for Research in Astronomy (AURA) under a cooperative agreement with the National Science Foundation on behalf of the Gemini partnership: the National Science Foundation (United States), the National Research Council (Canada), Agencia Nacional de Investigaci\'{o}n y Desarrollo (Chile), Ministerio de Ciencia, Tecnolog\'{i}a e Innovaci\'{o}n (Argentina), Minist\'{e}rio da Ci\^{e}ncia, Tecnologia, Inova\c{c}\~{o}es e Comunica\c{c}\~{o}es (Brazil), and Korea Astronomy and Space Science Institute (Republic of Korea).
Based also on data obtained from the ESO Science Archive Facility under request number 550148 and on observations collected at the European Southern Observatory under ESO programme 267.A-5698.

This work was enabled by observations made using the CFHT and the Gemini North telescope, both of which are located within the Maunakea Science Reserve and adjacent to the summit of Maunakea. We are grateful for the privilege of observing the Universe from a place that is unique in both its astronomical quality and its cultural significance.

\section{Data Availability}

The data underlying this article will be shared on reasonable request to the corresponding author.

\bibliographystyle{mnras}
\bibliography{pathall}

\begin{thebibliography}{}
\makeatletter
\relax
\def\mn@urlcharsother{\let\do\@makeother \do\$\do\&\do\#\do\^\do\_\do\%\do\~}
\def\mn@doi{\begingroup\mn@urlcharsother \@ifnextchar [ {\mn@doi@} {\mn@doi@[]}}
\def\mn@doi@[#1]#2{\def\@tempa{#1}\ifx\@tempa\@empty \href {http://dx.doi.org/#2} {doi:#2}\else \href {http://dx.doi.org/#2} {#1}\fi \endgroup}
\def\mn@eprint#1#2{\mn@eprint@#1:#2::\@nil}
\def\mn@eprint@arXiv#1{\href {http://arxiv.org/abs/#1} {{\tt arXiv:#1}}}
\def\mn@eprint@dblp#1{\href {http://dblp.uni-trier.de/rec/bibtex/#1.xml} {dblp:#1}}
\def\mn@eprint@#1:#2:#3:#4\@nil{\def\@tempa {#1}\def\@tempb {#2}\def\@tempc {#3}\ifx \@tempc \@empty \let \@tempc \@tempb \let \@tempb \@tempa \fi \ifx \@tempb \@empty \def\@tempb {arXiv}\fi \@ifundefined {mn@eprint@\@tempb}{\@tempb:\@tempc}{\expandafter \expandafter \csname mn@eprint@\@tempb\endcsname \expandafter{\@tempc}}}

\bibitem[\protect\citeauthoryear{{Allen}, {Hewett}, {Maddox}, {Richards}  \& {Belokurov}}{{Allen} et~al.}{2011}]{allenbal}
{Allen} J.~T.,  {Hewett} P.~C.,  {Maddox} N.,  {Richards} G.~T.,   {Belokurov} V.,  2011, \mn@doi [MNRAS] {10.1111/j.1365-2966.2010.17489.x}, \href {http://ui.adsabs.harvard.edu/abs/2011MNRAS.410..860A} {410, 860}

\bibitem[\protect\citeauthoryear{{Appleton} et~al.,}{{Appleton} et~al.}{2023}]{Appleton+23}
{Appleton} P.~N.,  et~al., 2023, \mn@doi [\apj] {10.3847/1538-4357/accc2a}, \href {https://ui.adsabs.harvard.edu/abs/2023ApJ...951..104A} {951, 104}

\bibitem[\protect\citeauthoryear{{Arav}, {Korista}, {de Kool}, {Junkkarinen}  \& {Begelman}}{{Arav} et~al.}{1999}]{aea99}
{Arav} N.,  {Korista} K.~T.,  {de Kool} M.,  {Junkkarinen} V.~T.,   {Begelman} M.~C.,  1999, ApJ, \href {http://ui.adsabs.harvard.edu/abs/1999ApJ...516...27A} {516, 27}

\bibitem[\protect\citeauthoryear{{Arav}, {Liu}, {Xu}, {Stidham}, {Benn}  \& {Chamberlain}}{{Arav} et~al.}{2018}]{2018ApJ...857...60A}
{Arav} N.,  {Liu} G.,  {Xu} X.,  {Stidham} J.,  {Benn} C.,   {Chamberlain} C.,  2018, \mn@doi [\apj] {10.3847/1538-4357/aab494}, \href {https://ui.adsabs.harvard.edu/#abs/2018ApJ...857...60A} {857, 60}

\bibitem[\protect\citeauthoryear{{Asplund}, {Grevesse}, {Sauval}  \& {Scott}}{{Asplund} et~al.}{2009}]{2009ARA&A..47..481A}
{Asplund} M.,  {Grevesse} N.,  {Sauval} A.~J.,   {Scott} P.,  2009, \mn@doi [\araa] {10.1146/annurev.astro.46.060407.145222}, \href {http://ui.adsabs.harvard.edu/abs/2009ARA%26A..47..481A} {47, 481}

\bibitem[\protect\citeauthoryear{{Bautista} \& {Dunn}}{{Bautista} \& {Dunn}}{2010}]{2010ApJ...717L..98Bautista}
{Bautista} M.~A.,  {Dunn} J.~P.,  2010, \mn@doi [\apjl] {10.1088/2041-8205/717/2/L98}, \href {https://ui.adsabs.harvard.edu/abs/2010ApJ...717L..98B} {717, L98}

\bibitem[\protect\citeauthoryear{{Blackburne}, {Kochanek}, {Chen}, {Dai}  \& {Chartas}}{{Blackburne} et~al.}{2015}]{2015ApJ...798...95B}
{Blackburne} J.~A.,  {Kochanek} C.~S.,  {Chen} B.,  {Dai} X.,   {Chartas} G.,  2015, \mn@doi [\apj] {10.1088/0004-637X/798/2/95}, \href {http://ui.adsabs.harvard.edu/abs/2015ApJ...798...95B} {798, 95}

\bibitem[\protect\citeauthoryear{{Cavaliere}, {Lapi}  \& {Menci}}{{Cavaliere} et~al.}{2002}]{clm02}
{Cavaliere} A.,  {Lapi} A.,   {Menci} N.,  2002, \mn@doi [\apjl] {10.1086/345890}, \href {https://ui.adsabs.harvard.edu/abs/2002ApJ...581L...1C} {581, L1}

\bibitem[\protect\citeauthoryear{Chao \& Wiskerchen}{Chao \& Wiskerchen}{1974}]{ChaoWiskerchen1974}
Chao J.~K.,  Wiskerchen M.~J.,  1974, \mn@doi [Journal of Geophysical Research (1896-1977)] {https://doi.org/10.1029/JA079i031p04769}, 79, 4769

\bibitem[\protect\citeauthoryear{{Chen} \& {Oh}}{{Chen} \& {Oh}}{2023}]{ChenOh23}
{Chen} Z.,  {Oh} S.~P.,  2023, arXiv e-prints, \href {https://ui.adsabs.harvard.edu/abs/2023arXiv231104275C} {p. arXiv:2311.04275}

\bibitem[\protect\citeauthoryear{{Chene} et~al.,}{{Chene} et~al.}{2014}]{GRACES}
{Chene} A.-N.,  et~al., 2014, SPIE Conference Series.
SPIE, p. 915147, \mn@doi{10.1117/12.2057417}

\bibitem[\protect\citeauthoryear{{Choi}, {Leighly}, {Terndrup}, {Dabbieri}, {Gallagher}  \& {Richards}}{{Choi} et~al.}{2022}]{felobal1}
{Choi} H.,  {Leighly} K.~M.,  {Terndrup} D.~M.,  {Dabbieri} C.,  {Gallagher} S.~C.,   {Richards} G.~T.,  2022, \mn@doi [\apj] {10.3847/1538-4357/ac61d9}, \href {https://ui.adsabs.harvard.edu/abs/2022ApJ...937...74C} {937, 74}

\bibitem[\protect\citeauthoryear{{Clavijo-Boh{\'o}rquez}, {de Gouveia Dal Pino}  \& {Melioli}}{{Clavijo-Boh{\'o}rquez} et~al.}{2023}]{cb23}
{Clavijo-Boh{\'o}rquez} W.~E.,  {de Gouveia Dal Pino} E.~M.,   {Melioli} C.,  2023, \mn@doi [arXiv:2306.11494] {10.48550/arXiv.2306.11494}, \href {https://ui.adsabs.harvard.edu/abs/2023arXiv230611494C} {}

\bibitem[\protect\citeauthoryear{{Di Matteo}, {Springel}  \& {Hernquist}}{{Di Matteo} et~al.}{2005}]{2005Natur.433..604D}
{Di Matteo} T.,  {Springel} V.,   {Hernquist} L.,  2005, \mn@doi [\nat] {10.1038/nature03335}, \href {https://ui.adsabs.harvard.edu/abs/2005Natur.433..604D} {433, 604}

\bibitem[\protect\citeauthoryear{{Donati}, {Semel}, {Carter}, {Rees}  \& {Collier Cameron}}{{Donati} et~al.}{1997}]{1997MNRAS.291..658D}
{Donati} J.~F.,  {Semel} M.,  {Carter} B.~D.,  {Rees} D.~E.,   {Collier Cameron} A.,  1997, \mn@doi [\mnras] {10.1093/mnras/291.4.658}, \href {https://ui.adsabs.harvard.edu/abs/1997MNRAS.291..658D} {291, 658}

\bibitem[\protect\citeauthoryear{{Donati}, {Catala}, {Landstreet}  \& {Petit}}{{Donati} et~al.}{2006}]{2006ASPC..358..362D}
{Donati} J.~F.,  {Catala} C.,  {Landstreet} J.~D.,   {Petit} P.,  2006, {ESPaDOnS: The New Generation Stellar Spectro-Polarimeter. Performances and First Results}.
San Francisco: ASP, p.~362

\bibitem[\protect\citeauthoryear{{Dyda}, {Davis}  \& {Proga}}{{Dyda} et~al.}{2023}]{dydadavisproga23}
{Dyda} S.,  {Davis} S.~W.,   {Proga} D.,  2023, \mn@doi [arXiv e-prints] {10.48550/arXiv.2310.18557}, \href {https://ui.adsabs.harvard.edu/abs/2023arXiv231018557D} {p. arXiv:2310.18557}

\bibitem[\protect\citeauthoryear{{Emmering}, {Blandford}  \& {Shlosman}}{{Emmering} et~al.}{1992}]{ebs92}
{Emmering} R.~T.,  {Blandford} R.~D.,   {Shlosman} I.,  1992, ApJ, 385, 460

\bibitem[\protect\citeauthoryear{{Falle}}{{Falle}}{1975a}]{1975A&A....43..323F}
{Falle} S.~A.~E.~G.,  1975a, \aap, \href {https://ui.adsabs.harvard.edu/abs/1975A&A....43..323F} {43, 323}

\bibitem[\protect\citeauthoryear{{Falle}}{{Falle}}{1975b}]{1975MNRAS.172...55F}
{Falle} S.~A.~E.~G.,  1975b, \mn@doi [\mnras] {10.1093/mnras/172.1.55}, \href {https://ui.adsabs.harvard.edu/abs/1975MNRAS.172...55F} {172, 55}

\bibitem[\protect\citeauthoryear{{Faucher-Gigu{\`e}re} \& {Quataert}}{{Faucher-Gigu{\`e}re} \& {Quataert}}{2012}]{fq12}
{Faucher-Gigu{\`e}re} C.-A.,  {Quataert} E.,  2012, \mn@doi [\mnras] {10.1111/j.1365-2966.2012.21512.x}, \href {http://ui.adsabs.harvard.edu/abs/2012MNRAS.425..605F} {425, 605 (FGQ12)}

\bibitem[\protect\citeauthoryear{{Faucher-Gigu{\`e}re}, {Quataert}  \& {Murray}}{{Faucher-Gigu{\`e}re} et~al.}{2012}]{fqm12}
{Faucher-Gigu{\`e}re} C.-A.,  {Quataert} E.,   {Murray} N.,  2012, \mn@doi [\mnras] {10.1111/j.1365-2966.2011.20120.x}, \href {http://ui.adsabs.harvard.edu/abs/2012MNRAS.420.1347F} {420, 1347}

\bibitem[\protect\citeauthoryear{{Ferrara} \& {Scannapieco}}{{Ferrara} \& {Scannapieco}}{2016}]{2016ApJ...833...46Ferrara}
{Ferrara} A.,  {Scannapieco} E.,  2016, \mn@doi [\apj] {10.3847/1538-4357/833/1/46}, \href {https://ui.adsabs.harvard.edu/abs/2016ApJ...833...46F} {833, 46}

\bibitem[\protect\citeauthoryear{{Ferri{\`e}re}}{{Ferri{\`e}re}}{2001}]{f01}
{Ferri{\`e}re} K.~M.,  2001, Reviews of Modern Physics, \href {http://ui.adsabs.harvard.edu/abs/2001RvMP...73.1031F} {73, 1031}

\bibitem[\protect\citeauthoryear{{Gabel} et~al.,}{{Gabel} et~al.}{2003}]{gea03}
{Gabel} J.~R.,  et~al., 2003, \mn@doi [ApJ] {10.1086/377342}, \href {http://ui.adsabs.harvard.edu/cgi-bin/nph-bib_query?bibcode=2003ApJ...595..120G&db_key=AST} {595, 120}

\bibitem[\protect\citeauthoryear{{Giustini} \& {Proga}}{{Giustini} \& {Proga}}{2019}]{2019A&A...630A..94G}
{Giustini} M.,  {Proga} D.,  2019, \mn@doi [\aap] {10.1051/0004-6361/201833810}, \href {https://ui.adsabs.harvard.edu/abs/2019A&A...630A..94G} {630, A94}

\bibitem[\protect\citeauthoryear{{Grier} et~al.,}{{Grier} et~al.}{2016}]{2016ApJ...824..130G}
{Grier} C.~J.,  et~al., 2016, \mn@doi [\apj] {10.3847/0004-637X/824/2/130}, \href {http://ui.adsabs.harvard.edu/abs/2016ApJ...824..130G} {824, 130}

\bibitem[\protect\citeauthoryear{{Gronke} \& {Oh}}{{Gronke} \& {Oh}}{2018}]{2018MNRAS.480L.111Gronke}
{Gronke} M.,  {Oh} S.~P.,  2018, \mn@doi [\mnras] {10.1093/mnrasl/sly131}, \href {https://ui.adsabs.harvard.edu/abs/2018MNRAS.480L.111G} {480, L111}

\bibitem[\protect\citeauthoryear{{Gronke} \& {Oh}}{{Gronke} \& {Oh}}{2020}]{2020MNRAS.492.1970G}
{Gronke} M.,  {Oh} S.~P.,  2020, \mn@doi [\mnras] {10.1093/mnras/stz3332}, \href {https://ui.adsabs.harvard.edu/abs/2020MNRAS.492.1970G} {492, 1970}

\bibitem[\protect\citeauthoryear{{Guillard}, {Boulanger}, {Pineau Des For{\^e}ts}  \& {Appleton}}{{Guillard} et~al.}{2009}]{Guillard+2009}
{Guillard} P.,  {Boulanger} F.,  {Pineau Des For{\^e}ts} G.,   {Appleton} P.~N.,  2009, \mn@doi [\aap] {10.1051/0004-6361/200811263}, \href {https://ui.adsabs.harvard.edu/abs/2009A&A...502..515G} {502, 515}

\bibitem[\protect\citeauthoryear{{Hall} \& {Hutsem{\'e}kers}}{{Hall} \& {Hutsem{\'e}kers}}{2004}]{pbhmhd}
{Hall} P.~B.,  {Hutsem{\'e}kers} D.,  2004, in {Richards} G.~T.,  {Hall} P.~B.,  eds, AGN Physics with the Sloan Digital Sky Survey. p.~227

\bibitem[\protect\citeauthoryear{{Hall} et~al.,}{{Hall} et~al.}{2002}]{sdss123}
{Hall} P.~B.,  et~al., 2002, ApJS, \href {http://ui.adsabs.harvard.edu/abs/2002ApJS..141..267H} {141, 267}

\bibitem[\protect\citeauthoryear{{Hall}, {Hutsem{\'e}kers}, {Anderson}, {Brinkmann}, {Fan}, {Schneider}  \& {York}}{{Hall} et~al.}{2003}]{sb2}
{Hall} P.~B.,  {Hutsem{\'e}kers} D.,  {Anderson} S.~F.,  {Brinkmann} J.,  {Fan} X.,  {Schneider} D.~P.,   {York} D.~G.,  2003, \mn@doi [ApJ] {10.1086/376409}, \href {http://ui.adsabs.harvard.edu/cgi-bin/nph-bib_query?bibcode=2003ApJ...593..189H&db_key=AST} {593, 189}

\bibitem[\protect\citeauthoryear{{Hall}, {Sadavoy}, {Hutsemekers}, {Everett}  \& {Rafiee}}{{Hall} et~al.}{2007}]{q0242}
{Hall} P.~B.,  {Sadavoy} S.~I.,  {Hutsemekers} D.,  {Everett} J.~E.,   {Rafiee} A.,  2007, \mn@doi [ApJ] {10.1086/519273}, \href {http://ui.adsabs.harvard.edu/abs/2007ApJ...665..174H} {665, 174}

\bibitem[\protect\citeauthoryear{{Hall}, {Noordeh}, {Chajet}, {Weiss}  \& {Nixon}}{{Hall} et~al.}{2014}]{ad}
{Hall} P.~B.,  {Noordeh} E.~S.,  {Chajet} L.~S.,  {Weiss} E.,   {Nixon} C.~J.,  2014, \mn@doi [\mnras] {10.1093/mnras/stu890}, \href {http://ui.adsabs.harvard.edu/abs/2014MNRAS.442.1090H} {442, 1090}

\bibitem[\protect\citeauthoryear{{Hamann}, {Herbst}, {Paris}  \& {Capellupo}}{{Hamann} et~al.}{2019}]{2019MNRAS.483.1808Hamann}
{Hamann} F.,  {Herbst} H.,  {Paris} I.,   {Capellupo} D.,  2019, \mn@doi [\mnras] {10.1093/mnras/sty2900}, \href {https://ui.adsabs.harvard.edu/abs/2019MNRAS.483.1808H} {483, 1808}

\bibitem[\protect\citeauthoryear{{Hartwig}, {Volonteri}  \& {Dashyan}}{{Hartwig} et~al.}{2018}]{2018MNRAS.476.2288Hartwig}
{Hartwig} T.,  {Volonteri} M.,   {Dashyan} G.,  2018, \mn@doi [\mnras] {10.1093/mnras/sty229}, \href {https://ui.adsabs.harvard.edu/abs/2018MNRAS.476.2288H} {476, 2288}

\bibitem[\protect\citeauthoryear{{He} et~al.,}{{He} et~al.}{2022}]{2022SciA....8.3291H}
{He} Z.,  et~al., 2022, \mn@doi [Science Advances] {10.1126/sciadv.abk3291}, \href {https://ui.adsabs.harvard.edu/abs/2022SciA....8.3291H} {8, eabk3291}

\bibitem[\protect\citeauthoryear{{Hopkins} \& {Elvis}}{{Hopkins} \& {Elvis}}{2010}]{he10}
{Hopkins} P.~F.,  {Elvis} M.,  2010, \mn@doi [MNRAS] {10.1111/j.1365-2966.2009.15643.x}, \href {http://ui.adsabs.harvard.edu/abs/2010MNRAS.401....7H} {401, 7}

\bibitem[\protect\citeauthoryear{{Hopkins} \& {Quataert}}{{Hopkins} \& {Quataert}}{2010}]{hq10}
{Hopkins} P.~F.,  {Quataert} E.,  2010, \mn@doi [MNRAS] {10.1111/j.1365-2966.2010.17064.x}, \href {http://ui.adsabs.harvard.edu/abs/2010MNRAS.407.1529H} {407, 1529}

\bibitem[\protect\citeauthoryear{{Hopkins}, {Torrey}, {Faucher-Gigu{\`e}re}, {Quataert}  \& {Murray}}{{Hopkins} et~al.}{2016}]{hopkins15}
{Hopkins} P.~F.,  {Torrey} P.,  {Faucher-Gigu{\`e}re} C.-A.,  {Quataert} E.,   {Murray} N.,  2016, \mn@doi [\mnras] {10.1093/mnras/stw289}, \href {http://ui.adsabs.harvard.edu/abs/2016MNRAS.458..816H} {458, 816}

\bibitem[\protect\citeauthoryear{{Joshi}, {Chand}, {Srianand}  \& {Majumdar}}{{Joshi} et~al.}{2014}]{2014MNRAS.442..862J}
{Joshi} R.,  {Chand} H.,  {Srianand} R.,   {Majumdar} J.,  2014, \mn@doi [\mnras] {10.1093/mnras/stu840}, \href {https://ui.adsabs.harvard.edu/abs/2014MNRAS.442..862J} {442, 862}

\bibitem[\protect\citeauthoryear{{Joshi}, {Srianand}, {Chand}, {Wu}, {Noterdaeme}, {Petitjean}  \& {Ho}}{{Joshi} et~al.}{2019}]{2019ApJ...871...43J}
{Joshi} R.,  {Srianand} R.,  {Chand} H.,  {Wu} X.-B.,  {Noterdaeme} P.,  {Petitjean} P.,   {Ho} L.~C.,  2019, \mn@doi [\apj] {10.3847/1538-4357/aaf500}, \href {https://ui.adsabs.harvard.edu/abs/2019ApJ...871...43J} {871, 43}

\bibitem[\protect\citeauthoryear{{Joung}, {Bryan}  \& {Putman}}{{Joung} et~al.}{2012}]{2012ApJ...745..148Joung}
{Joung} M.~R.,  {Bryan} G.~L.,   {Putman} M.~E.,  2012, \mn@doi [\apj] {10.1088/0004-637X/745/2/148}, \href {https://ui.adsabs.harvard.edu/abs/2012ApJ...745..148J} {745, 148}

\bibitem[\protect\citeauthoryear{{King} \& {Pounds}}{{King} \& {Pounds}}{2015}]{2015ARA&A..53..115K}
{King} A.,  {Pounds} K.,  2015, \mn@doi [\araa] {10.1146/annurev-astro-082214-122316}, \href {https://ui.adsabs.harvard.edu/abs/2015ARA&A..53..115K} {53, 115}

\bibitem[\protect\citeauthoryear{{King}, {Zubovas}  \& {Power}}{{King} et~al.}{2011}]{2011MNRAS.415L...6K}
{King} A.~R.,  {Zubovas} K.,   {Power} C.,  2011, \mn@doi [\mnras] {10.1111/j.1745-3933.2011.01067.x}, \href {https://ui.adsabs.harvard.edu/abs/2011MNRAS.415L...6K} {415, L6}

\bibitem[\protect\citeauthoryear{{Koo} \& {McKee}}{{Koo} \& {McKee}}{1992}]{1992ApJ...388..103K}
{Koo} B.-C.,  {McKee} C.~F.,  1992, \mn@doi [\apj] {10.1086/171133}, \href {https://ui.adsabs.harvard.edu/abs/1992ApJ...388..103K} {388, 103 (KM92)}

\bibitem[\protect\citeauthoryear{{Kriss} et~al.,}{{Kriss} et~al.}{2019}]{2019A&A...621A..12Kriss}
{Kriss} G.~A.,  et~al., 2019, \mn@doi [\aap] {10.1051/0004-6361/201834326}, \href {https://ui.adsabs.harvard.edu/abs/2019A&A...621A..12K} {621, A12}

\bibitem[\protect\citeauthoryear{{Larson}, {Tinsley}  \& {Caldwell}}{{Larson} et~al.}{1980}]{1980ApJ...237..692L}
{Larson} R.~B.,  {Tinsley} B.~M.,   {Caldwell} C.~N.,  1980, \mn@doi [\apj] {10.1086/157917}, \href {https://ui.adsabs.harvard.edu/abs/1980ApJ...237..692L} {237, 692}

\bibitem[\protect\citeauthoryear{{Lawther}, {Vestergaard}  \& {Fan}}{{Lawther} et~al.}{2018}]{2018MNRAS.475.3213Lawther}
{Lawther} D.,  {Vestergaard} M.,   {Fan} X.,  2018, \mn@doi [\mnras] {10.1093/mnras/stx3203}, \href {https://ui.adsabs.harvard.edu/abs/2018MNRAS.475.3213L} {475, 3213}

\bibitem[\protect\citeauthoryear{{Leighly}, {Terndrup}, {Baron}, {Lucy}, {Dietrich}  \& {Gallagher}}{{Leighly} et~al.}{2014}]{2014ApJ...788..123L}
{Leighly} K.~M.,  {Terndrup} D.~M.,  {Baron} E.,  {Lucy} A.~B.,  {Dietrich} M.,   {Gallagher} S.~C.,  2014, \mn@doi [\apj] {10.1088/0004-637X/788/2/123}, \href {http://ui.adsabs.harvard.edu/abs/2014ApJ...788..123L} {788, 123}

\bibitem[\protect\citeauthoryear{{Leighly}, {Terndrup}, {Gallagher}, {Richards}  \& {Dietrich}}{{Leighly} et~al.}{2018}]{simbal}
{Leighly} K.~M.,  {Terndrup} D.~M.,  {Gallagher} S.~C.,  {Richards} G.~T.,   {Dietrich} M.,  2018, \mn@doi [\apj] {10.3847/1538-4357/aadee6}, \href {https://ui.adsabs.harvard.edu/abs/2018ApJ...866....7L} {866, 7}

\bibitem[\protect\citeauthoryear{{Martioli}, {Teeple}, {Manset}, {Devost}, {Withington}, {Venne}  \& {Tannock}}{{Martioli} et~al.}{2012}]{OPERA}
{Martioli} E.,  {Teeple} D.,  {Manset} N.,  {Devost} D.,  {Withington} K.,  {Venne} A.,   {Tannock} M.,  2012, SPIE Conference Series.
SPIE, p. 84512B, \mn@doi{10.1117/12.926627}

\bibitem[\protect\citeauthoryear{{Mayer}}{{Mayer}}{2007}]{Mayer2007}
{Mayer} M.,  2007, \mn@doi [\aap] {10.1051/0004-6361:20066245}, \href {https://ui.adsabs.harvard.edu/abs/2007A&A...461..381M} {461, 381}

\bibitem[\protect\citeauthoryear{McCourt, Oh, O'Leary  \& Madigan}{McCourt et~al.}{2017}]{McCourt+17}
McCourt M.,  Oh S.~P.,  O'Leary R.,   Madigan A.-M.,  2017, \mn@doi [MNRAS] {10.1093/mnras/stx2687}, 473, 5407

\bibitem[\protect\citeauthoryear{{Mercedes-Feliz} et~al.,}{{Mercedes-Feliz} et~al.}{2023}]{FIREclumps23}
{Mercedes-Feliz} J.,  et~al., 2023, \mn@doi [arXiv e-prints] {10.48550/arXiv.2310.19863}, \href {https://ui.adsabs.harvard.edu/abs/2023arXiv231019863M} {p. arXiv:2310.19863}

\bibitem[\protect\citeauthoryear{{Misawa}, {Eracleous}, {Charlton}  \& {Kashikawa}}{{Misawa} et~al.}{2019}]{2019ApJ...870...68M}
{Misawa} T.,  {Eracleous} M.,  {Charlton} J.~C.,   {Kashikawa} N.,  2019, \mn@doi [\apj] {10.3847/1538-4357/aaf0fe}, \href {https://ui.adsabs.harvard.edu/\#abs/2019ApJ...870...68M} {870, 68}

\bibitem[\protect\citeauthoryear{{Morganti}}{{Morganti}}{2017}]{Morganti2017}
{Morganti} R.,  2017, \mn@doi [Frontiers in Astronomy and Space Sciences] {10.3389/fspas.2017.00042}, \href {https://ui.adsabs.harvard.edu/abs/2017FrASS...4...42M} {4, 42}

\bibitem[\protect\citeauthoryear{{Morton}}{{Morton}}{1991}]{mor91}
{Morton} D.~C.,  1991, ApJS, \href {http://ui.adsabs.harvard.edu/abs/1991ApJS...77..119M} {77, 119}

\bibitem[\protect\citeauthoryear{{Murphy}, {Kacprzak}, {Savorgnan}  \& {Carswell}}{{Murphy} et~al.}{2019}]{SQUAD1}
{Murphy} M.~T.,  {Kacprzak} G.~G.,  {Savorgnan} G. A.~D.,   {Carswell} R.~F.,  2019, \mn@doi [\mnras] {10.1093/mnras/sty2834}, \href {https://ui.adsabs.harvard.edu/abs/2019MNRAS.482.3458M} {482, 3458}

\bibitem[\protect\citeauthoryear{{Murray} \& {Chiang}}{{Murray} \& {Chiang}}{1997}]{mc97}
{Murray} N.,  {Chiang} J.,  1997, ApJ, 474, 91

\bibitem[\protect\citeauthoryear{{Murray}, {Chiang}, {Grossman}  \& {Voit}}{{Murray} et~al.}{1995}]{mcgv}
{Murray} N.,  {Chiang} J.,  {Grossman} S.~A.,   {Voit} G.~M.,  1995, ApJ, \href {http://ui.adsabs.harvard.edu/abs/1995ApJ...451..498M} {451, 498}

\bibitem[\protect\citeauthoryear{{Naddaf}, {Martinez-Aldama}, {Marziani}, {Panda}, {Sniegowska}  \& {Czerny}}{{Naddaf} et~al.}{2023}]{Naddaf23}
{Naddaf} M.~H.,  {Martinez-Aldama} M.~L.,  {Marziani} P.,  {Panda} S.,  {Sniegowska} M.,   {Czerny} B.,  2023, \mn@doi [\aap] {10.1051/0004-6361/202245698}, \href {https://ui.adsabs.harvard.edu/abs/2023A&A...675A..43N} {675, A43}

\bibitem[\protect\citeauthoryear{{Nguyen}, {Thompson}, {Schneider}  \& {Tarrant}}{{Nguyen} et~al.}{2023}]{Nguyen+23}
{Nguyen} D.~D.,  {Thompson} T.~A.,  {Schneider} E.~E.,   {Tarrant} A.~P.,  2023, \mn@doi [arXiv:2307.11930] {10.48550/arXiv.2307.11930}, \href {https://ui.adsabs.harvard.edu/abs/2023arXiv230711930N} {}

\bibitem[\protect\citeauthoryear{{Peng}, {Maiolino}  \& {Cochrane}}{{Peng} et~al.}{2015}]{Peng+2015}
{Peng} Y.,  {Maiolino} R.,   {Cochrane} R.,  2015, \mn@doi [\nat] {10.1038/nature14439}, \href {https://ui.adsabs.harvard.edu/abs/2015Natur.521..192P} {521, 192}

\bibitem[\protect\citeauthoryear{{Proga} \& {Kallman}}{{Proga} \& {Kallman}}{2004}]{pk04}
{Proga} D.,  {Kallman} T.~R.,  2004, \mn@doi [ApJ] {10.1086/425117}, \href {http://ui.adsabs.harvard.edu/cgi-bin/nph-bib_query?bibcode=2004ApJ...616..688P&db_key=AST} {616, 688}

\bibitem[\protect\citeauthoryear{{Proga} \& {Waters}}{{Proga} \& {Waters}}{2015}]{2015ApJ...804..137P}
{Proga} D.,  {Waters} T.,  2015, \mn@doi [\apj] {10.1088/0004-637X/804/2/137}, \href {http://ui.adsabs.harvard.edu/abs/2015ApJ...804..137P} {804, 137}

\bibitem[\protect\citeauthoryear{{Rafiee} \& {Hall}}{{Rafiee} \& {Hall}}{2011}]{2011ApJS..194...42R}
{Rafiee} A.,  {Hall} P.~B.,  2011, \mn@doi [ApJS] {10.1088/0067-0049/194/2/42}, \href {http://ui.adsabs.harvard.edu/abs/2011ApJS..194...42R} {194, 42}

\bibitem[\protect\citeauthoryear{{Richings} \& {Faucher-Gigu{\`e}re}}{{Richings} \& {Faucher-Gigu{\`e}re}}{2018a}]{2018MNRAS.474.3673Richings}
{Richings} A.~J.,  {Faucher-Gigu{\`e}re} C.-A.,  2018a, \mn@doi [\mnras] {10.1093/mnras/stx3014}, \href {https://ui.adsabs.harvard.edu/abs/2018MNRAS.474.3673R} {474, 3673}

\bibitem[\protect\citeauthoryear{{Richings} \& {Faucher-Gigu{\`e}re}}{{Richings} \& {Faucher-Gigu{\`e}re}}{2018b}]{2018MNRAS.478.3100Richings}
{Richings} A.~J.,  {Faucher-Gigu{\`e}re} C.-A.,  2018b, \mn@doi [\mnras] {10.1093/mnras/sty1285}, \href {https://ui.adsabs.harvard.edu/abs/2018MNRAS.478.3100R} {478, 3100}

\bibitem[\protect\citeauthoryear{{Rogerson}, {Hall}, {Snedden}, {Brotherton}  \& {Anderson}}{{Rogerson} et~al.}{2011}]{2011NewA...16..128R}
{Rogerson} J.~A.,  {Hall} P.~B.,  {Snedden} S.~A.,  {Brotherton} M.~S.,   {Anderson} S.~F.,  2011, \mn@doi [New Astronomy] {10.1016/j.newast.2010.07.002}, \href {http://ui.adsabs.harvard.edu/abs/2011NewA...16..128R} {16, 128}

\bibitem[\protect\citeauthoryear{{Shapiro}}{{Shapiro}}{2005}]{Shapiro2005}
{Shapiro} S.~L.,  2005, \mn@doi [\apj] {10.1086/427065}, \href {https://ui.adsabs.harvard.edu/abs/2005ApJ...620...59S} {620, 59}

\bibitem[\protect\citeauthoryear{{Sharma}, {McCourt}, {Quataert}  \& {Parrish}}{{Sharma} et~al.}{2012}]{2012MNRAS.420.3174Sharma}
{Sharma} P.,  {McCourt} M.,  {Quataert} E.,   {Parrish} I.~J.,  2012, \mn@doi [\mnras] {10.1111/j.1365-2966.2011.20246.x}, \href {https://ui.adsabs.harvard.edu/abs/2012MNRAS.420.3174S} {420, 3174}

\bibitem[\protect\citeauthoryear{{Slone} \& {Netzer}}{{Slone} \& {Netzer}}{2012}]{2012MNRAS.426..656S}
{Slone} O.,  {Netzer} H.,  2012, \mn@doi [\mnras] {10.1111/j.1365-2966.2012.21699.x}, \href {http://ui.adsabs.harvard.edu/abs/2012MNRAS.426..656S} {426, 656}

\bibitem[\protect\citeauthoryear{{Tan}, {Oh}  \& {Gronke}}{{Tan} et~al.}{2023}]{TanOhGronke}
{Tan} B.,  {Oh} S.~P.,   {Gronke} M.,  2023, \mn@doi [\mnras] {10.1093/mnras/stad236}, \href {https://ui.adsabs.harvard.edu/abs/2023MNRAS.520.2571T} {520, 2571}

\bibitem[\protect\citeauthoryear{{Tonry} \& {Davis}}{{Tonry} \& {Davis}}{1979}]{1979AJ.....84.1511T}
{Tonry} J.,  {Davis} M.,  1979, \mn@doi [\aj] {10.1086/112569}, \href {https://ui.adsabs.harvard.edu/abs/1979AJ.....84.1511T} {84, 1511}

\bibitem[\protect\citeauthoryear{{Vayner} et~al.,}{{Vayner} et~al.}{2021}]{2021ApJ...919..122Vayner}
{Vayner} A.,  et~al., 2021, \mn@doi [\apj] {10.3847/1538-4357/ac0f56}, \href {https://ui.adsabs.harvard.edu/abs/2021ApJ...919..122V} {919, 122}

\bibitem[\protect\citeauthoryear{{Villforth}, {Herbst}, {Hamann}, {Hamilton}, {Bertemes}, {Efthymiadou}  \& {Hewlett}}{{Villforth} et~al.}{2019}]{2019MNRAS.483.2441Villforth}
{Villforth} C.,  {Herbst} H.,  {Hamann} F.,  {Hamilton} T.,  {Bertemes} C.,  {Efthymiadou} A.,   {Hewlett} T.,  2019, \mn@doi [\mnras] {10.1093/mnras/sty3271}, \href {https://ui.adsabs.harvard.edu/abs/2019MNRAS.483.2441V} {483, 2441}

\bibitem[\protect\citeauthoryear{{Waters}, {Proga}, {Dannen}  \& {Dyda}}{{Waters} et~al.}{2022}]{2022ApJ...931..134Waters}
{Waters} T.,  {Proga} D.,  {Dannen} R.,   {Dyda} S.,  2022, \mn@doi [\apj] {10.3847/1538-4357/ac6612}, \href {https://ui.adsabs.harvard.edu/abs/2022ApJ...931..134W} {931, 134}

\bibitem[\protect\citeauthoryear{{Weaver}, {McCray}, {Castor}, {Shapiro}  \& {Moore}}{{Weaver} et~al.}{1977}]{1977ApJ...218..377W}
{Weaver} R.,  {McCray} R.,  {Castor} J.,  {Shapiro} P.,   {Moore} R.,  1977, \mn@doi [\apj] {10.1086/155692}, \href {https://ui.adsabs.harvard.edu/abs/1977ApJ...218..377W} {218, 377 (W77)}

\bibitem[\protect\citeauthoryear{Xu, Arav, Miller, Korista  \& Benn}{Xu et~al.}{2021}]{XuQ0059}
Xu X.,  Arav N.,  Miller T.,  Korista K.~T.,   Benn C.,  2021, \mn@doi [MNRAS] {10.1093/mnras/stab1866}, 506, 2725

\bibitem[\protect\citeauthoryear{{Xu} et~al.,}{{Xu} et~al.}{2023}]{Xu+23}
{Xu} X.,  et~al., 2023, \mn@doi [\apj] {10.3847/1538-4357/acbf46}, \href {https://ui.adsabs.harvard.edu/abs/2023ApJ...948...28X} {948, 28}

\bibitem[\protect\citeauthoryear{{Yao} \& {Gan}}{{Yao} \& {Gan}}{2020}]{2020MNRAS.492..444Y}
{Yao} Z.,  {Gan} Z.,  2020, \mn@doi [\mnras] {10.1093/mnras/stz3474}, \href {https://ui.adsabs.harvard.edu/abs/2020MNRAS.492..444Y} {492, 444}

\bibitem[\protect\citeauthoryear{{Zeilig-Hess}, {Levinson}, {Xu}  \& {Arav}}{{Zeilig-Hess} et~al.}{2020}]{2020MNRAS.491.4325Z}
{Zeilig-Hess} M.,  {Levinson} A.,  {Xu} X.,   {Arav} N.,  2020, \mn@doi [\mnras] {10.1093/mnras/stz3352}, \href {https://ui.adsabs.harvard.edu/abs/2020MNRAS.491.4325Z} {491, 4325}

\bibitem[\protect\citeauthoryear{{Zubovas} \& {King}}{{Zubovas} \& {King}}{2012}]{2012ApJ...745L..34Z}
{Zubovas} K.,  {King} A.,  2012, \mn@doi [\apjl] {10.1088/2041-8205/745/2/L34}, \href {https://ui.adsabs.harvard.edu/abs/2012ApJ...745L..34Z} {745, L34}

\bibitem[\protect\citeauthoryear{{Zubovas} \& {King}}{{Zubovas} \& {King}}{2014}]{2014MNRAS.439..400Zubovas}
{Zubovas} K.,  {King} A.~R.,  2014, \mn@doi [\mnras] {10.1093/mnras/stt2472}, \href {https://ui.adsabs.harvard.edu/abs/2014MNRAS.439..400Z} {439, 400}

\bibitem[\protect\citeauthoryear{{di Francesco}, {Evans}, {Caselli}, {Myers}, {Shirley}, {Aikawa}  \& {Tafalla}}{{di Francesco} et~al.}{2007}]{2007prpl.conf...17D}
{di Francesco} J.,  {Evans} II N.~J.,  {Caselli} P.,  {Myers} P.~C.,  {Shirley} Y.,  {Aikawa} Y.,   {Tafalla} M.,  2007, Protostars and Planets V, \href {http://ui.adsabs.harvard.edu/abs/2007prpl.conf...17D} {pp 17--32}

\makeatother
\end{thebibliography}

\end{document}